%% file: main.tex
\documentclass[preprint,review,10pt]{elsarticle}

\input{my_packages}

\input{my_definitions}

\biboptions{sort&compress}

\begin{document}

\begin{frontmatter}



   \title{Modeling of progressive high-cycle fatigue in composite laminates accounting for local stress ratios}

{}



\author[]{P. Hofman}
\author[]{F. P. van der Meer}
\author[]{L. J. Sluys}

%
\affiliation[1]{organization={Delft University of Technology},
             city={Delft},
             country={Netherlands}}


\ead[1]{p.hofman@tudelft.nl}

\begin{abstract}

A numerical framework for simulating progressive failure under high-cycle fatigue loading is validated against experiments of composite quasi-isotropic open-hole laminates. 
Transverse matrix cracking and delamination are modeled with a mixed-mode fatigue cohesive zone model, covering crack initiation and propagation. 
Furthermore, XFEM is used for simulating transverse matrix cracks and splits at arbitrary locations. 
An adaptive cycle jump approach is employed for efficiently simulating high-cycle fatigue while accounting for local stress ratio variations in the presence of thermal residual stresses. The cycle jump scheme is integrated in the XFEM framework, where the local stress ratio is used to determine the insertion of cracks and to propagate fatigue damage.
The fatigue cohesive zone model is based on S-N curves and requires static material properties and only a few fatigue parameters, calibrated on simple fracture testing specimens. 
The simulations demonstrate a good correspondence with experiments in terms of fatigue life and damage evolution. 

\end{abstract}

\begin{graphicalabstract}

\vspace{2cm}
  \hspace{-0.7cm}
    \begin{tikzpicture}[]
 
    \node at (-0.2,-4.5) [rectangle] {\def\svgwidth{1.5\textwidth}\scalebox{0.70}{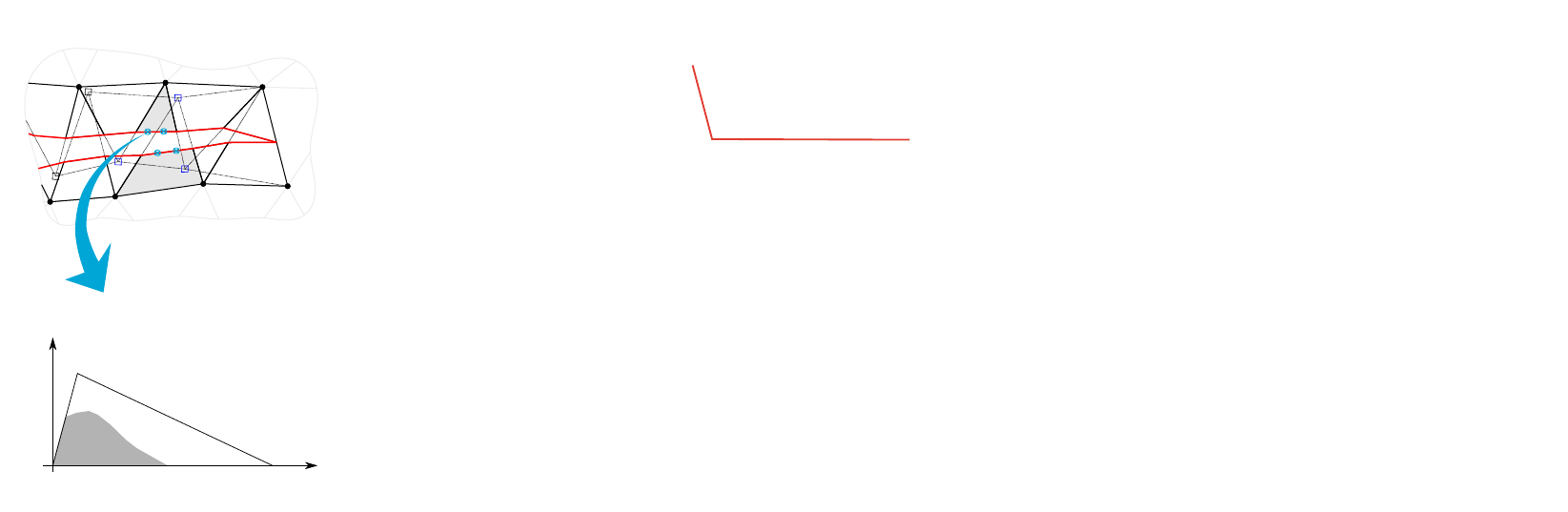}};

    \end{tikzpicture}

\end{graphicalabstract}



\end{frontmatter}



\include{sections/introduction}

\include{sections/methods}

\include{sections/openholes}
\include{sections/conclusion}

\section*{Declaration of competing interest}
The authors declare that they have no known competing financial interests or personal relationships that could have appeared to influence the work reported in this paper.

\section*{Acknowledgement}
This research was carried out as part of the project ENLIGHTEN (project number N21010h) in the framework of the Partnership Program of the Materials innovation institute M2i (www.m2i.nl) and the Netherlands Organization for Scientific Research (www.nwo.nl).




\bibliographystyle{elsarticle-num} 
\bibliography{my_library_manual,my_library}





\end{document}

%% file: my_packages.tex
\usepackage{setspace}
\usepackage[a4paper, portrait, margin=3cm]{geometry}
\usepackage[english]{babel}
\usepackage[utf8]{inputenc}
\usepackage[T1]{fontenc}
\usepackage{helvet}
\usepackage[retainorgcmds]{IEEEtrantools}
\usepackage{etoolbox}
\usepackage{graphicx}
\usepackage{titlesec}
\usepackage{caption}
\usepackage{csquotes}
\usepackage{tabularx} 
\usepackage{booktabs}
\usepackage{xcolor} 
\usepackage[most]{tcolorbox}
\usepackage{tikz}
\usetikzlibrary{3d}
\usetikzlibrary{external}   
\usetikzlibrary{calc, decorations.pathreplacing, arrows.meta}

\usepackage{pgfplots}
\usepgfplotslibrary{}
\pgfplotsset{compat=newest}
\usepgfplotslibrary{fillbetween}

\usepackage{flafter} 
\usepackage[obeyFinal]{todonotes}
\usepackage{subcaption}
\usepackage{stmaryrd}
\usepackage{bm}
\usepackage{amsmath,amssymb, amsthm,mathtools,mathrsfs,stmaryrd,titletoc}
\usepackage{wasysym}
\usepackage[norefs,nocites,ignoreunlbld]{refcheck} 
\usepackage[acronym]{glossaries}
\usepackage{xfrac} 
\usepackage{nicefrac} 
\usepackage{numprint} 
\usepackage{physics}
\usepackage{listings} 
\usepackage{algorithm} 
\usepackage{algpseudocode} 
\usepackage{siunitx} 
\usepackage{threeparttable}
\usepackage{xspace}

\usepackage[colorlinks, citecolor=cyan]{hyperref}
\hypersetup{
	colorlinks=true,
	linkcolor=blue,
	filecolor=magenta,      
	urlcolor=blue,
	citecolor=blue 
}

\usepackage[nameinlink]{cleveref}

 \AtBeginDocument{%
    \crefname{equation}{equation}{equations}%
    \crefname{chapter}{chapter}{chapters}%
    \crefname{section}{section}{sections}%
    \crefname{appendix}{appendix}{appendices}%
    \crefname{enumi}{item}{items}%
    \crefname{footnote}{footnote}{footnotes}%
    \crefname{figure}{figure}{figures}%
    \crefname{table}{table}{tables}%
    \crefname{theorem}{theorem}{theorems}%
    \crefname{lemma}{lemma}{lemmas}%
    \crefname{corollary}{corollary}{corollaries}%
    \crefname{proposition}{proposition}{propositions}%
    \crefname{definition}{definition}{definitions}%
    \crefname{result}{result}{results}%
    \crefname{example}{example}{examples}%
    \crefname{remark}{remark}{remarks}%
    \crefname{note}{note}{notes}%
}

\usepackage{caption}
\graphicspath{ {./images/} }
\graphicspath{{inkscape/}}
\usepackage{fancyhdr}
\usepackage{graphicx}

\fancypagestyle{plain}{
	\fancyhf{}

	\lhead{\color{cyan}\small \textbf{~}\\ \color{black}
	\textit{~}\\ }	
}



%% file: my_definitions.tex
\newcommand{\apriori}{\emph{a priori} }
\newcommand{\Davila}{D\'{a}vila}

\newcommand{\jump}[1]{\ensuremath{\llbracket#1\rrbracket}}
\newcommand{\mc}[1]{\ensuremath{\left<#1\right>}}
\newcommand{\Dam}{\mathcal{D}}
\newcommand{\B}{\mathcal{B}}

\newcommand{\dispjump}{\ensuremath{\llbracket\textbf{u}\rrbracket}}

\newcommand{\Svec}{\ensuremath{\textbf{S}}}
\newcommand{\SvecMin}{\ensuremath{\textbf{S}_{\mathrm{min}}}}
\newcommand{\SvecMax}{\ensuremath{\textbf{S}_{\mathrm{max}}}}

\definecolor{TUDblue}{RGB}{0,166,214} 
\definecolor{Donkerblauw}{RGB}{12,35,64}
\definecolor{Turkoois}{RGB}{0,184,200} 
\definecolor{Blauw}{RGB}{0,118,194} 
\definecolor{Paars}{RGB}{111,29,119} 
\definecolor{Roze}{RGB}{165,0,52} 
\definecolor{Rood}{RGB}{224,60,49} 
\definecolor{Oranje}{RGB}{255,184,28} 
\definecolor{Lichtgroen}{RGB}{108,194,74} 
\definecolor{Donkergroen}{RGB}{0,155,119}

\newcommand{\layup}[2]{\ensuremath{\big[#1\big]_{\mathrm{#2}}}}

\algblockdefx{Start}{End}[1]{\textbf{#1}}{\textbf{end}}

\newtcolorbox[auto counter,crefname={Box}{Boxes}]{MyBox}[2][]{%
  colback=gray!5!white,colframe=blue!75!black,fonttitle=\bfseries,before skip=20pt plus
  2pt,after skip=20pt,
  title=Box~\thetcbcounter: #2, #1}

%% file: inkscape/xfem_v4.pdf_tex
\begingroup%
  \makeatletter%
  \providecommand\color[2][]{%
    \errmessage{(Inkscape) Color is used for the text in Inkscape, but the package 'color.sty' is not loaded}%
    \renewcommand\color[2][]{}%
  }%
  \providecommand\transparent[1]{%
    \errmessage{(Inkscape) Transparency is used (non-zero) for the text in Inkscape, but the package 'transparent.sty' is not loaded}%
    \renewcommand\transparent[1]{}%
  }%
  \providecommand\rotatebox[2]{#2}%
  \newcommand*\fsize{\dimexpr\f@size pt\relax}%
  \newcommand*\lineheight[1]{\fontsize{\fsize}{#1\fsize}\selectfont}%
  \ifx\svgwidth\undefined%
    \setlength{\unitlength}{789.49960327bp}%
    \ifx\svgscale\undefined%
      \relax%
    \else%
      \setlength{\unitlength}{\unitlength * \real{\svgscale}}%
    \fi%
  \else%
    \setlength{\unitlength}{\svgwidth}%
  \fi%
  \global\let\svgwidth\undefined%
  \global\let\svgscale\undefined%
  \makeatother%
  \begin{picture}(1,0.3258776)%
    \lineheight{1}%
    \setlength\tabcolsep{0pt}%
    \put(0,0){\includegraphics[width=\unitlength,page=1]{xfem_v4.pdf}}%
    \put(0.09031816,0.04756425){\color[rgb]{0,0,0}\makebox(0,0)[lt]{\lineheight{1.25}\smash{\begin{tabular}[t]{l}$\sigma_{\mathrm{max}}$\end{tabular}}}}%
    \put(0.01630935,0.10691133){\color[rgb]{0,0,0}\makebox(0,0)[lt]{\lineheight{1.25}\smash{\begin{tabular}[t]{l}$\sigma$\end{tabular}}}}%
    \put(0.20892732,0.0238989){\color[rgb]{0,0,0}\makebox(0,0)[lt]{\lineheight{1.25}\smash{\begin{tabular}[t]{l}$\Delta$\end{tabular}}}}%
    \put(0,0){\includegraphics[width=\unitlength,page=2]{xfem_v4.pdf}}%
    \put(0.04157631,0.04469081){\color[rgb]{0,0,0}\makebox(0,0)[lt]{\lineheight{1.25}\smash{\begin{tabular}[t]{l}$\small\Delta N$\end{tabular}}}}%
    \put(0,0){\includegraphics[width=\unitlength,page=3]{xfem_v4.pdf}}%
    \put(0.31850841,-0.03854758){\color[rgb]{0,0,0}\makebox(0,0)[lt]{\begin{minipage}{0.01786955\unitlength}\raggedright \end{minipage}}}%
    \put(-0.11461531,0.17503601){\color[rgb]{0,0,0}\makebox(0,0)[lt]{\begin{minipage}{0.30378225\unitlength}\raggedright \end{minipage}}}%
    \put(0.399888,0.13252957){\color[rgb]{0,0,0}\makebox(0,0)[lt]{\lineheight{1.25}\smash{\begin{tabular}[t]{l}$F_{\mathrm{min}}$\end{tabular}}}}%
    \put(0,0){\includegraphics[width=\unitlength,page=4]{xfem_v4.pdf}}%
    \put(0.42994937,0.29596764){\color[rgb]{0,0,0}\makebox(0,0)[lt]{\lineheight{1.25}\smash{\begin{tabular}[t]{l}$T$\end{tabular}}}}%
    \put(0.59596129,0.21958904){\color[rgb]{0,0,0}\makebox(0,0)[lt]{\lineheight{1.25}\smash{\begin{tabular}[t]{l}$t$\end{tabular}}}}%
    \put(0.59737066,0.02287625){\color[rgb]{0,0,0}\makebox(0,0)[lt]{\lineheight{1.25}\smash{\begin{tabular}[t]{l}$t$\end{tabular}}}}%
    \put(0.59590485,0.12121989){\color[rgb]{0,0,0}\makebox(0,0)[lt]{\lineheight{1.25}\smash{\begin{tabular}[t]{l}$t$\end{tabular}}}}%
    \put(0.48775588,0.18594178){\color[rgb]{0,0,0}\makebox(0,0)[lt]{\lineheight{1.25}\smash{\begin{tabular}[t]{l}$\Delta N$\end{tabular}}}}%
    \put(0.53784445,0.18624262){\color[rgb]{0,0,0}\makebox(0,0)[lt]{\lineheight{1.25}\smash{\begin{tabular}[t]{l}$\Delta N$\end{tabular}}}}%
    \put(0.42778328,0.19700428){\color[rgb]{0,0,0}\makebox(0,0)[lt]{\lineheight{1.25}\smash{\begin{tabular}[t]{l}$F$\end{tabular}}}}%
    \put(0.40111666,0.17076276){\color[rgb]{0,0,0}\makebox(0,0)[lt]{\lineheight{1.25}\smash{\begin{tabular}[t]{l}$F_{\mathrm{max}}$\end{tabular}}}}%
    \put(0,0){\includegraphics[width=\unitlength,page=5]{xfem_v4.pdf}}%
    \put(0.47101462,0.0736247){\color[rgb]{0,0,0}\makebox(0,0)[lt]{\lineheight{1.25}\smash{\begin{tabular}[t]{l}$R, \Delta N$\end{tabular}}}}%
    \put(0.52698689,0.08002347){\color[rgb]{0,0,0}\makebox(0,0)[lt]{\lineheight{1.25}\smash{\begin{tabular}[t]{l}$R, \Delta N$\end{tabular}}}}%
    \put(0.58393295,0.04823115){\color[rgb]{0,0,0}\makebox(0,0)[lt]{\lineheight{1.25}\smash{\begin{tabular}[t]{l}$\sigma_{\mathrm{min}}$\end{tabular}}}}%
    \put(0.58218152,0.07439727){\color[rgb]{0,0,0}\makebox(0,0)[lt]{\lineheight{1.25}\smash{\begin{tabular}[t]{l}$\sigma_{\mathrm{max}}$\end{tabular}}}}%
    \put(0.42680065,0.09896378){\color[rgb]{0,0,0}\makebox(0,0)[lt]{\lineheight{1.25}\smash{\begin{tabular}[t]{l}$\sigma$\end{tabular}}}}%
    \put(0,0){\includegraphics[width=\unitlength,page=6]{xfem_v4.pdf}}%
    \put(0.17093306,0.06266689){\color[rgb]{0,0,0}\makebox(0,0)[lt]{\lineheight{1.25}\smash{\begin{tabular}[t]{l}delamination\end{tabular}}}}%
    \put(0,0){\includegraphics[width=\unitlength,page=7]{xfem_v4.pdf}}%
    \put(0.24715449,0.18558691){\color[rgb]{0,0,0}\makebox(0,0)[lt]{\lineheight{1.25}\smash{\begin{tabular}[t]{l}transverse cracks\end{tabular}}}}%
    \put(0,0){\includegraphics[width=\unitlength,page=8]{xfem_v4.pdf}}%
    \put(0.37980489,0.0820805){\color[rgb]{0,0,0}\makebox(0,0)[lt]{\lineheight{1.25}\smash{\begin{tabular}[t]{l}$F$\end{tabular}}}}%
    \put(0.17446604,0.32135443){\color[rgb]{0,0,0}\makebox(0,0)[lt]{\lineheight{1.25}\smash{\begin{tabular}[t]{l}XFEM-based progressive failure framework\end{tabular}}}}%
    \put(0.72829272,0.32071511){\color[rgb]{0,0,0}\makebox(0,0)[lt]{\lineheight{1.25}\smash{\begin{tabular}[t]{l}Open-hole laminate simulation\end{tabular}}}}%
    \put(0.73359358,0.26741219){\color[rgb]{0,0,0}\makebox(0,0)[lt]{\lineheight{1.25}\smash{\begin{tabular}[t]{l}ply damage\end{tabular}}}}%
    \put(0.8644,0.26749098){\color[rgb]{0,0,0}\makebox(0,0)[lt]{\lineheight{1.25}\smash{\begin{tabular}[t]{l}delamination\end{tabular}}}}%
    \put(0.67768234,0.21872391){\color[rgb]{0,0,0}\makebox(0,0)[lt]{\lineheight{1.25}\smash{\begin{tabular}[t]{l}$45^\circ$\end{tabular}}}}%
    \put(0.6776438,0.17312157){\color[rgb]{0,0,0}\makebox(0,0)[lt]{\lineheight{1.25}\smash{\begin{tabular}[t]{l}$90^\circ$\end{tabular}}}}%
    \put(0.67201991,0.12691907){\color[rgb]{0,0,0}\makebox(0,0)[lt]{\lineheight{1.25}\smash{\begin{tabular}[t]{l}-$45^\circ$\end{tabular}}}}%
    \put(0.6827561,0.08068204){\color[rgb]{0,0,0}\makebox(0,0)[lt]{\lineheight{1.25}\smash{\begin{tabular}[t]{l}$0^\circ$\end{tabular}}}}%
    \put(0,0){\includegraphics[width=\unitlength,page=9]{xfem_v4.pdf}}%
  \end{picture}%
\endgroup%

%% file: sections/introduction.tex
\section{Introduction}

Fiber reinforced polymer (FRP) composites have many advantages over traditional (metallic) engineering materials and are increasingly used in the aerospace and automotive industries. These materials possess high-strength-to-weight ratios, good corrosion resistance and the material can be tailored to meet specific requirements by altering the stacking-sequence, fiber material, matrix material and ply thickness. 

In order to speed-up and improve the design and certification process of FRP laminated structures, numerical models must be developed to predict the behavior under critical loading conditions. Cyclic loading is an important loading condition and is often governing for the design of composite structures.

In previous years, many high-cycle fatigue models have been developed, capable of accurately simulating fatigue crack growth in specimens with pre-existing cracks \cite{Peerlings2000,Turon2007b,Harper2010c,Kawashita2012,Latifi2015a,Bak2016, LATIFI201772, Latifi2017AnComposites,Amiri-Rad2017a,Amiri-Rad2017,Latifi2020,Carreras2019a,Trabal2022a}. However, there are still challenges in progressive fatigue modeling of multidirectional laminates, where both intra- and inter-laminar cracks can initiate, propagate and interact and the final failure mode is a combination of multiple complex failure processes \cite{Spearing1992,Nixon-Pearson2013a,nixon-pearsonInvestigationDamageDevelopment2015,Aymerich2000}. 

A few authors have developed a progressive failure modeling methodology to predict fatigue failure in composite laminates \cite{Nixon-Pearson2013b, Iarve2016a,Tao2018, zhengReliable2022,luFatigue2022, Tao2023,Llobet2021a, manevalProgressive2023}.
In most of the methods, Paris-type fatigue cohesive zone models are used to describe fatigue crack propagation, while fatigue crack initiation is simulated with criteria based on S-N curves. Therefore, these two stages of fatigue damage are treated separately and a large amount of material characterization tests is required to obtain Paris' and S-N curves for different mode-mixities and stress ratios. 

Recently, \Davila ~\cite{Davila2020, Davila2020Nasa, Davila2018Nasa} proposed a fatigue cohesive zone model, covering fatigue crack initiation and propagation in a unified formulation. The model requires the static material properties and a limited number of fatigue parameters that can be obtained by calibration with typical fracture characterization tests. The effects of mode-mixity and stress ratio are taken into account in the constitutive equations through empirical relations and engineering assumptions. The fatigue cohesive zone model has been validated against simple fracture tests with thermosets \cite{Davila2020, Davila2018Nasa, Davila2020Nasa, lecinanaRobust2023,davilaCohesive2023a}, tests exhibiting $\mathcal{R}$-curve effects with thermoplastics \cite{lecinanaCharacterization2023}, a \textit{reinforced double cantilever beam} specimen with changing crack front shapes \cite{Raimondo2022, lecinanaRobust2023} and a \textit{clamped tapered beam} specimen with delamination-migration \cite{liangReducedInput2021}. Furthermore, the model successfully predicted initiation of transverse matrix cracks in $[0_2/90_4]_\mathrm{s}$-laminates with thermal residual stresses \cite{Joosten2022}.  Recently, the model has been used to simulate progressive fatigue failure in an open-hole $[0/90]_\mathrm{s}$-laminate with predefined splits \cite{manevalProgressive2023} and in an open-hole $[\pm45]_\mathrm{s}$-laminate with multiple transverse matrix cracks using XFEM \cite{hofmanNumerical2024}. However, the applicability of the fatigue cohesive zone model to simulate progressive failure in quasi-isotropic open-hole laminates with a complex interaction of distributed transverse matrix cracks and delamination has not yet been demonstrated. 

Moreover, most of the progressive fatigue failure models employ predefined discrete cracks to simulate transverse matrix cracks and splits \cite{Nixon-Pearson2013b, Tao2018, Tao2023, manevalProgressive2023}, or a continuum damage mechanics model in combination with fiber-aligned meshes \cite{Llobet2021a}. With XFEM, fiber-aligned meshes are not required and unstructured meshes can be used, thereby reducing meshing efforts. Furthermore, XFEM captures the discrete nature of a transverse (mesoscale) crack and allows for multiple cracks at arbitrary locations, which makes it possible to simulate the complete failure process from distributed cracking to localized failure, including the interaction between discrete matrix cracks and delamination.

This paper builds on previous work \cite{hofmanNumerical2024}, where a robust and efficient XFEM-based progressive failure framework for tensile static loading  \cite{Meer2009,Meer2010} has been extended for simulating high-cycle fatigue.
In this framework, both intra- and inter-laminar cracking are modeled with  \Davila's mixed-mode fatigue cohesive zone model for simulating fatigue crack initiation and propagation with only a few input parameters. Furthermore, the fatigue cohesive zone model has been enhanced with an implicit time integration scheme of the damage variable and a fully consistent tangent to improve efficiency of the full-laminate analyses. The progressive fatigue failure framework is further extended in this work by using an adaptive cycle jump strategy that can capture local stress ratios. This is particularly important in multidirectional laminates, where due to the presence of non-uniform thermal residual stresses after curing, the local stress ratio varies in the laminate. 
Simulations of two quasi-isotropic open-hole laminates, with different stacking sequences, are performed with various tensile cyclic loadings and results are compared against experimental data from literature. 

This paper is organized as follows. First, the fatigue cohesive zone model with implicit fatigue damage update is summarized, followed by the formulation of XFEM for simulating intra-laminar cracking. Then, the extension with an efficient adaptive cycle jump scheme for simulating high-cycle fatigue, while accounting for local stress ratios, is addressed. Finally, the simulation results of two quasi-isotropic open-hole laminates are presented and discussed.

%% file: sections/methods.tex
\section{Progressive failure framework}

\subsection{Fatigue cohesive zone model}
\label{sec:fatigue-czm}

\begin{figure}
  \centering
  \begin{tikzpicture}
    \node at (0,0) {\def\svgwidth{0.45\columnwidth}\scalebox{1.00}{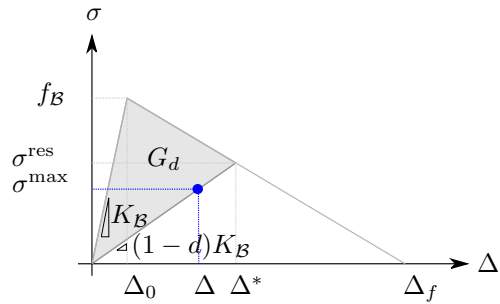}};
  \end{tikzpicture}
  \caption{Fatigue cohesive zone model: the traction-separation response under fatigue loading ($\textcolor{blue}{\bullet}$) is inside the quasi-static envelope}
  \label{fig:nomenclature-fczm}
\end{figure}

The high-cycle fatigue cohesive zone model by \Davila\, \cite{Davila2020} builds upon Turon's static mixed-mode cohesive zone model \cite{Turon2006ALoading, Turon2010, Turon2018b} and is formulated in terms of an equivalent 1D traction-separation relation:
\begin{equation}
  \sigma = (1-d)K_\B \Delta
  \label{eq:eq-traction-separation}
\end{equation}
where $\sigma$ is the equivalent traction, $K_\B$ is the mode-dependent dummy stiffness and $\Delta$ is the equivalent displacement jump (see \Cref{fig:nomenclature-fczm}).

The damage variable $d$ is related to an energy-based damage variable $\Dam$, defined as the ratio of dissipated energy $G_d$ over the critical mixed-mode energy release rate $G_c$

\begin{equation}
  \Dam \equiv \frac{G_d}{G_c} = \frac{\Delta^*-\Delta_0}{\Delta_f - \Delta_0}
  \label{eq:bigd-def}
\end{equation}
where $\Delta_0$ and $\Delta_f$ are the initiation and ultimate equivalent displacements, respectively. Furthermore, the reference displacement (corresponding to the displacement at which static damage develops) is defined as
\begin{equation}
  \Delta^* = \Dam(\Delta_f - \Delta_0) + \Delta_0
  \label{eq:refdisp}
\end{equation}

The energy-based damage variable $\Dam$ is the state variable and can only increase in \emph{pseudo} time $t$, such that for current time step $t_n$

\begin{equation}
  \Dam(t_n) = \max \limits_{0 \leq \tau \leq t_n} \big(\Dam(\tau)\big)
  \label{eq:smalld-int}
\end{equation}

\begin{figure}
  \centering
  \begin{tikzpicture}[>=stealth]
    \node(aa) at (9.5,1.0) [rectangle] {\def\svgwidth{0.5\textwidth}\scalebox{0.85}{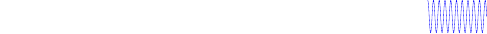}};
      \node at (10,-1.5) [rectangle] {\def\svgwidth{1.0\columnwidth}\scalebox{0.85}{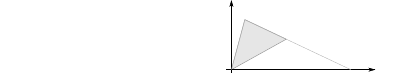}};
      \node (a) at (9.49,0.9) {};
      \node (b) at (11.5,-0.5) {};
      \node (c) at (11.9,-1.4) {};
      \node (d) at (12.9,-1.0) {\small$G_{d}$};%
      \draw [->,blue] (a.north) to [out=230, in=90] (b.north);
      \draw [-,black] (c.north) to [out=30, in=180] (d.west);
  \end{tikzpicture}
\caption{S-N-based fatigue cohesive zone model}
\label{fig:fczm}
\end{figure}

The stiffness-based damage variable $d$ in \Cref{eq:eq-traction-separation} is related to the energy-based damage variable with the following equation

\begin{equation}
  d= 1-\frac{(1-\Dam)\Delta_0}{\Dam \Delta_f + (1-\Dam) \Delta_0}
  \label{eq:bigd-smalld-rel}
\end{equation}
The evolution of the energy-based damage variable is such that, at constant stress amplitudes and mode-mixities, the number of cycles to failure is described by an S-N curve (see \Cref{fig:fczm}). The rate of change of the damage variable $\nicefrac{\mathrm{d}\Dam}{\mathrm{d}N}$  is described with a nonlinear differential equation
\begin{align}
  &\dv{\Dam}{N} = \frac{1}{\gamma} \frac{(1-\Dam)^{\beta-p}}{E^\beta (p+1)}\bigg(\frac{\Delta}{\Delta^*}\bigg)^\beta
 \label{eq:CF20} 
\end{align}
where the right-hand side is the CF20 damage accumulation function \cite{Davila2020Nasa}. In this function, $\gamma$ is the number of cycles to failure at the endurance limit, $p$ can be calibrated to Paris curves and $\beta$ is the exponent in the S-N curve, expressed as

\begin{equation}
\beta = \frac{-7 \eta}{\mathrm{log} E}
\end{equation}
where $\eta$ is a \emph{brittleness} parameter to account for the low-cycle fatigue response. For a given stress ratio $R$, the relative endurance limit $E$, defined as the ratio of equivalent endurance limit $\sigma_{\mathrm{end}}$ and mode-dependent static strength $f_{\B}$, is computed from the endurance limit $\epsilon$ at full load reversal ($R=-1$) with the Goodmann diagram:

\begin{equation}
  E = \frac{2 C_\mathrm{l} \epsilon}{C_\mathrm{l}\epsilon + 1 + R(C_\mathrm{l} \epsilon - 1)}
  \label{eq:endurance-R-B-dependence}
\end{equation}

In this expression, $C_\mathrm{l}$ is an empirical relation which accounts for the effect of mode-mixity \cite{juvinallFundamentalsMachineComponent2012}

\begin{equation}
  C_\mathrm{l} = 1-0.42 \B
  \label{eq:juvinall-relation}
\end{equation}
where $\B$ is a displacement-based measure for mode-mixity and is defined as

\begin{equation}
  \B = \frac{K_{sh} \jump{u}_{sh}^2}{K_{n} \mc{\jump{u}_n}^2 + K_{sh} \jump{u}_{sh}^2 } 
  \label{eq:mode-mixity-def}
\end{equation}
which is a function of the normal and shear dummy stiffnessess ($K_n$ and $K_{sh}$) and the normal and shear displacement jumps ($\jump{u}_{n}$ and $\jump{u}_{sh}$). 
Furthermore, the ratio of shear and normal stiffness must satisfy the constraint equation in Ref. \cite{Turon2010} to ensure correct energy dissipation during static mixed-mode fracture.

Following the previous work \cite{hofmanNumerical2024}, the fatigue damage at current \emph{pseudo} time $t_{n}$ is computed with an implicit time integration scheme, using the trapezoidal rule:

\begin{equation}
  \Dam^{(n)}_{f} = \Dam^{(n-1)} + \frac{\Delta N}{2} \left(\dv{\Dam}{N}^{(n-1)} + \dv{\Dam}{N}^{(n)} \right)
  \label{eq:implicit-update}
\end{equation}
where the damage rates $\nicefrac{\mathrm{d}\Dam}{\mathrm{d}N}^{(n-1)}$ and $\nicefrac{\mathrm{d}\Dam}{\mathrm{d}N}^{(n)}$ are determined by evaluating \Cref{eq:CF20} with values of the previous and the current \emph{psuedo} time step, respectively. This nonlinear equation is iteratively solved at \emph{local} integration point level with Newton's method. The quasi-static damage $\Dam_s$ is computed as

\begin{equation}
  \Dam_s = \frac{\Delta - \Delta_0 }{ \Delta_f - \Delta_0 }
  \label{eq:dam-static}
\end{equation}
and the updated damage is determined as the maximum of the static and the fatigue damage

\begin{equation}
  \Dam = \max\big(\Dam_s, \Dam_f\big)
  \label{eq:dam-static-or-fatigue}
\end{equation}

The formulation is completed with a consistent tangent stiffness matrix, which is derived and presented in Ref. \cite{hofmanNumerical2024}.

\subsection{Intra-laminar cracking}
\label{sec:xfem-formulation}

\begin{figure}
  \centering
  \begin{tikzpicture}
    \node at (0,0) {
      {\def\svgwidth{1.0\textwidth}{\scalebox{1.0}{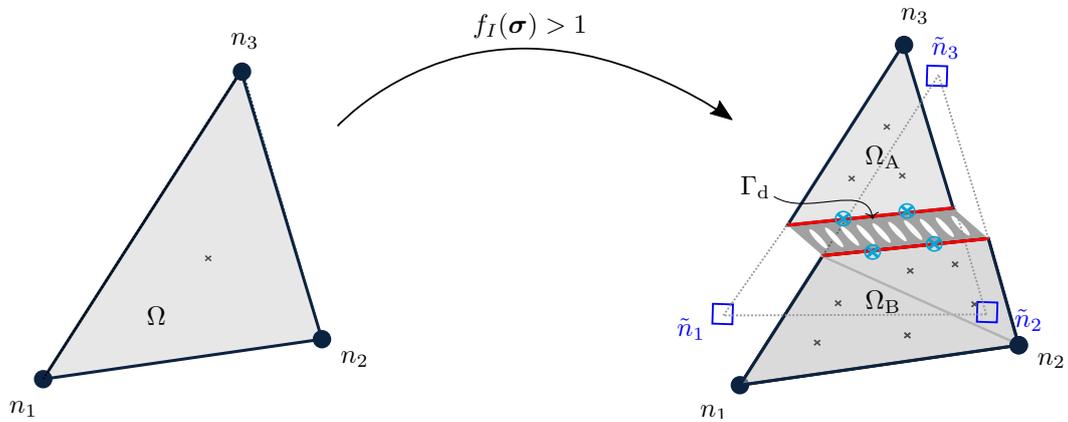}}}};
    \draw[->, bend left=45, in=110, out=-20] (3.0,0.1) to node[pos=-0.25, below] {$\Gamma_\mathrm{d}$} (4.3,-.2);
 \end{tikzpicture}
 \caption{XFEM crack insertion. The mixed-mode fatigue cohesive zone model is used in each cohesive integration point (\textcolor{TUDblue}{$\bm{\bigotimes}$}) for describing fatigue damage}
  \label{fig:cracked-elem}
\end{figure}

The \textit{phantom node version} of XFEM \cite{Hansbo2004} is used for simulating distributed matrix cracks at arbitrary locations \cite{Meer2009, Meer2010, VanderMeer2012a}. 
A crack segment is inserted in a continuum element as soon as the stress in the plies $\bm{\sigma}$ reaches a critical envelope $f_I(\bm{\sigma})=1$. 
The discontinuity in the displacement field is resolved by duplicating the original element and defining the connectivity of each sub-element $\Omega_\mathrm{A}$ and $\Omega_\mathrm{B}$ (see \Cref{fig:cracked-elem}) as
\begin{align}
  &\Omega^\mathrm{conn}_\mathrm{A} = \{\textcolor{blue}{\tilde{n}_1},\textcolor{blue}{\tilde{n}_2},n_3\} \\
  &\Omega^\mathrm{conn}_\mathrm{B} = \{n_1,n_2,\textcolor{blue}{\tilde{n}_3}\} 
\end{align}
where $\{n_i\}$ and $\{\textcolor{blue}{\tilde{n}_i}\}$ are the set of original and \emph{phantom} nodes, respectively.

The displacement field of the XFEM element is expressed in terms of the independent displacement fields of the two sub-elements that are overlapping:

\begin{equation}
\textbf{u}(\textbf{x}) =  
\left\{
\begin{array}{ll}
  \textbf{N}(\textbf{x})\textbf{u}_{\mathrm{A}},~\textbf{x}\in \Omega_{\mathrm{A}} \\
  \textbf{N}(\textbf{x})\textbf{u}_{\mathrm{B}},~\textbf{x}\in \Omega_{\mathrm{B}} \\
\end{array} 
\right.
\end{equation}
where $\textbf{u}_{\mathrm{A}}$ and $\textbf{u}_{\mathrm{B}}$ are the vectors with nodal degrees of freedom of each sub-element and $\textbf{N}(\textbf{x})$ is the matrix containing the shape functions. The displacement jump vector along the crack segment $\Gamma_\mathrm{d}$ is defined as

\begin{equation}
  \dispjump (\textbf{x}) = \textbf{N}(\textbf{x})(\textbf{u}_{\mathrm{A}}-\textbf{u}_{\mathrm{B}}),~\textbf{x} \in \Gamma_\mathrm{d}
\end{equation}

The crack segment is inserted parallel to the direction of the fibers in the ply to enforce transverse cracks to propagate in fiber direction. The traction-separation relation of the cohesive integration points that are located on the crack segment is described by the fatigue cohesive zone model (\Cref{sec:fatigue-czm}).

In order to retain well-posedness of the problem, a predefined crack-spacing parameter $l_\mathrm{c}$ is used \cite{Meer2009}. XFEM crack segments can be inserted either at zero orthogonal distance of existing cracks (propagation), or initiate as new cracks at a distance that is at least equal to $l_\mathrm{c}$. The objectivity of the crack-spacing parameter is discussed in earlier publications for static loading \cite{Meer2010} and for fatigue loading \cite{hofmanNumerical2024}.

XFEM crack segments are inserted when a fatigue crack insertion criterion ($f_I(\bm{\sigma})>1.0$) is satisfied. The criterion is based on the endurance limit to maintain consistency with the fatigue damage formulation \cite{hofmanNumerical2024}. The failure index function is defined as the ratio of the equivalent traction $\sigma$ to the endurance limit $\sigma_\mathrm{end}$:

\begin{equation}
  f_I(\bm{\sigma})\equiv \frac{\sigma(\bm{\sigma})}{\sigma_{\mathrm{end}}(\bm{\sigma})} 
  \label{eq:failure-index-function}
\end{equation}
with $\sigma_{\mathrm{end}}= E f_\B$. The relative endurance limit $E$ is determined from \Cref{eq:juvinall-relation,eq:endurance-R-B-dependence} and the mode-dependent static strength $f_\B$ is computed as

\begin{equation}
  f_\B = \sqrt{(K_n(1-\B)+\B K_{sh}) \big[f^2_{n}/K_n + ( f_{sh}^2/K_{sh} - f^2_{n}/K_n) \B^{\eta}\big]}
  \label{eq:mode-dependent-strength-expanded}
\end{equation}
where $f_n$ and $f_{sh}$ are the normal and shear static strengths.

\begin{figure}
  \centering
   {\def\svgwidth{1.0\textwidth}{\scalebox{1.0}{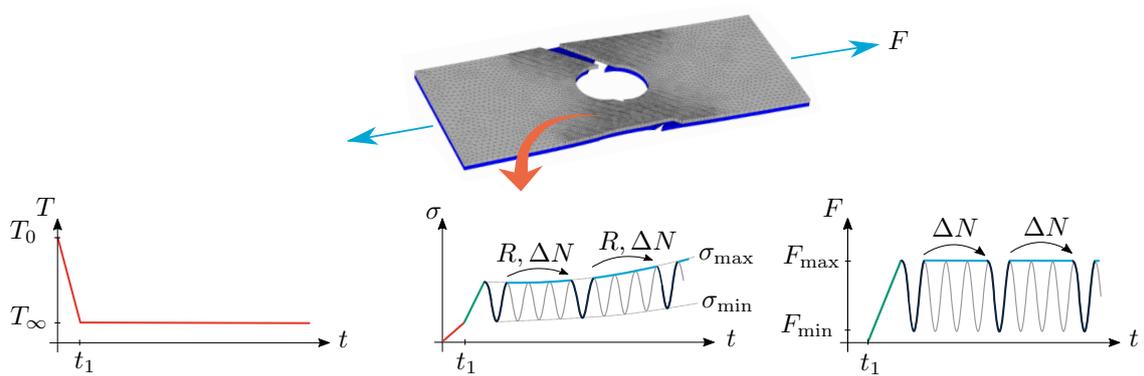}}}
   \caption{Cycle jump scheme with four phases: thermal load phase (in \emph{red}), static ramp-up phase (in \emph{green}), control cycle phase (in \emph{dark blue}) and cycle jump phase (in \emph{light blue}) }
  \label{fig:cycle-jump-scheme}
\end{figure}

\subsection{Cycle jump scheme}
\label{sec:cycle-jump-scheme}

A cycle jump approach, which takes into account the effects of the cyclic load in the constitutive relation, is often used in high-cycle fatigue analyses since simulating each cycle explicitly would be computationally intractable. A load envelope approach can be used to apply the loading in a simplified way, for example when a constant amplitude cyclic load is applied and all inelastic behavior is assumed to be of damage-type (with secant unloading), which makes the unloading-reloading a linear problem. Because of this linearity, the \emph{global} load ratio is equal to the \emph{local} stress ratio in each integration point. Consequently, the \emph{local} stress ratio is \apriori known and can be provided as an input parameter of the constitutive equations. This approach is used by many authors for the simulation of high-cycle fatigue
\cite{Peerlings2000,Turon2007b,Harper2010c,Kawashita2012,Latifi2015a,Bak2016, LATIFI201772, Latifi2017AnComposites,Amiri-Rad2017a,Amiri-Rad2017,Carreras2019a,Davila2020,Davila2020Nasa,Latifi2020,liangReducedInput2021,Trabal2022a,May2010a,May2011}.

In the presence of multiple unsynchronized load signals, plasticity, geometric nonlinearities or thermal residual stresses, the \emph{local} stress ratio is not equal to the \emph{global} load ratio.
However, the load envelope approach does not give access to the \emph{local} stress ratios, even though these should govern the \emph{local} material response.

One approach to overcome this issue is to use a \emph{min-max} technique \cite{Raimondo2020a}, where two models, one with the minimum and the other with the maximum applied load, are used and information is exchanged between them. 
Another, more general approach, is a cycle jump strategy with explicit load cycles, also called \emph{control cycles}, before each cycle increment $\Delta N$, as previously applied to composite materials by various authors \cite{vanpaepegemFinite2001,Nojavan2016a,Nojavan2016,sallyEfficient2020, Joosten2022}.
During these control cycles, the minimum and maximum stresses are monitored such that after the control cycle, the \emph{local} stress ratio can be computed before a cycle jump $\Delta N$ is applied. During the cycle jump, the \emph{local} stress ratio is then available for use in the calculation of fatigue damage.

Joosten et al.~\cite{Joosten2022} combined this cycle jump approach with \Davila's fatigue CZM and proposed a \emph{local} stress ratio definition, computed during the control cycles with a projection of the minimum and maximum severities:

\begin{equation}
  R\equiv \frac{\SvecMin \cdot \SvecMax}{\norm{\SvecMax}^2}
\label{eq:stress-ratio-definition}
\end{equation}
where $\Svec=\big[\nicefrac{t_n}{f_n}, \nicefrac{t_{sh}}{f_{sh}}\big]^\mathrm{T}$ is the stress severity vector. The components of this vector contain the normal ($t_n$) and shear traction ($t_{sh}$) components, scaled with the pure-mode strengths. 

In this work, a similar approach is followed to capture the varying \emph{local} stress ratio in quasi-isotropic laminates with non-uniform thermal residual stresses. Four different loading phases are defined.
First, a \emph{thermal load phase} is applied to simulate a temperature drop from the curing temperature $T_0$ to the cooling temperature $T_\infty$, resulting in a residual stress in every material point due to the mismatch of thermal constants between plies. After the thermal load phase, the \emph{static-ramp-up phase} is simulated to reach the maximum load $F_{\mathrm{max}}$. Subsequently, a \emph{control cycle} is applied in order to determine the \emph{local} stress ratio in every integration point. 
When the control cycle has finished, a \emph{cycle jump phase} is entered in which fatigue damage accumulates according to the formulation with \emph{local} $R$ in \Cref{eq:endurance-R-B-dependence}. After each cycle jump phase, a control cycle is re-entered to update the stress ratios for the next cycle jump phase. This process is repeated until all fatigue cycles have been simulated. An overview of these phases is shown in \Cref{fig:cycle-jump-scheme}. 

\subparagraph{remark} It should be mentioned that the use of control cycles to compute the \emph{local} stress ratio in each integration point is an explicit procedure.
Since fatigue damage accumulates during cycle jumps, resulting in stress redistribution in the laminate, the \emph{local} stress ratio that is used at the beginning of the step is different from the \emph{local} stress ratio at the end of the step. Although time step dependence has been reduced strongly with the implicit time integration scheme \cite{hofmanNumerical2024}, a re-assessment of the time step dependence is required. This is carried out in the numerical examples later in this article (\Cref{sec:efficiency-accuracy}).

\subsection{Monitoring the local stress ratio in the bulk integration points}

At the end of each cycle jump, the failure index function \Cref{eq:failure-index-function} must be evaluated in every integration point. For this purpose, the \emph{local} stress ratio must be available for computing the endurance limit with \Cref{eq:endurance-R-B-dependence}. The local stress ratio in \Cref{eq:stress-ratio-definition} is defined with traction components in local coordinate frame, aligned with the crack segment. 
Since crack segments are inserted parallel to the fiber direction, the traction vector $\textbf{t}$ can be computed from the bulk stress $\bm{\sigma}$ and the normal vector $\textbf{n}$, that is perpendicular to the fibers, with $\textbf{t}=\bm{\sigma} \textbf{n}$. The \emph{local} stress ratio can then be computed during control cycles in each bulk integration point, such that it is available when \Cref{eq:failure-index-function} is evaluated.

\subsection{Transferring local stress ratio from bulk to cohesive integration points}

Once new crack segments are inserted, the solution must be re-equilibrated by re-entering the Newton-Raphson solver. Upon insertion of new crack segments, the \emph{local} stress ratio $R$ must be known to evaluate fatigue damage in each new cohesive integration point. 
To achieve this, the bulk \emph{local} stress ratio is passed to the cohesive integration points on the newly inserted crack segments. With linear (constant stress) elements, this is straightforward. With other types of elements where the stress, and thus the \emph{local} stress ratio, varies across an element, the history transfer approach by Wells and Sluys \cite{wellsNew2001} is a suitable method.

\subsection{Adaptive stepping strategy}
\label{sec:adaptive-stepping}

Since an implicit time integration scheme is employed for updating the fatigue damage variable during cycle jumps, the cycle increment $\Delta N$ can be determined based on \emph{global} convergence behavior \cite{hofmanNumerical2024}. 
An adaptive stepping scheme is used during each loading phase of the analysis. The number of iterations $n_{\mathrm{iter}}$, required to reach convergence in the previous time step, is used to determine the next step size:

\begin{equation}
  \Delta t^{(n+1)} = C^{-\big(\frac{n_{\mathrm{iter}}-n_{\mathrm{iter}}^{\mathrm{opt}}}{\xi}\big)}  \Delta t^{(n)}
  \label{eq:adaptive-step}
\end{equation}
where $C$, $\xi$ and $n_{\mathrm{iter}}^{\mathrm{opt}}$ are model parameters and $\Delta t$ is an increment in \emph{pseudo} time which translates to a cycle increment during cycle jumps and to a load increment during static ramp up and control cycles. If convergence is not reached within a specified maximum number of iterations $n_{\mathrm{iter}}^{\mathrm{max}}$, the step is cancelled and restarted with a reduced time step increment $\Delta t^{(n)} \leftarrow c_{\mathrm{red}} \Delta t^{(n)}$. 

At the start of the first cycle jump, the step size $\Delta N_{\mathrm{init}}$ is initialized and used to compute fatigue damage. This step size is adapted during the subsequent cycle jumps according to \Cref{eq:adaptive-step}. After the cycle jump phase and before a new control-phase is entered, the new cycle increment $\Delta N$ is computed and stored as the initial cycle increment for the next cycle jump phase. This ensures that the control cycle increments and the cycle jumps are separately adapted.

When control cycles are entered, it is first tried to find the solution at the minimum load at once. Since fatigue damage is de-activated during control cycles and the analysis reduces to a linear problem, only two steps, corresponding to the minimum and maximum load, are sufficient in most cases. Sometimes, more steps during control cycles are necessary, in which case the time step is adapted as described above.

\subsection{Modeling thermal residual stresses}
\label{sec:methods_thermal}

Multidirectional laminates develop residual stresses after curing due to a mismatch in thermal constants between plies. The effect of a temperature change $\Delta T$ is taken into account in the ply constitutive model. The orthotropic linear elastic stress-strain relation in 2D is

\begin{equation}
   \bm{\sigma} = \mathbf{D}(\bm{\varepsilon} - \bm{\varepsilon}^{th})
\end{equation}
where 
\begin{equation}
    \bm{\varepsilon^{th}} = \big[\alpha_1 \Delta T, \alpha_2 \Delta T, 0\big]^T
\end{equation}
is the thermal strain, with $\alpha_i$ the coefficients of thermal expansion in longitudinal (fiber) direction ($\alpha_1$) and in transverse direction ($\alpha_2$).

%% file: inkscape/nomenclature-fczm.pdf_tex
\begingroup%
  \makeatletter%
  \providecommand\color[2][]{%
    \errmessage{(Inkscape) Color is used for the text in Inkscape, but the package 'color.sty' is not loaded}%
    \renewcommand\color[2][]{}%
  }%
  \providecommand\transparent[1]{%
    \errmessage{(Inkscape) Transparency is used (non-zero) for the text in Inkscape, but the package 'transparent.sty' is not loaded}%
    \renewcommand\transparent[1]{}%
  }%
  \providecommand\rotatebox[2]{#2}%
  \newcommand*\fsize{\dimexpr\f@size pt\relax}%
  \newcommand*\lineheight[1]{\fontsize{\fsize}{#1\fsize}\selectfont}%
  \ifx\svgwidth\undefined%
    \setlength{\unitlength}{70.06611321bp}%
    \ifx\svgscale\undefined%
      \relax%
    \else%
      \setlength{\unitlength}{\unitlength * \real{\svgscale}}%
    \fi%
  \else%
    \setlength{\unitlength}{\svgwidth}%
  \fi%
  \global\let\svgwidth\undefined%
  \global\let\svgscale\undefined%
  \makeatother%
  \begin{picture}(1,0.56494113)%
    \lineheight{1}%
    \setlength\tabcolsep{0pt}%
    \put(0,0){\includegraphics[width=\unitlength,page=1]{nomenclature-fczm.pdf}}%
    \put(0.23223757,0.08772521){\color[rgb]{0,0,0}\makebox(0,0)[lt]{\lineheight{1.25}\smash{\begin{tabular}[t]{l}$(1-d)K_{\B}$\end{tabular}}}}%
    \put(0,0){\includegraphics[width=\unitlength,page=2]{nomenclature-fczm.pdf}}%
    \put(0.04692014,0.38521538){\color[rgb]{0,0,0}\makebox(0,0)[lt]{\lineheight{1.25}\smash{\begin{tabular}[t]{l}$f_\B$\end{tabular}}}}%
    \put(-0.00176794,0.25833726){\color[rgb]{0,0,0}\makebox(0,0)[lt]{\lineheight{1.25}\smash{\begin{tabular}[t]{l}$\sigma^{\mathrm{res}}$\end{tabular}}}}%
    \put(-0.00042243,0.20758639){\color[rgb]{0,0,0}\makebox(0,0)[lt]{\lineheight{1.25}\smash{\begin{tabular}[t]{l}$\sigma^{\mathrm{max}}$\end{tabular}}}}%
    \put(0.14360896,0.54437653){\color[rgb]{0,0,0}\makebox(0,0)[lt]{\lineheight{1.25}\smash{\begin{tabular}[t]{l}$\sigma$\end{tabular}}}}%
    \put(0.90579153,0.05043442){\color[rgb]{0,0,0}\makebox(0,0)[lt]{\lineheight{1.25}\smash{\begin{tabular}[t]{l}$\Delta$\end{tabular}}}}%
    \put(0,0){\includegraphics[width=\unitlength,page=3]{nomenclature-fczm.pdf}}%
    \put(0.42268091,0.0063837){\color[rgb]{0,0,0}\makebox(0,0)[lt]{\lineheight{1.25}\smash{\begin{tabular}[t]{l}$\Delta^{*}$\end{tabular}}}}%
    \put(0.21566377,0.0063837){\color[rgb]{0,0,0}\makebox(0,0)[lt]{\lineheight{1.25}\smash{\begin{tabular}[t]{l}$\Delta_0$\end{tabular}}}}%
    \put(0.76186123,0.0063837){\color[rgb]{0,0,0}\makebox(0,0)[lt]{\lineheight{1.25}\smash{\begin{tabular}[t]{l}$\Delta_f$\end{tabular}}}}%
    \put(0.35340927,0.0063837){\color[rgb]{0,0,0}\makebox(0,0)[lt]{\lineheight{1.25}\smash{\begin{tabular}[t]{l}$\Delta$\end{tabular}}}}%
    \put(0.26235802,0.25832843){\color[rgb]{0,0,0}\makebox(0,0)[lt]{\lineheight{1.25}\smash{\begin{tabular}[t]{l}$G_d$\end{tabular}}}}%
    \put(0,0){\includegraphics[width=\unitlength,page=4]{nomenclature-fczm.pdf}}%
    \put(0.19119428,0.14671388){\color[rgb]{0,0,0}\makebox(0,0)[lt]{\lineheight{1.25}\smash{\begin{tabular}[t]{l}$K_{\B}$\end{tabular}}}}%
    \put(0,0){\includegraphics[width=\unitlength,page=5]{nomenclature-fczm.pdf}}%
  \end{picture}%
\endgroup%

%% file: inkscape/1Dbar.pdf_tex
\begingroup%
  \makeatletter%
  \providecommand\color[2][]{%
    \errmessage{(Inkscape) Color is used for the text in Inkscape, but the package 'color.sty' is not loaded}%
    \renewcommand\color[2][]{}%
  }%
  \providecommand\transparent[1]{%
    \errmessage{(Inkscape) Transparency is used (non-zero) for the text in Inkscape, but the package 'transparent.sty' is not loaded}%
    \renewcommand\transparent[1]{}%
  }%
  \providecommand\rotatebox[2]{#2}%
  \newcommand*\fsize{\dimexpr\f@size pt\relax}%
  \newcommand*\lineheight[1]{\fontsize{\fsize}{#1\fsize}\selectfont}%
  \ifx\svgwidth\undefined%
    \setlength{\unitlength}{233.68890958bp}%
    \ifx\svgscale\undefined%
      \relax%
    \else%
      \setlength{\unitlength}{\unitlength * \real{\svgscale}}%
    \fi%
  \else%
    \setlength{\unitlength}{\svgwidth}%
  \fi%
  \global\let\svgwidth\undefined%
  \global\let\svgscale\undefined%
  \makeatother%
  \begin{picture}(1,0.06879384)%
    \lineheight{1}%
    \setlength\tabcolsep{0pt}%
    \put(0,0){\includegraphics[width=\unitlength,page=1]{1Dbar.pdf}}%
    \put(0.83910602,0.01940553){\color[rgb]{0,0,0}\makebox(0,0)[lt]{\lineheight{1.25}\smash{\begin{tabular}[t]{l}$\sigma$\end{tabular}}}}%
    \put(0,0){\includegraphics[width=\unitlength,page=2]{1Dbar.pdf}}%
    \put(0.1383028,0.01866251){\color[rgb]{0,0,0}\makebox(0,0)[lt]{\lineheight{1.25}\smash{\begin{tabular}[t]{l}$\sigma$\end{tabular}}}}%
    \put(0,0){\includegraphics[width=\unitlength,page=3]{1Dbar.pdf}}%
  \end{picture}%
\endgroup%

%% file: inkscape/davila.pdf_tex
\begingroup%
  \makeatletter%
  \providecommand\color[2][]{%
    \errmessage{(Inkscape) Color is used for the text in Inkscape, but the package 'color.sty' is not loaded}%
    \renewcommand\color[2][]{}%
  }%
  \providecommand\transparent[1]{%
    \errmessage{(Inkscape) Transparency is used (non-zero) for the text in Inkscape, but the package 'transparent.sty' is not loaded}%
    \renewcommand\transparent[1]{}%
  }%
  \providecommand\rotatebox[2]{#2}%
  \newcommand*\fsize{\dimexpr\f@size pt\relax}%
  \newcommand*\lineheight[1]{\fontsize{\fsize}{#1\fsize}\selectfont}%
  \ifx\svgwidth\undefined%
    \setlength{\unitlength}{190.3518028bp}%
    \ifx\svgscale\undefined%
      \relax%
    \else%
      \setlength{\unitlength}{\unitlength * \real{\svgscale}}%
    \fi%
  \else%
    \setlength{\unitlength}{\svgwidth}%
  \fi%
  \global\let\svgwidth\undefined%
  \global\let\svgscale\undefined%
  \makeatother%
  \begin{picture}(1,0.20166119)%
    \lineheight{1}%
    \setlength\tabcolsep{0pt}%
    \put(0,0){\includegraphics[width=\unitlength,page=1]{davila.pdf}}%
    \put(0.87968223,-0.00540218){\color[rgb]{0,0,0}\makebox(0,0)[lt]{\lineheight{1.25}\smash{\begin{tabular}[t]{l}$\Delta_f$\end{tabular}}}}%
    \put(0.53961933,0.14985647){\color[rgb]{0,0,0}\makebox(0,0)[lt]{\lineheight{1.25}\smash{\begin{tabular}[t]{l}$f_\B$\end{tabular}}}}%
    \put(0.52096261,0.10060542){\color[rgb]{0,0,0}\makebox(0,0)[lt]{\lineheight{1.25}\smash{\begin{tabular}[t]{l}$\sigma^{\mathrm{max}}$\end{tabular}}}}%
    \put(0,0){\includegraphics[width=\unitlength,page=2]{davila.pdf}}%
    \put(0.54664051,0.19042263){\color[rgb]{0,0,0}\makebox(0,0)[lt]{\lineheight{1.25}\smash{\begin{tabular}[t]{l}$\sigma$\end{tabular}}}}%
    \put(0.96688121,0.02053453){\color[rgb]{0,0,0}\makebox(0,0)[lt]{\lineheight{1.25}\smash{\begin{tabular}[t]{l}$\Delta$\end{tabular}}}}%
    \put(0,0){\includegraphics[width=\unitlength,page=3]{davila.pdf}}%
    \put(0.70936462,-0.0058246){\color[rgb]{0,0,0}\makebox(0,0)[lt]{\lineheight{1.25}\smash{\begin{tabular}[t]{l}$\Delta^{*}$\end{tabular}}}}%
    \put(0,0){\includegraphics[width=\unitlength,page=4]{davila.pdf}}%
    \put(0.62843452,0.07648403){\color[rgb]{0,0,0}\makebox(0,0)[lt]{\lineheight{1.25}\smash{\begin{tabular}[t]{l}$\small\Delta N$\end{tabular}}}}%
    \put(0,0){\includegraphics[width=\unitlength,page=5]{davila.pdf}}%
    \put(0.05891912,0.18351577){\color[rgb]{0,0,0}\makebox(0,0)[lt]{\lineheight{1.25}\smash{\begin{tabular}[t]{l}$\frac{\sigma^{\mathrm{max}}}{f_\B}$\end{tabular}}}}%
    \put(0.09451347,0.14621724){\color[rgb]{0,0,0}\makebox(0,0)[lt]{\lineheight{1.25}\smash{\begin{tabular}[t]{l}1\end{tabular}}}}%
    \put(0.06686322,0.04760983){\color[rgb]{0,0,0}\makebox(0,0)[lt]{\lineheight{1.25}\smash{\begin{tabular}[t]{l}$\frac{\sigma_{\mathrm{end}}}{f_\B}$\end{tabular}}}}%
    \put(0.46037351,0.01767337){\color[rgb]{0,0,0}\makebox(0,0)[lt]{\lineheight{1.25}\smash{\begin{tabular}[t]{l}$N$\end{tabular}}}}%
    \put(0,0){\includegraphics[width=\unitlength,page=6]{davila.pdf}}%
  \end{picture}%
\endgroup%

%% file: inkscape/xfem-insertion.pdf_tex
\begingroup%
  \makeatletter%
  \providecommand\color[2][]{%
    \errmessage{(Inkscape) Color is used for the text in Inkscape, but the package 'color.sty' is not loaded}%
    \renewcommand\color[2][]{}%
  }%
  \providecommand\transparent[1]{%
    \errmessage{(Inkscape) Transparency is used (non-zero) for the text in Inkscape, but the package 'transparent.sty' is not loaded}%
    \renewcommand\transparent[1]{}%
  }%
  \providecommand\rotatebox[2]{#2}%
  \newcommand*\fsize{\dimexpr\f@size pt\relax}%
  \newcommand*\lineheight[1]{\fontsize{\fsize}{#1\fsize}\selectfont}%
  \ifx\svgwidth\undefined%
    \setlength{\unitlength}{385.45687938bp}%
    \ifx\svgscale\undefined%
      \relax%
    \else%
      \setlength{\unitlength}{\unitlength * \real{\svgscale}}%
    \fi%
  \else%
    \setlength{\unitlength}{\svgwidth}%
  \fi%
  \global\let\svgwidth\undefined%
  \global\let\svgscale\undefined%
  \makeatother%
  \begin{picture}(1,0.363704)%
    \lineheight{1}%
    \setlength\tabcolsep{0pt}%
    \put(0,0){\includegraphics[width=\unitlength,page=1]{xfem-insertion.pdf}}%
    \put(0.15006288,0.07533791){\color[rgb]{0,0,0}\rotatebox{1.139739}{\makebox(0,0)[lt]{\lineheight{1.25}\smash{\begin{tabular}[t]{l}$\Omega$\end{tabular}}}}}%
    \put(0,0){\includegraphics[width=\unitlength,page=2]{xfem-insertion.pdf}}%
    \put(0.43577866,0.3334801){\color[rgb]{0,0,0}\makebox(0,0)[lt]{\lineheight{1.25}\smash{\begin{tabular}[t]{l}$f_{I}(\bm{\sigma})>1$\end{tabular}}}}%
    \put(0,0){\includegraphics[width=\unitlength,page=3]{xfem-insertion.pdf}}%
    \put(0.78041146,0.08878471){\color[rgb]{0,0,0}\rotatebox{1.6308919}{\makebox(0,0)[lt]{\lineheight{1.25}\smash{\begin{tabular}[t]{l}$\Omega_\mathrm{B}$\end{tabular}}}}}%
    \put(0,0){\includegraphics[width=\unitlength,page=4]{xfem-insertion.pdf}}%
    \put(0.78009893,0.21664674){\color[rgb]{0,0,0}\rotatebox{1.139739}{\makebox(0,0)[lt]{\lineheight{1.25}\smash{\begin{tabular}[t]{l}$\Omega_\mathrm{A}$\end{tabular}}}}}%
    \put(0,0){\includegraphics[width=\unitlength,page=5]{xfem-insertion.pdf}}%
    \put(0.93127631,0.04009988){\color[rgb]{0,0,0}\rotatebox{1.139739}{\makebox(0,0)[lt]{\lineheight{1.25}\smash{\begin{tabular}[t]{l}$n_2$\end{tabular}}}}}%
    \put(0.81152002,0.34483891){\color[rgb]{0,0,0}\rotatebox{1.139739}{\makebox(0,0)[lt]{\lineheight{1.25}\smash{\begin{tabular}[t]{l}$n_3$\end{tabular}}}}}%
    \put(0.83843587,0.31349781){\color[rgb]{0,0,1}\rotatebox{1.139739}{\makebox(0,0)[lt]{\lineheight{1.25}\smash{\begin{tabular}[t]{l}$\tilde{n}_3$\end{tabular}}}}}%
    \put(0.61528013,0.06503264){\color[rgb]{0,0,1}\rotatebox{1.139739}{\makebox(0,0)[lt]{\lineheight{1.25}\smash{\begin{tabular}[t]{l}$\tilde{n}_1$\end{tabular}}}}}%
    \put(0.91134314,0.07284692){\color[rgb]{0,0,1}\rotatebox{1.139739}{\makebox(0,0)[lt]{\lineheight{1.25}\smash{\begin{tabular}[t]{l}$\tilde{n}_2$\end{tabular}}}}}%
    \put(0.32013604,0.04150711){\color[rgb]{0,0,0}\rotatebox{1.139739}{\makebox(0,0)[lt]{\lineheight{1.25}\smash{\begin{tabular}[t]{l}$n_2$\end{tabular}}}}}%
    \put(0.2237007,0.32234407){\color[rgb]{0,0,0}\rotatebox{1.139739}{\makebox(0,0)[lt]{\lineheight{1.25}\smash{\begin{tabular}[t]{l}$n_3$\end{tabular}}}}}%
    \put(0.02952379,-0.00288201){\color[rgb]{0,0,0}\rotatebox{1.139739}{\makebox(0,0)[lt]{\lineheight{1.25}\smash{\begin{tabular}[t]{l}$n_1$\end{tabular}}}}}%
    \put(0.63552154,-0.00652716){\color[rgb]{0,0,0}\rotatebox{1.139739}{\makebox(0,0)[lt]{\lineheight{1.25}\smash{\begin{tabular}[t]{l}$n_1$\end{tabular}}}}}%
  \end{picture}%
\endgroup%

%% file: inkscape/cycle-jump.pdf_tex
\begingroup%
  \makeatletter%
  \providecommand\color[2][]{%
    \errmessage{(Inkscape) Color is used for the text in Inkscape, but the package 'color.sty' is not loaded}%
    \renewcommand\color[2][]{}%
  }%
  \providecommand\transparent[1]{%
    \errmessage{(Inkscape) Transparency is used (non-zero) for the text in Inkscape, but the package 'transparent.sty' is not loaded}%
    \renewcommand\transparent[1]{}%
  }%
  \providecommand\rotatebox[2]{#2}%
  \newcommand*\fsize{\dimexpr\f@size pt\relax}%
  \newcommand*\lineheight[1]{\fontsize{\fsize}{#1\fsize}\selectfont}%
  \ifx\svgwidth\undefined%
    \setlength{\unitlength}{493.78207397bp}%
    \ifx\svgscale\undefined%
      \relax%
    \else%
      \setlength{\unitlength}{\unitlength * \real{\svgscale}}%
    \fi%
  \else%
    \setlength{\unitlength}{\svgwidth}%
  \fi%
  \global\let\svgwidth\undefined%
  \global\let\svgscale\undefined%
  \makeatother%
  \begin{picture}(1,0.4052614)%
    \lineheight{1}%
    \setlength\tabcolsep{0pt}%
    \put(0,0){\includegraphics[width=\unitlength,page=1]{cycle-jump.pdf}}%
    \put(0.02204848,0.13925182){\color[rgb]{0,0,0}\makebox(0,0)[lt]{\lineheight{1.25}\smash{\begin{tabular}[t]{l}$T$\end{tabular}}}}%
    \put(0.28748206,0.02320694){\color[rgb]{0,0,0}\makebox(0,0)[lt]{\lineheight{1.25}\smash{\begin{tabular}[t]{l}$t$\end{tabular}}}}%
    \put(-0.00100864,0.1190869){\color[rgb]{0,0,0}\makebox(0,0)[lt]{\lineheight{1.25}\smash{\begin{tabular}[t]{l}$T_0$\end{tabular}}}}%
    \put(-0.00065666,0.04335221){\color[rgb]{0,0,0}\makebox(0,0)[lt]{\lineheight{1.25}\smash{\begin{tabular}[t]{l}$T_{\infty}$\end{tabular}}}}%
    \put(0.62755123,0.02411654){\color[rgb]{0,0,0}\makebox(0,0)[lt]{\lineheight{1.25}\smash{\begin{tabular}[t]{l}$t$\end{tabular}}}}%
    \put(0.98078254,0.0240311){\color[rgb]{0,0,0}\makebox(0,0)[lt]{\lineheight{1.25}\smash{\begin{tabular}[t]{l}$t$\end{tabular}}}}%
    \put(0.80786498,0.12143824){\color[rgb]{0,0,0}\makebox(0,0)[lt]{\lineheight{1.25}\smash{\begin{tabular}[t]{l}$\Delta N$\end{tabular}}}}%
    \put(0.88795073,0.12191926){\color[rgb]{0,0,0}\makebox(0,0)[lt]{\lineheight{1.25}\smash{\begin{tabular}[t]{l}$\Delta N$\end{tabular}}}}%
    \put(0.71197583,0.13912588){\color[rgb]{0,0,0}\makebox(0,0)[lt]{\lineheight{1.25}\smash{\begin{tabular}[t]{l}$F$\end{tabular}}}}%
    \put(0.6769066,0.09733985){\color[rgb]{0,0,0}\makebox(0,0)[lt]{\lineheight{1.25}\smash{\begin{tabular}[t]{l}$F_{\mathrm{max}}$\end{tabular}}}}%
    \put(0.67494214,0.03620943){\color[rgb]{0,0,0}\makebox(0,0)[lt]{\lineheight{1.25}\smash{\begin{tabular}[t]{l}$F_{\mathrm{min}}$\end{tabular}}}}%
    \put(0,0){\includegraphics[width=\unitlength,page=2]{cycle-jump.pdf}}%
    \put(0.42552281,0.09918178){\color[rgb]{0,0,0}\makebox(0,0)[lt]{\lineheight{1.25}\smash{\begin{tabular}[t]{l}$R, \Delta N$\end{tabular}}}}%
    \put(0.51501589,0.10941267){\color[rgb]{0,0,0}\makebox(0,0)[lt]{\lineheight{1.25}\smash{\begin{tabular}[t]{l}$R, \Delta N$\end{tabular}}}}%
    \put(0.60606596,0.0585805){\color[rgb]{0,0,0}\makebox(0,0)[lt]{\lineheight{1.25}\smash{\begin{tabular}[t]{l}$\sigma_{\mathrm{min}}$\end{tabular}}}}%
    \put(0.60326561,0.10041706){\color[rgb]{0,0,0}\makebox(0,0)[lt]{\lineheight{1.25}\smash{\begin{tabular}[t]{l}$\sigma_{\mathrm{max}}$\end{tabular}}}}%
    \put(0.5649756,-0.12360967){\color[rgb]{0,0,0}\makebox(0,0)[lt]{\begin{minipage}{0.02857131\unitlength}\raggedright \end{minipage}}}%
    \put(-0.1275384,0.21788543){\color[rgb]{0,0,0}\makebox(0,0)[lt]{\begin{minipage}{0.48571218\unitlength}\raggedright \end{minipage}}}%
    \put(0.05546261,0.00493652){\color[rgb]{0,0,0}\makebox(0,0)[lt]{\lineheight{1.25}\smash{\begin{tabular}[t]{l}$t_1$\end{tabular}}}}%
    \put(0.74500403,0.00286573){\color[rgb]{0,0,0}\makebox(0,0)[lt]{\lineheight{1.25}\smash{\begin{tabular}[t]{l}$t_1$\end{tabular}}}}%
    \put(0.39053566,0.0035846){\color[rgb]{0,0,0}\makebox(0,0)[lt]{\lineheight{1.25}\smash{\begin{tabular}[t]{l}$t_1$\end{tabular}}}}%
    \put(0,0){\includegraphics[width=\unitlength,page=3]{cycle-jump.pdf}}%
    \put(0.36394321,0.13665809){\color[rgb]{0,0,0}\makebox(0,0)[lt]{\lineheight{1.25}\smash{\begin{tabular}[t]{l}$\sigma$\end{tabular}}}}%
    \put(0,0){\includegraphics[width=\unitlength,page=4]{cycle-jump.pdf}}%
    \put(0.76938211,0.28639267){\color[rgb]{0,0,0}\makebox(0,0)[lt]{\lineheight{1.25}\smash{\begin{tabular}[t]{l}$F$\end{tabular}}}}%
  \end{picture}%
\endgroup%

%% file: sections/openholes.tex
\section{Quasi-isotropic open-hole laminate simulations}

Two quasi-isotropic open-hole laminates, experimentally tested in Refs.~\cite{Nixon-Pearson2013a,nixon-pearsonInvestigationDamageDevelopment2015}, have been simulated and results are presented in this section. The laminates have the same number of plies and thicknesses but different lay-ups, leading to distinct failure modes and fatigue lifes.
The first laminate has lay-up \layup{45_2/90_2/\mathrm{-}45_2/0_2}{s} and is \emph{ply-level scaled}, where two plies with the same fiber direction are stacked, effectively increasing the ply thickness.
The second laminate has lay-up \layup{45/90/\mathrm{-}45/0}{2s} and is \emph{sub-laminate scaled}, in which the laminate is created by repeating \emph{sub-laminates}.
In the following, the first laminate is denoted as \emph{ply-level scaled specimen} and the second as \emph{sub-laminate scaled specimen}.

\subsection{Model preliminaries}

\begin{figure}
  \hspace{2cm}
   {\def\svgwidth{1.0\columnwidth}{\scalebox{1.0}{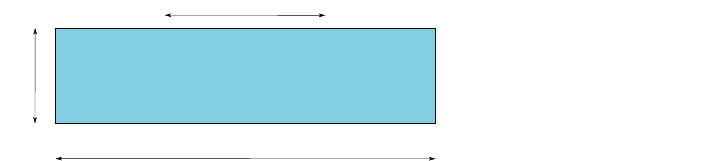}}}
   \caption{Specimen dimensions (in mm) of the quasi-isotropic open-hole laminates. The fine mesh region is indicated in dark blue}
  \label{fig:open-hole-dimensions}
\end{figure}

The open-hole laminates are made of carbon fiber/epoxy plies (prepreg system IM7/8552).
The dimensions of the specimens are $\SI{64}{mm} \times \SI{16}{mm} \times \SI{2}{mm}$ with a hole diameter of \SI{3.175}{mm} (see \Cref{fig:open-hole-dimensions}). 
The thickness of each ply is \SI{0.125}{mm}. Thermal residual stresses arise by accounting for the temperature change $\Delta T$ from processing temperature ($180~^\circ\mathrm{C}$) to room temperature ($20~^\circ\mathrm{C}$) in the linear elastic constitutive relation (\Cref{sec:methods_thermal}), while deformations are freely allowed to occur. 

The static versions of these cases, experimentally tested in Ref.~\cite{greenExperimental2007}, were previously simulated with the same XFEM progressive failure framework for static loading \cite{vandermeerComputationalModelingComplex2012}. In order to aid convergence of the complex full-laminate simulations, the interface strength was reduced to \SI{45}{MPa}, while the ply fracture energy in mode-I was increased to the value of mode-II, resulting in larger cohesive zone lengths. Since the fatigue cohesive zone model is an extension of the static formulation, the same set of static material properties as in Ref. \cite{vandermeerComputationalModelingComplex2012} is used for the present fatigue simulations (see \Cref{tab:open-hole-properties}).
For the fatigue-related parameters of CF20, recommended values are used from Ref.~\cite{Davila2020Nasa}, which predicted excellent results with IM7/8552 carbon fiber/epoxy in a \textit{double cantilever beam} test~\cite{Davila2020}, \textit{mixed-mode bending} test~\cite{liangReducedInput2021, Joosten2022}, \textit{double notch shear} test~\cite{Joosten2022}, \emph{clamped tapered beam} specimen \cite{liangReducedInput2021} and $[0_2/90_4]_\mathrm{s}$-laminate~\cite{Joosten2022}.

The dummy stiffness between the plies is related to the in-plane shear modulus and ply thickness $t_\mathrm{p}$ through $K_\mathrm{d}=\nicefrac{G_{12}}{\frac{1}{2}t_\mathrm{p}}$ \cite{Meer2010}. Furthermore,  the crack-spacing parameter (see \Cref{sec:xfem-formulation}) is set to $l_c=\SI{0.75}{mm}$.
Since the lay-ups are symmetric, only half of each laminate is modelled. The computational domain is discretized with an unstructured mesh and each ply is represented by a layer of plane-stress, constant strain, triangular elements. 
A rectangular region around the hole is defined where delamination is allowed by inserting zero-thickness interface elements between the plies (shown in \Cref{fig:open-hole-dimensions}). This region has a fine-mesh with a typical element size of \SI{0.4}{mm}. The typical element size outside this region is \SI{1.6}{mm}, where the plies are rigidly tied. A Newton-Cotes integration scheme is used for the interface elements for superior interaction with the neighboring elements containing the transverse matrix cracks \cite{Meer2010}.

The adaptive stepping scheme in \Cref{sec:adaptive-stepping} allows for efficiently adapting the time steps in the static, control cycle, and cycle jump phases. Three cycle jumps are simulated in each cycle jump phase, after which a control phase is entered to update the \emph{local} stress ratio in every integration point.

\begin{table}
  \centering
  \caption{Ply material properties used in the simulations} 
  \label{tab:open-hole-properties}
  \begin{tabular}{ll ll ll ll}
  \toprule
  \multicolumn{2}{c}{\textbf{elastic} \cite{jiangConcise2007}} & \multicolumn{2}{c}{\textbf{fracture} \cite{vandermeerComputationalModelingComplex2012}} & \multicolumn{2}{c}{\textbf{fatigue} \cite{Davila2020Nasa}} & \multicolumn{2}{c}{\textbf{thermal} \cite{jiangConcise2007}}  \\ 
  \midrule
  \multicolumn{1}{l}{$E_1$}    & \SI{161.}{GPa} & \multicolumn{1}{l}{$f_{n}$} & 95 MPa  & \multicolumn{1}{l}{$\eta$}     & 0.95  & $\alpha_1$ &  \SI{0}{^{\circ}C^{-1}} \\ 
  \multicolumn{1}{l}{$E_2$}    & \SI{11.38}{GPa}      & \multicolumn{1}{l}{$f_{s}$} & 107 MPa  & \multicolumn{1}{l}{$\epsilon$} & 0.2   & $\alpha_2$ & \SI{3.0e-5}{^{\circ}C^{-1}} \\ 
  \multicolumn{1}{l}{$G_{12}$} & \SI{5.17}{GPa}      & \multicolumn{1}{l}{$G_{Ic}$} & \SI{1.0}{\newton\per\milli\meter}   & \multicolumn{1}{l}{$p$}        & $\beta$ &  &\\ 
  \multicolumn{1}{l}{$\nu_{12}$}  & \SI{0.32}{}      & \multicolumn{1}{l}{$G_{IIc}$}  & \SI{1.0}{\newton\per\milli\meter}   & \multicolumn{1}{l}{}           &      &   &  \\
  \multicolumn{1}{l}{}        &          & \multicolumn{1}{l}{$\eta$}     & 2.1  & \multicolumn{1}{l}{}           &     &  &    \\ \bottomrule
  \end{tabular}
\end{table}

\subsection{Ply-level scaled specimen}

The \emph{ply-level scaled specimen} is simulated with four different maximum applied stress levels $\sigma^{\mathrm{max}}(\mathrm{MPa})=\{334.4, 292.6, 254.0, 209.0\}$, corresponding to $80$, $70$, $60$ and $50\%$ severity levels, respectively. Global severity is defined as the ratio of the maximum load over the static strength \cite{Nixon-Pearson2013a}. The \emph{global} load ratio $R_\mathrm{glob}$ is $0.1$.

\subsubsection{Fatigue life and damage evolution}

\begin{figure}[t]
   \centering
  \begin{tikzpicture}
  \node[] at (0,0)
    {
      \input{figures/ply-level/S-N-locR.tex}
    };
  \end{tikzpicture}
  \caption{S-N curve of the ply-level scaled specimen. Experimental values are extracted from \cite{nixon-pearsonInvestigationDamageDevelopment2015}}
  \label{fig:open-hole_S-N-locR}
\end{figure}
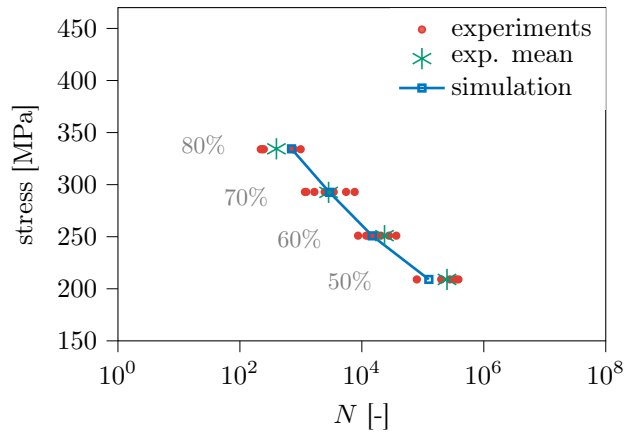

\begin{figure}
  \hspace{-1.5cm}
  \begin{tikzpicture}
    \node[] at (-0.5,4)
    {
      \input{figures/ply-level/k-N.tex}
    };
    \newlength{\myfigurewidth}
    \setlength{\myfigurewidth}{0.36\columnwidth} 
    
    \hspace{0.25cm}

    \node at (-6,-1.4) {\includegraphics[clip, width=\myfigurewidth, trim = 0 0.0cm 0 0cm]{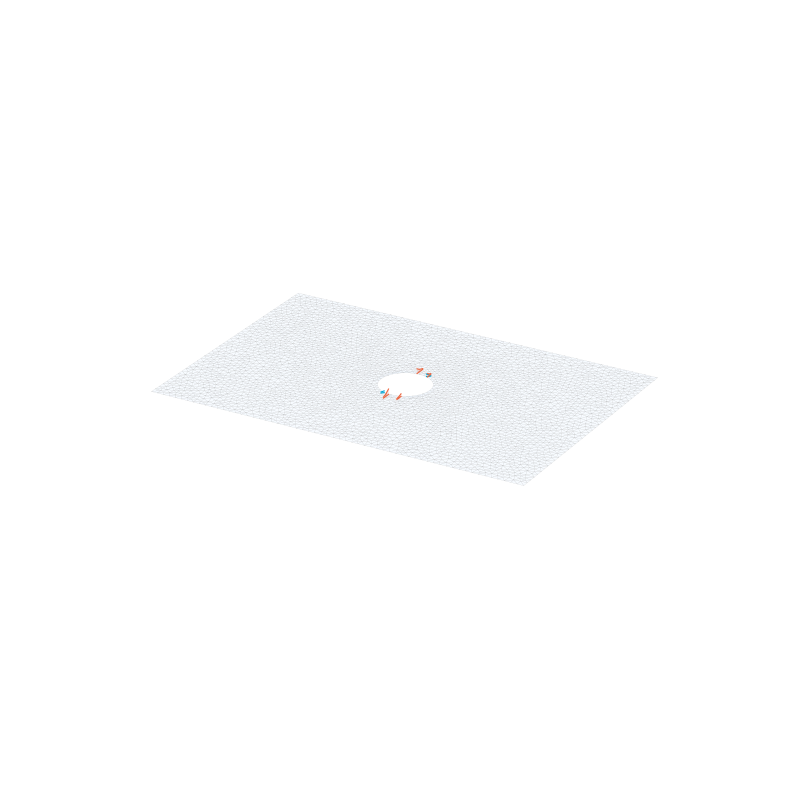}};
    \node at (-2,-1.4) {\includegraphics[clip, width=\myfigurewidth, trim = 0 0.0cm 0 0cm]{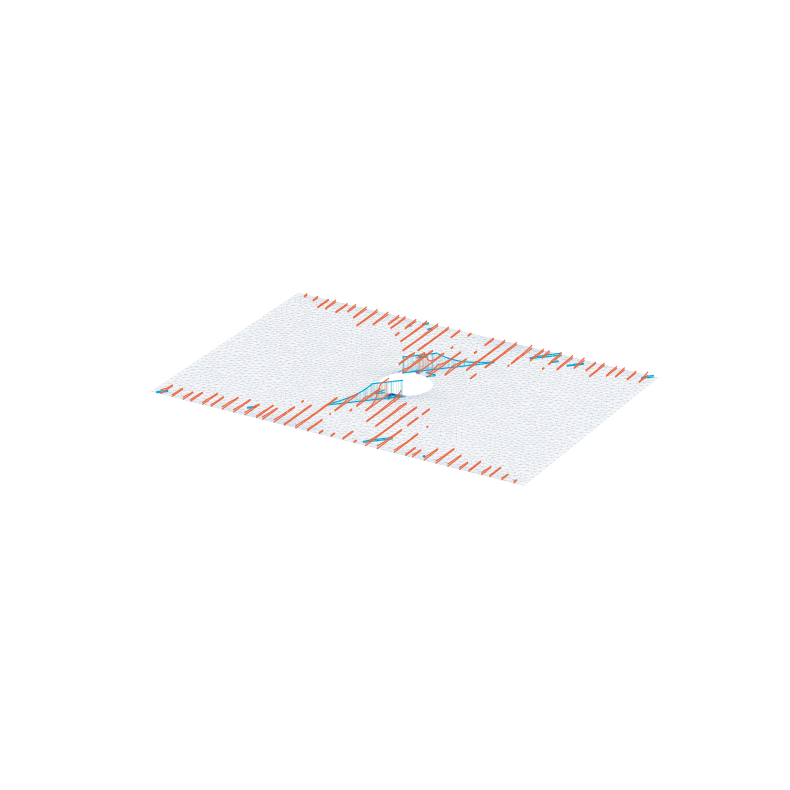}};
    \node at (2,-1.4) {\includegraphics[clip, width=\myfigurewidth, trim = 0 0.0cm 0 0cm]{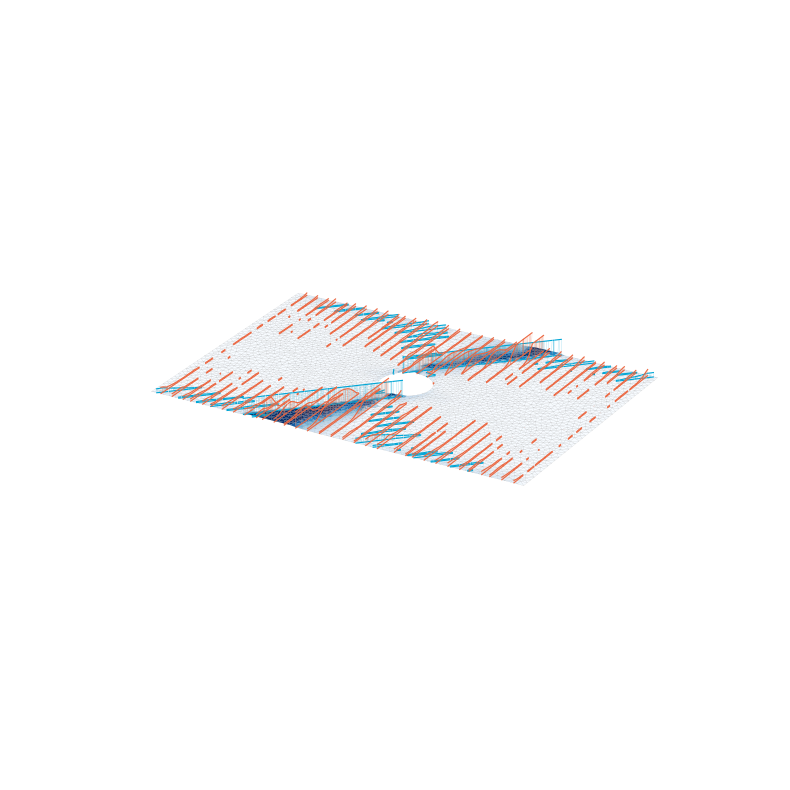}};
    \node at (6,-1.4) {\includegraphics[clip, width=\myfigurewidth, trim = 0 0.0cm 0 0cm]{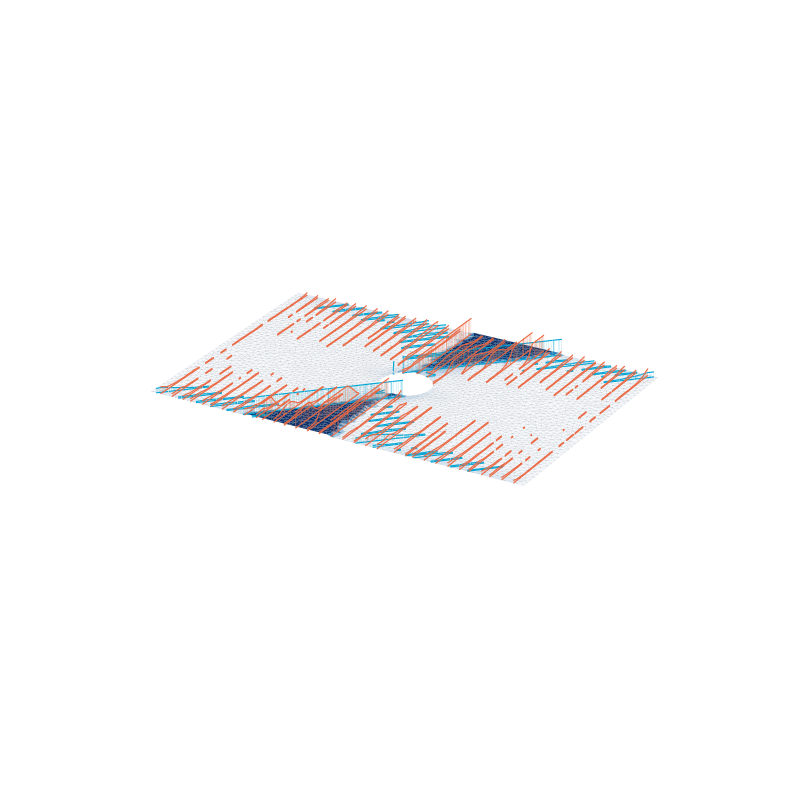}};

    \node at (-6,-2.5) {\includegraphics[clip, width=\myfigurewidth, trim = 0 0.0cm 0 0cm]{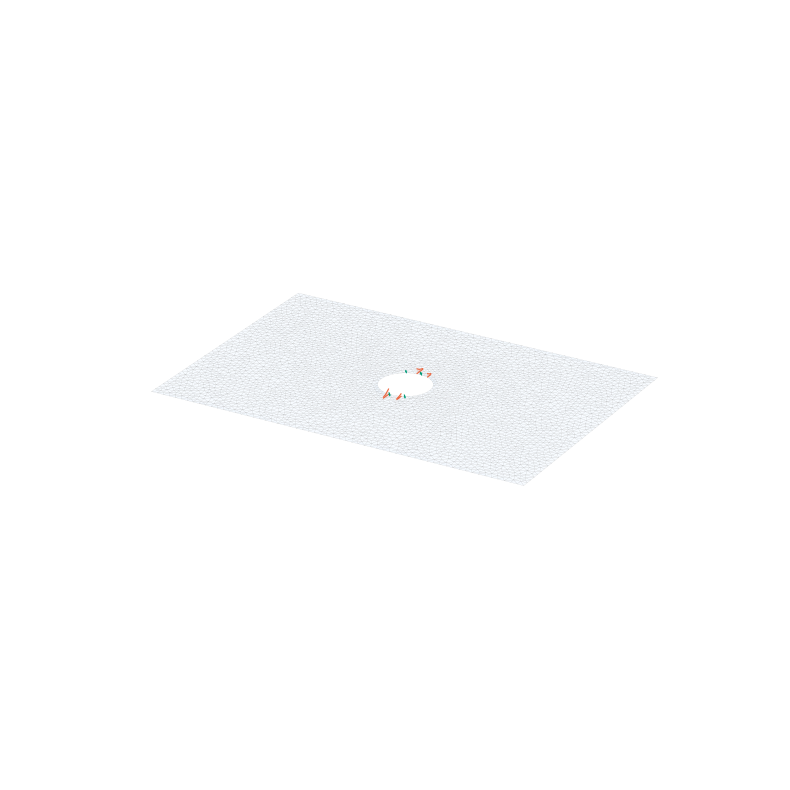}};
    \node at (-2,-2.5) {\includegraphics[clip, width=\myfigurewidth, trim = 0 0.0cm 0 0cm]{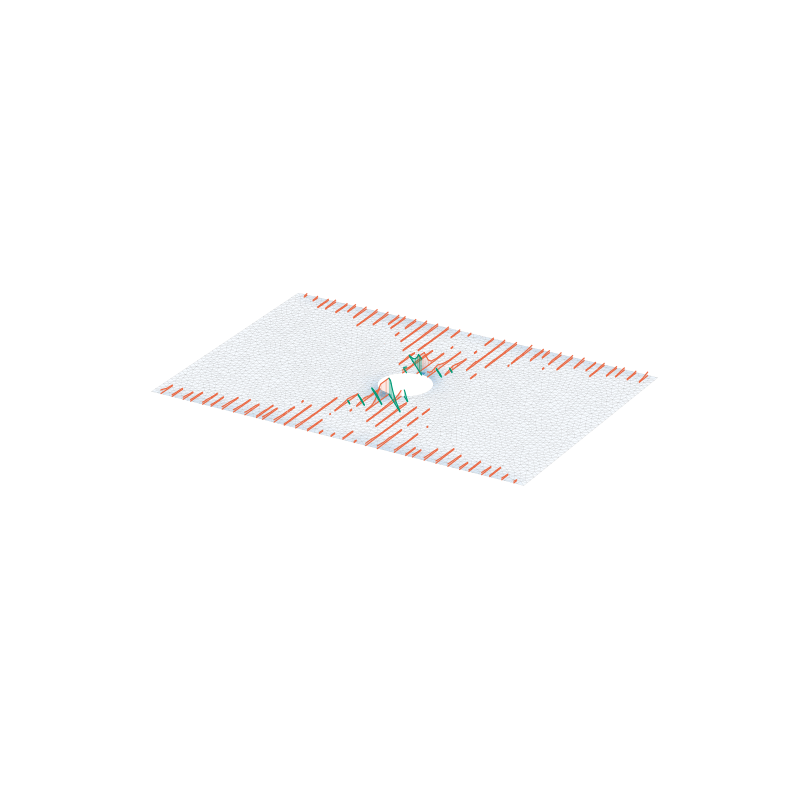}};
    \node at (2,-2.5) {\includegraphics[clip, width=\myfigurewidth, trim = 0 0.0cm 0 0cm]{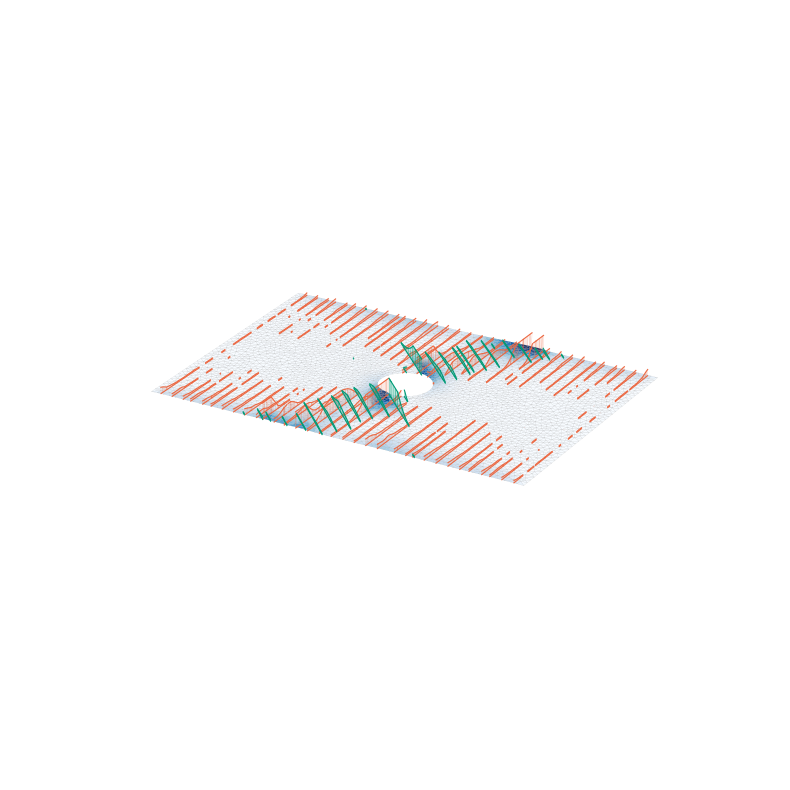}};
    \node at (6,-2.5) {\includegraphics[clip, width=\myfigurewidth, trim = 0 0.0cm 0 0cm]{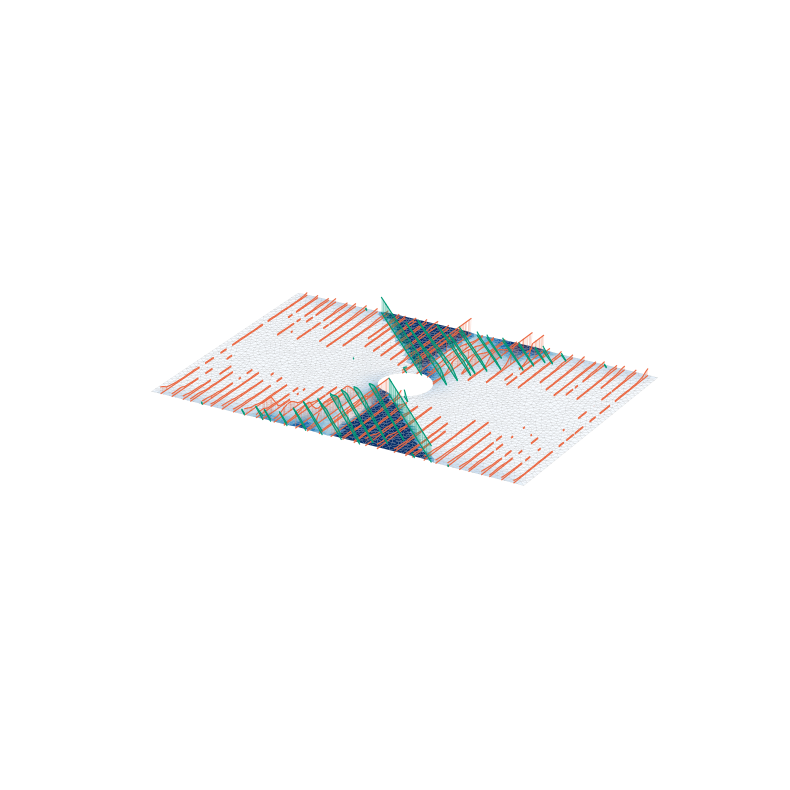}};

    \node at (-6,-3.6) {\includegraphics[clip, width=\myfigurewidth, trim = 0 0.0cm 0 0cm]{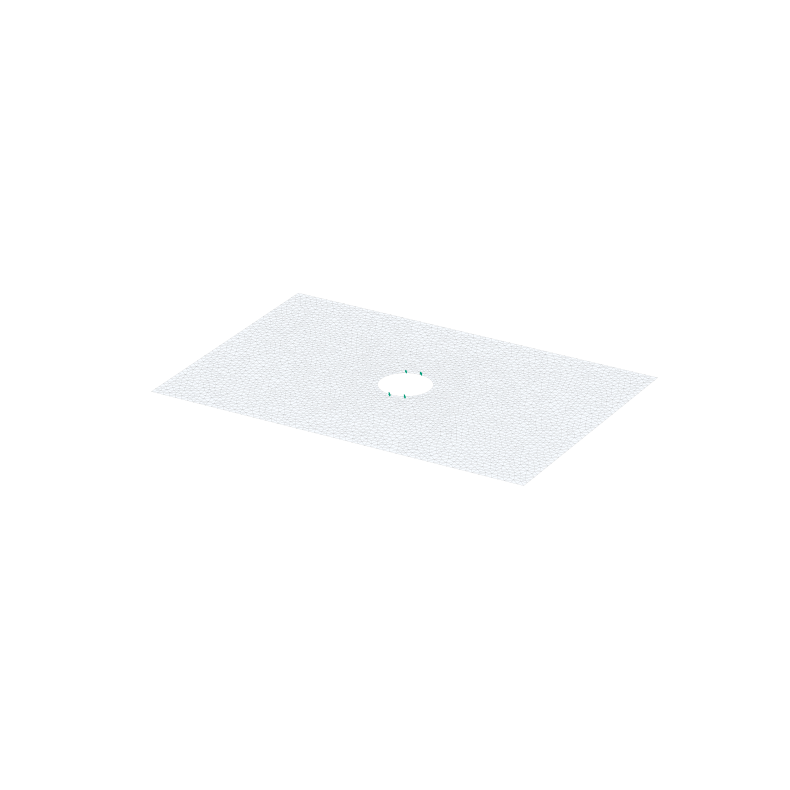}};
    \node at (-2,-3.6) {\includegraphics[clip, width=\myfigurewidth, trim = 0 0.0cm 0 0cm]{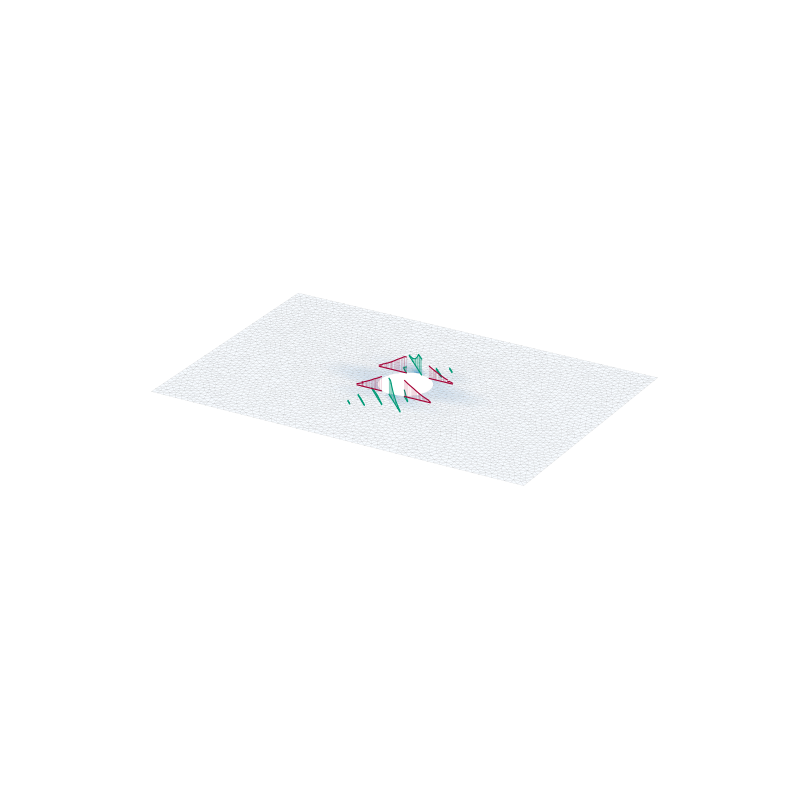}};
    \node at (2,-3.6) {\includegraphics[clip, width=\myfigurewidth, trim = 0 0.0cm 0 0cm]{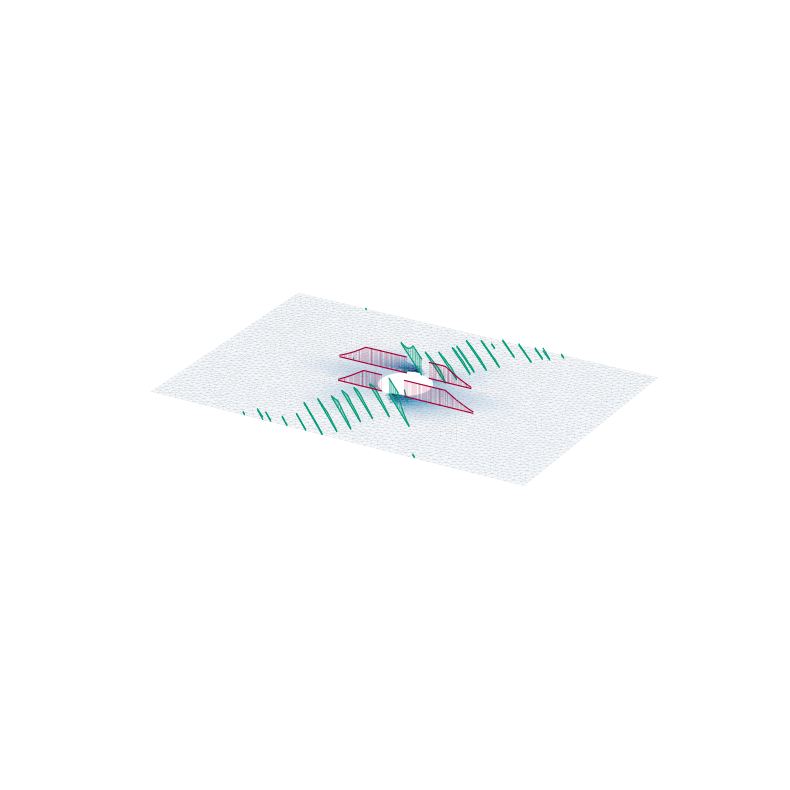}};
    \node at (6,-3.6) {\includegraphics[clip, width=\myfigurewidth, trim = 0 0.0cm 0 0cm]{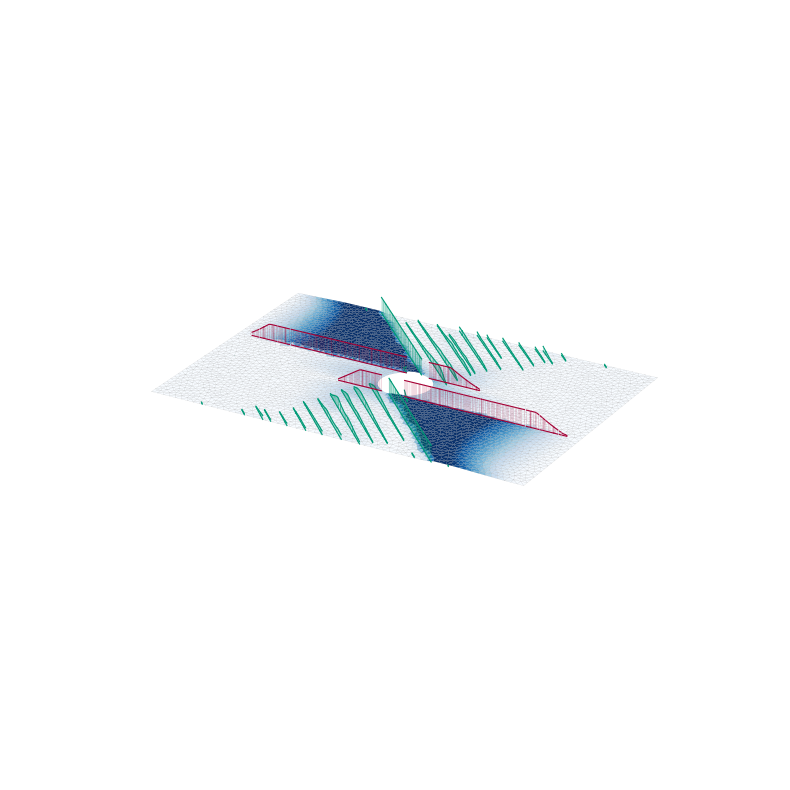}};

    \node[] at (-6,0){$N_1$};
    \node[] at (-2,0){$N_2$};
    \node[] at (2,0){$N_3$};
    \node[] at (6,0){$N_4$};

    \node[] at ( -8.8,-1.3){45/90};
    \node[] at ( -8.8,-2.4){90/-45};
    \node[] at ( -8.8,-3.5){-45/0};

  \end{tikzpicture}

  \caption{Stiffness reduction as a function of number of cycles $N$ (\emph{top}) and damage evolution at indicated time instances in the interfaces (in \emph{dark blue}) and XFEM matrix cracks in 45$^\circ$-ply (in \emph{light blue}), 90$^\circ$-ply (in \emph{orange}), -45$^\circ$-ply (in \emph{green}) and 0$^\circ$-ply (in \emph{red})}
  \label{fig:open-hole_damage-evol}
\end{figure}

S-N curves from experiments and simulations are shown in \Cref{fig:open-hole_S-N-locR}, where the fatigue life corresponds to a $15\%$ loss in normalized effective stiffness $\nicefrac{E_{\mathrm{eff}}}{{E_{\mathrm{eff,0}}}}$, associated with a steep drop in stiffness. The effective stiffness is computed consistently with the experiments as described in Ref. \cite{nixon-pearsonInvestigationDamageDevelopment2015}. It can be observed that a good match is obtained with the experimental fatigue lifes. The stiffness reduction and evolution of fatigue damage in the interfaces for four different time instances are shown in \Cref{fig:open-hole_damage-evol}. First, matrix cracks develop with a limited amount of stiffness loss, accompanied by delamination of small triangular areas near the hole ($N_1$ - $N_2$). A significant stiffness drop occurs due to delamination in the $45/90$ interface, starting from the transverse matrix cracks ($N_2$ - $N_3$). When the matrix cracks have fully developed in the $90^\circ$ and $-45^\circ$ plies, rapid delamination growth takes place in the 90/-45 and -45/0 interface, growing from the hole to the outer edges. This delamination corresponds to an almost vertical drop in the stiffness ($N_3$ - $N_4$). 
The final damage patterns at $15\%$ stiffness reduction match well with the experimental damage patterns \cite{Nixon-Pearson2013a}.

\subsubsection{Efficiency and accuracy}
\label{sec:efficiency-accuracy}

The efficiency and accuracy of the simulations have been investigated. An implicit time integration scheme of the damage variable is used which allows for larger cycle increments. The combination of the implicit scheme and the consistent tangent enables the use of an adaptive time stepping strategy where the number of global iterations to reach a converged solution is a good measure to determine the cycle increment for the next \emph{pseudo} time step \cite{hofmanNumerical2024}.

The time step dependence (see remark in \Cref{sec:cycle-jump-scheme}) and performance of the adaptive cycle jump scheme is assessed by repeating the $60\%$ severity analysis with a small step size. This limits the amount of stress redistribution between steps and increases the number of control cycles with more regular updates of the \emph{local} stress ratios. The maximum allowed cycle increment is set to $\Delta N = 10$ cycles.

The stiffness reduction curve is shown in \Cref{fig:time-step-dependence} with markers for every individual time step, from which it can be observed that the global response in terms of stiffness degradation as function of number of cycles is very similar for the two simulations with 154 and 2746 time steps, respectively. The accumulation of the cycle number $N$ with every time step is depicted in \Cref{fig:istep-N}, which shows that the adaptive stepping strategy effectively adapts the cycle increments throughout the simulation.

It can be concluded that the adaptive stepping strategy, in combination with the implicit fatigue damage update and consistent tangent, results in efficient and accurate analyses resulting in relatively short run times (\SI{5834}{s} on a laptop computer\footnote{Dell laptop with Intel Core i7 processor, 16 GB of RAM and operating system Linux Ubuntu 20.04.}).

\begin{figure}
      \centering
      \hspace{-1.5cm}
    \begin{minipage}{0.49\textwidth}
        \begin{tikzpicture}
          \node[] at (0,0)
          {
            \input{figures/ply-level/k-N-large-vs-small-steps.tex}
          };
        \end{tikzpicture}
        \caption{Stiffness reduction with number of cycles}
        \label{fig:time-step-dependence}
         \end{minipage}
    \hfill
    \begin{minipage}{0.49\textwidth}
        \begin{tikzpicture}
          \node[] at (0,0)
          {
            \input{figures/ply-level/istep-N.tex}
          };
       \draw[->, gray] (1.4,-0.95) -- (3.7,-0.75) node[pos=0.5,above, rotate=4] {\textcolor{gray}{2746 steps}};
       \draw[->, gray] (1.1,0.60) -- (0.35,2.24) node[pos=0.0, below, rotate=0] {\textcolor{gray}{$15\%$ stiffness loss}};
        \end{tikzpicture}
         \caption{Accumulation of cycles with time steps}
        \label{fig:istep-N}
    \end{minipage}
\end{figure}

\subsubsection{Effect of local stress ratio}

In order to investigate the effect of accounting for the \emph{local} stress ratio $R$, the analyses are repeated but this time with the \emph{local} stress ratio in \Cref{eq:endurance-R-B-dependence} set equal to the \emph{global} load ratio $R_{\mathrm{glob}}=\nicefrac{F^{\mathrm{min}}}{F^{\mathrm{max}}}=0.1$.

\begin{figure}
   \centering
  \begin{tikzpicture}
  \node[] at (0,0)
    {
      \input{figures/ply-level/S-N-locR-globR.tex}
    };
  \end{tikzpicture}
  \caption{S-N curve of the ply-level scaled specimen. \emph{Local} stress ratio vs \emph{global} load ratio}
  \label{fig:open-hole_S-N-locR-globR}
\end{figure}
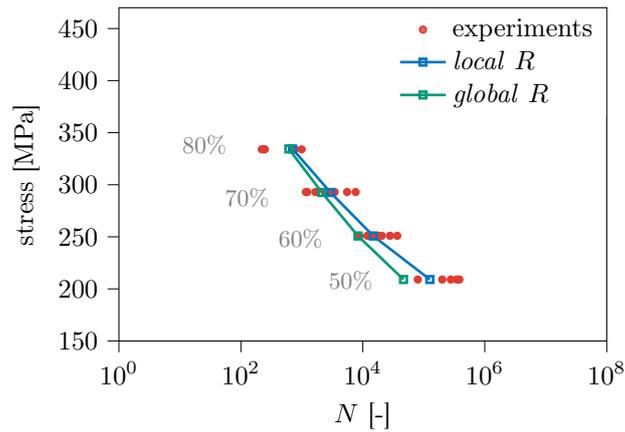

The stiffness reduction curves with \emph{global} and \emph{local} $R$ are shown in \Cref{fig:stiffness-reduction}. It can be observed that accounting for \emph{local} stress ratio results in a significantly slower development of fatigue damage. Furthermore, the use of the \emph{local} stress ratio affects the slope of the laminate S-N curve, as shown in \Cref{fig:open-hole_S-N-locR-globR}. The fatigue life prediction with the \emph{global} load ratio shows a mismatch with the experimental values for the lower load levels, whereas with the highest load level, the response is almost independent of the use of \emph{global} or \emph{local} $R$.

\begin{figure}
  \centering
   \begin{tikzpicture}

    \newlength{\myfigurewidth}
    \setlength{\myfigurewidth}{0.35\columnwidth} 

    \coordinate (50-intf0) at (-4.0,-1.0);

    \coordinate (50-intf0) at (-2.0,-1.0);
    \coordinate (50-intf1) at (-2.0,-4.0);
    \coordinate (50-intf2) at (-2.0,-7.0);

    \coordinate (80-intf0) at (2.0,-1.0);
    \coordinate (80-intf1) at (2.0,-4.0);
    \coordinate (80-intf2) at (2.0,-7.0);

    \coordinate (colorbar) at (4.2,-7.0);

    \node(aa) at (50-intf0) {\includegraphics[clip, trim=0 2.0cm 3cm 2.5cm, width=\myfigurewidth]{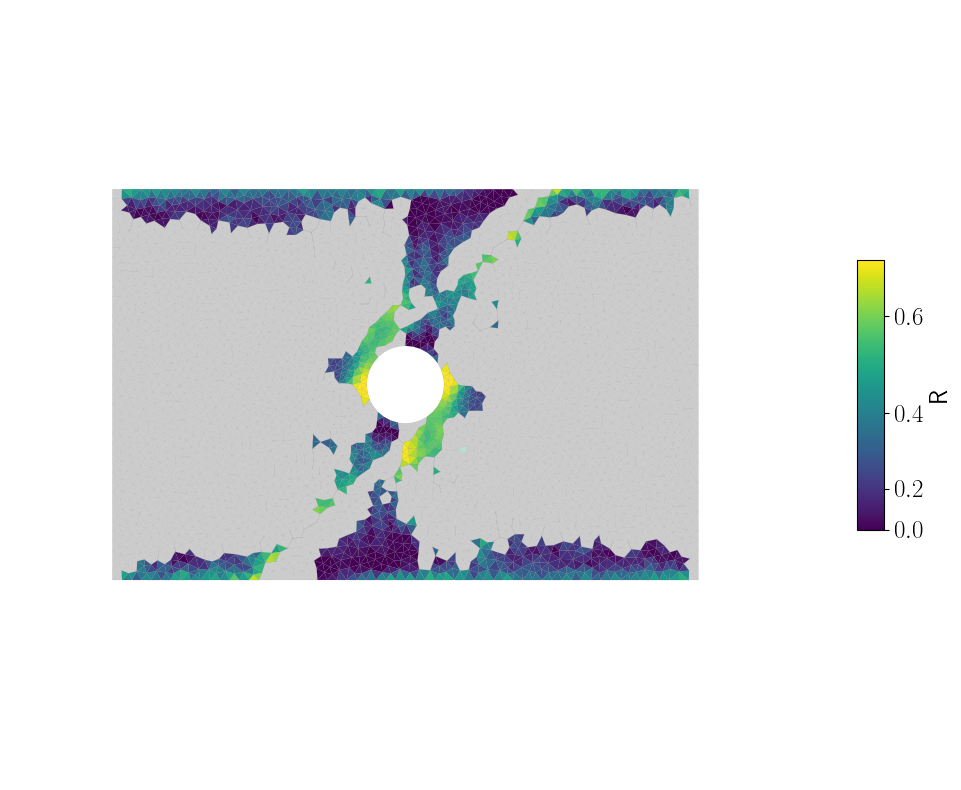}};
    \node(aa) at (50-intf1) {\includegraphics[clip, trim=0 2.0cm 3cm 2.5cm, width=\myfigurewidth]{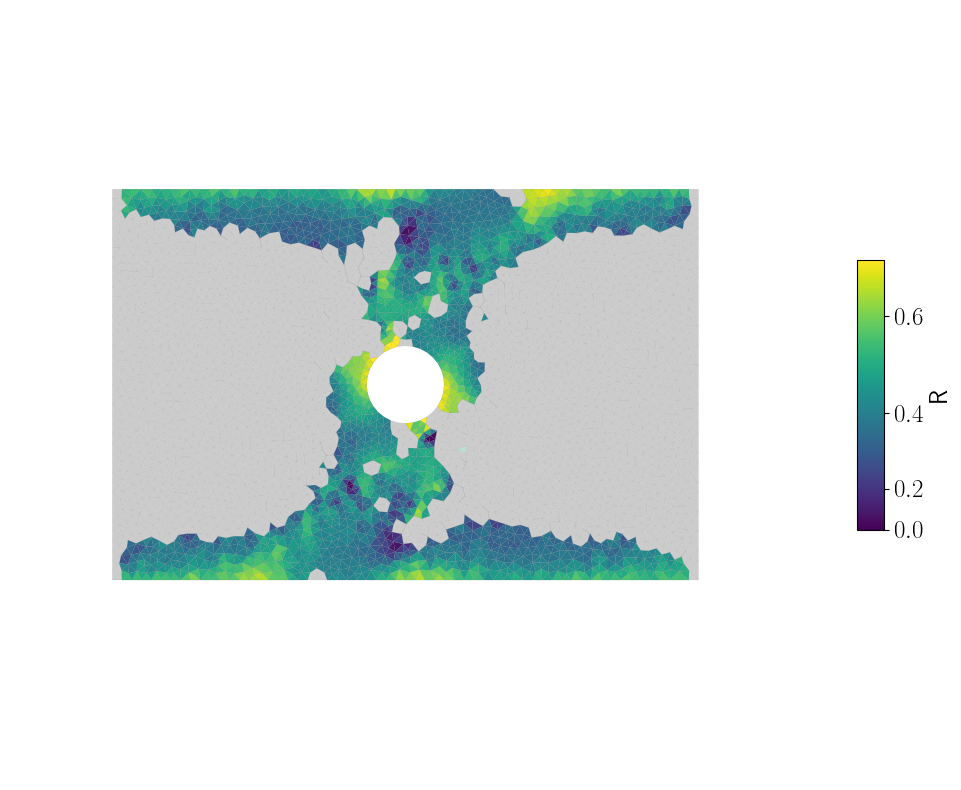}};
    \node(aa) at (50-intf2) {\includegraphics[clip, trim=0 2.0cm 3cm 2.5cm, width=\myfigurewidth]{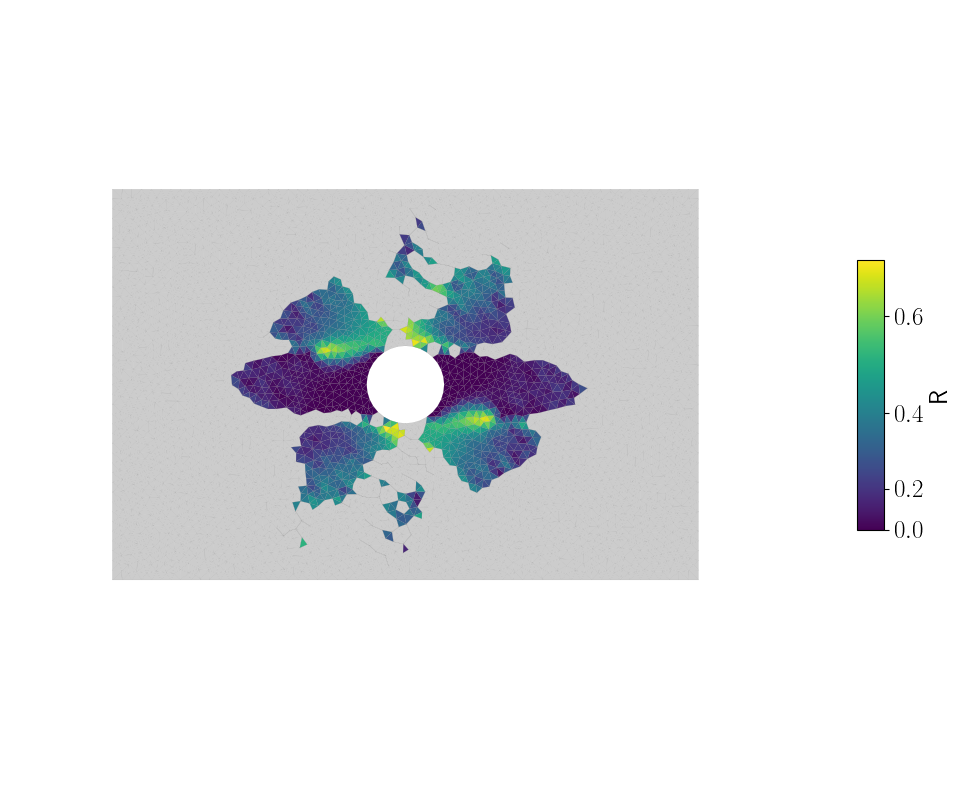}};

    \node(aa) at (80-intf0) {\includegraphics[clip, trim=0 2.0cm 3cm 2.5cm, width=\myfigurewidth]{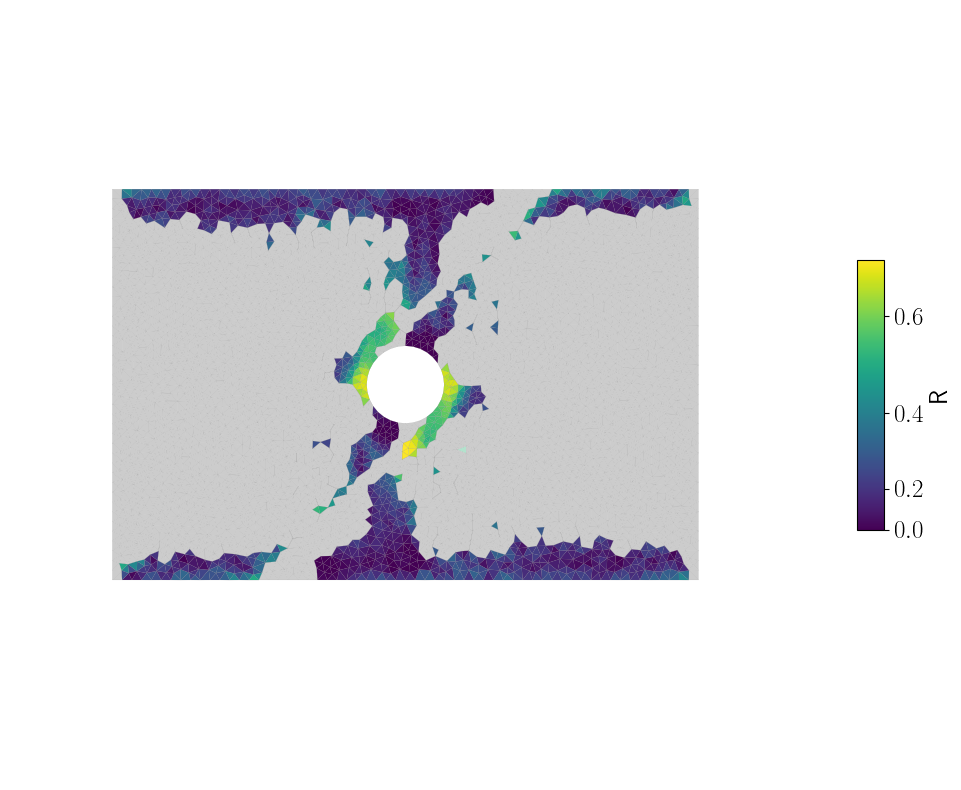}};
    \node(aa) at (80-intf1) {\includegraphics[clip, trim=0 2.0cm 3cm 2.5cm, width=\myfigurewidth]{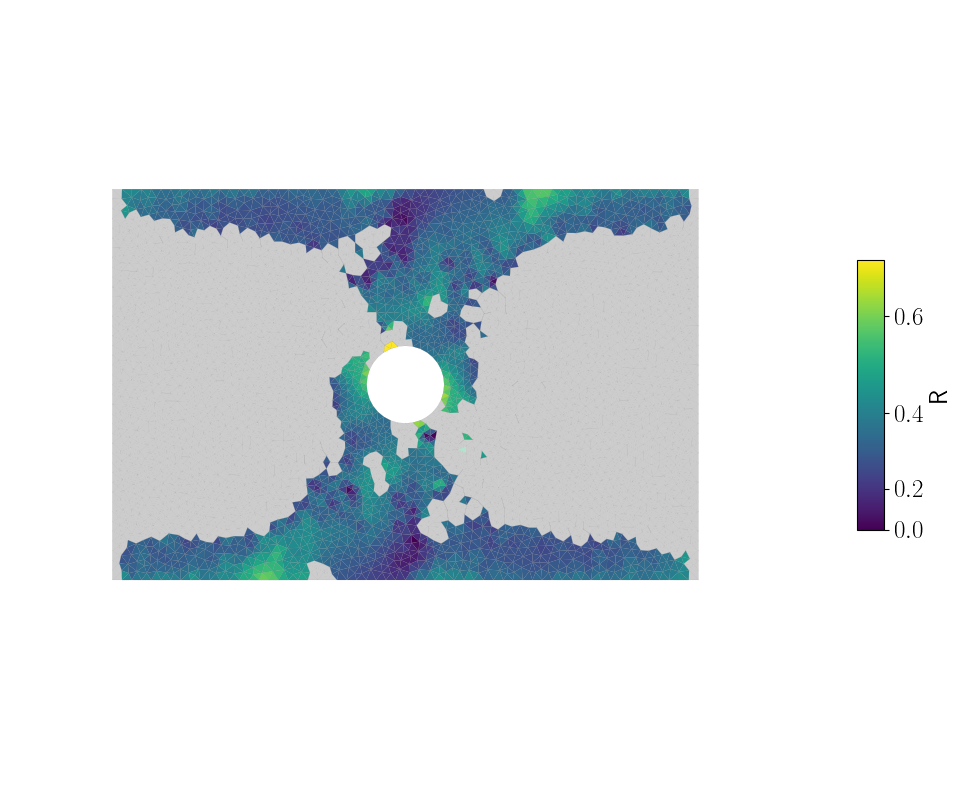}};
    \node(aa) at (80-intf2) {\includegraphics[clip, trim=0 2.0cm 3cm 2.5cm, width=\myfigurewidth]{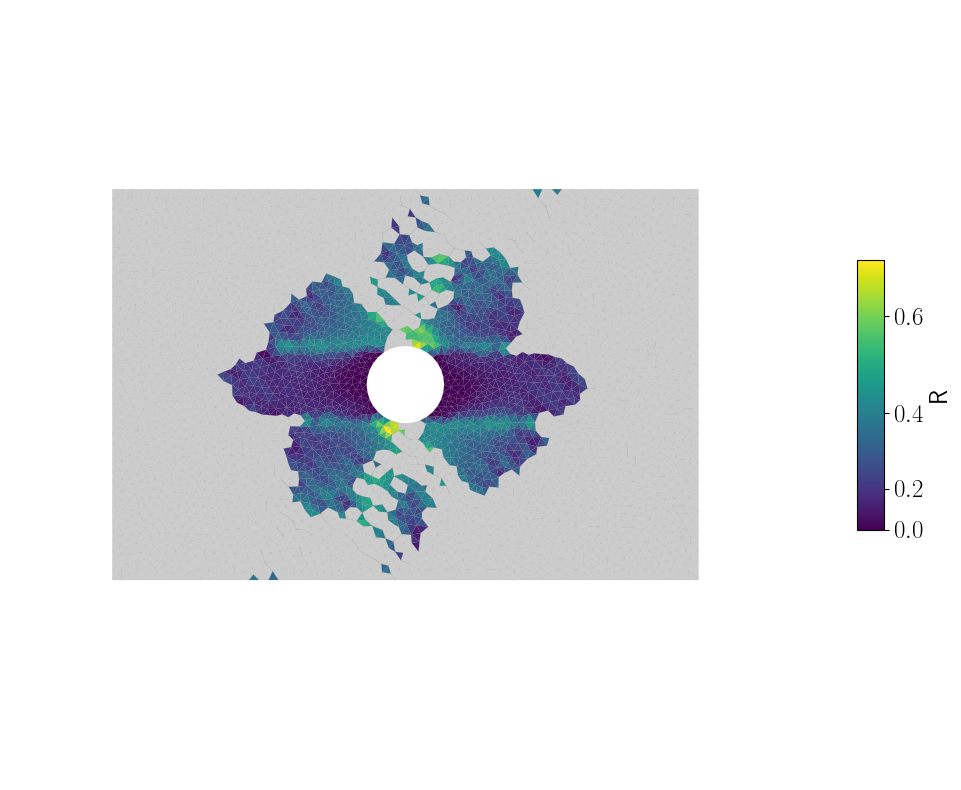}};

    \node(aa) at (colorbar) {\includegraphics[clip, trim=20cm 2.0cm 1.92cm 3.0cm, width=0.15\myfigurewidth]{figures/ply-level/0.80-intf2-locR-62.png}};

    \node[] at (-1.7,-9){\textbf{(a)} $50\%$ severity};
    \node[] at ( 1.7,-9){\textbf{(b)} $80\%$ severity};

    \node[] at ( -5,-1){45/90};
    \node[] at ( -5,-4){90/-45};
    \node[] at ( -5,-7){-45/0};

    \node[] at ( 4.5,-5.5){$R$};
    \node[] at ( 4.9,-7.9){$0.0$};
    \node[] at ( 4.9,-5.9){$0.7$};

  \end{tikzpicture}
  \caption{\emph{Local} stress ratio field in the cohesive zone for each interface at approximately $7\%$ stiffness loss}
  \label{fig:open-hole-local-stress-ratio}
\end{figure}

In \Cref{fig:open-hole-local-stress-ratio}, the \emph{local} $R$ values in the cohesive zone (where $\Dam \in (0,1)$) are plotted as a field for the lowest ($50\%$) and highest load level ($80\%$) in every interface at approximately $7\%$ stiffness loss. It can be observed that the \emph{local} stress ratio is generally higher for the lowest load level, while for the highest load level the \emph{local} stress ratio is close to the \emph{global} load ratio. This can be explained by looking at \Cref{fig:cycle-jump-scheme}, where the stress in a point is the superposition of the thermal residual stress and the stress due to the mechanical load.
With an increased maximum load and equal \emph{global} load ratio, the \emph{local} stress ratio reduces due to the diminishing relative magnitude of the residual stresses.

With an overall larger stress ratio, fatigue damage accumulates slower compared to simulations in which a \emph{global} load ratio is used in every integration point, leading to an increased discrepancy between the simulation results with decreasing maximum load level (\Cref{fig:open-hole_S-N-locR-globR}).

\begin{figure}
   \centering
  \begin{tikzpicture}
  \node[] at (0,0)
    {
      \input{figures/ply-level/all-k-N.tex}
    };

  \draw[->, gray] (1.8,2.3) -- (1.6,1.90) node[pos=0.1,above] {\textcolor{gray}{\emph{local} $R$}};
  \draw[->, gray] (3.1,1.9) -- (2.8,0.74) node[pos=0.1,above] {\textcolor{gray}{\emph{global} $R$}};

  \end{tikzpicture}
  \caption{Stiffness reduction with number of cycles for four different severity levels. Dashed and solid lines correspond to the response with \emph{global} and \emph{local} stress ratio, respectively}
  \label{fig:stiffness-reduction}
\end{figure}
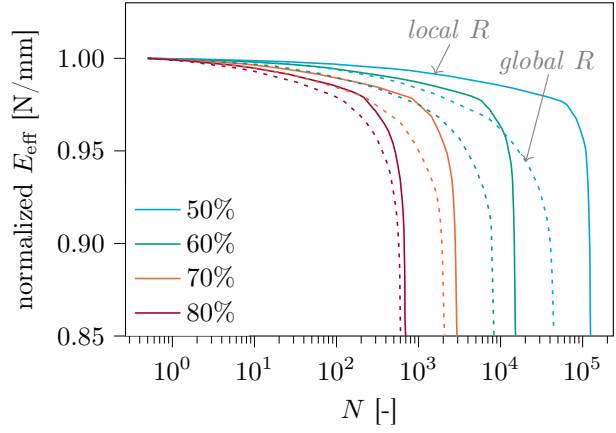

\subsection{Sensitivity study of static material properties}
The fatigue cohesive zone model requires the quasi-static fracture properties and a few fatigue-related parameters. However, strength measurements of IM7/8552 carbon fiber/epoxy vary with different testing methods 
\cite{arndtExperimental2020,May2010c}. 
Moreover, the strength measured with unidirectional laminates is smaller than the in-situ strength of embedded plies in a multidirectional laminate and depends on the fiber direction in the constraining plies, the ply thickness and the location of ply in the laminate \cite{dvorakAnalysis1987,arteiroMicromechanical2014,lauxPly2020,sunInsitu2021}. 

\begin{figure}
  \hspace{-0.5cm}
  \begin{tikzpicture}
  \node[] at (-4.2,0)
    {
      \input{figures/ply-level/sens-matrix_g1-S-N.tex}
    };

  \node[] at (4.2,0)
    {
      \input{figures/ply-level/sens-matrix_f2t-S-N.tex}
    };

  \node[] at (-4.2,-6)
    {
      \input{figures/ply-level/sens-matrix_g2-S-N.tex}
    };

  \node[] at (4.2,-6)
    {
      \input{figures/ply-level/sens-matrix_f6-S-N.tex}
    };

  \node[] at (-4.2,-12)
    {
      \input{figures/ply-level/sens-interface_g2-S-N.tex}
    };

  \node[] at (4.2,-12)
    {
      \input{figures/ply-level/sens-interface_f6-S-N.tex}
    };

  \end{tikzpicture}
 
  \caption{Sensitivity study of static material properties}
  \label{fig:open-hole_sensitivity}
\end{figure}
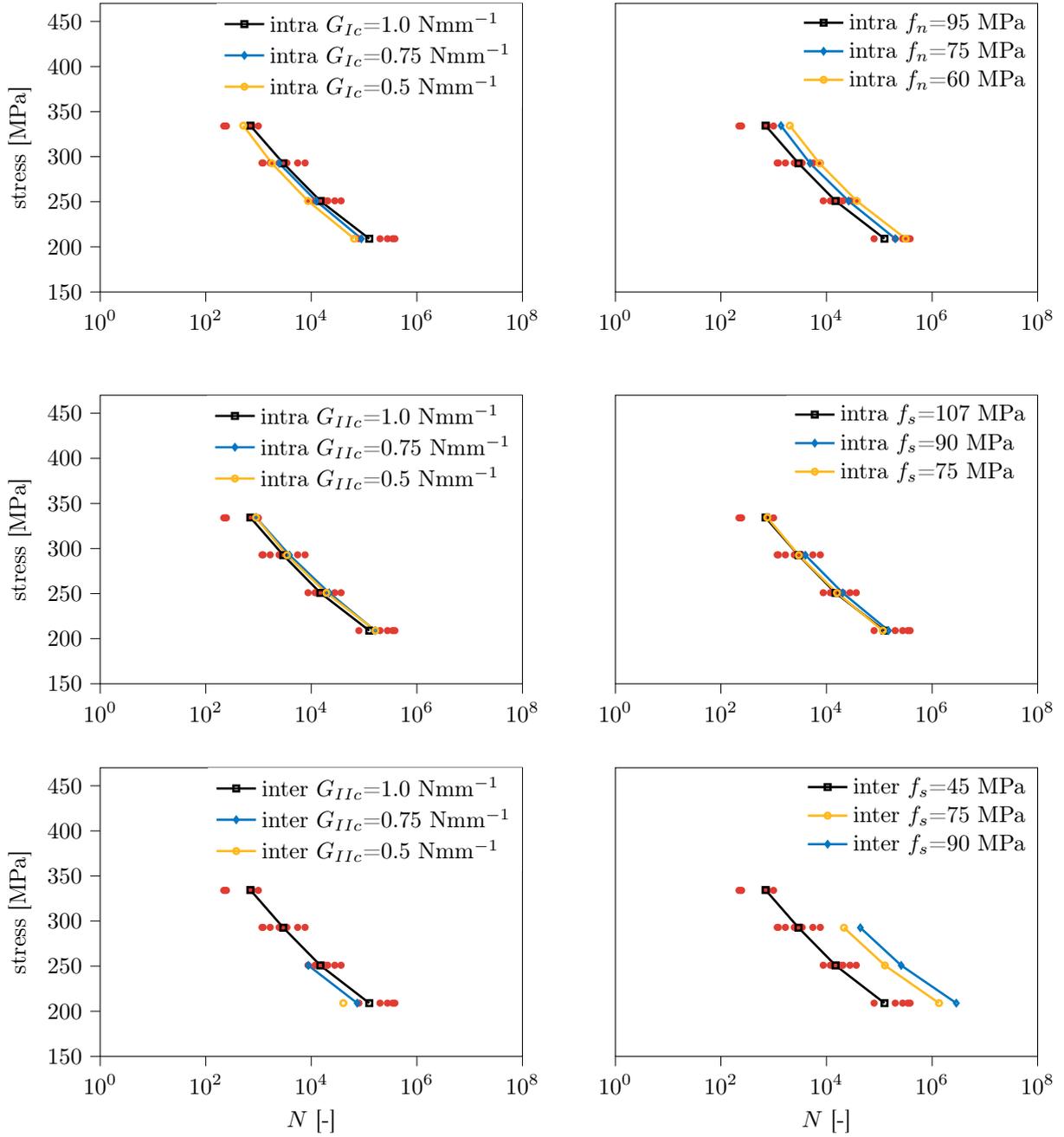

The effect of the static fracture properties on the quasi-isotropic laminate simulations is investigated by repeating the simulation of the \emph{base} case with properties tabulated in \Cref{tab:open-hole-properties}, changing one material property at a time. 
The effect of varying each static property on the laminate S-N curve is shown in \Cref{fig:open-hole_sensitivity}.\footnote{Some simulations did not reach a $15\%$ reduction in effective stiffness due to convergence issues, resulting in incomplete S-N plots.} The curve corresponding to the \emph{base} case is indicated with the black line. It can be observed that the response is only slightly affected by varying the intra-laminar fracture properties. The tensile strength seems to have the largest influence, where increasing the strength results in a shorter fatigue life. Decreasing the intra-laminar strength shifts the underlying \emph{local} S-N curve downwards and consequently leads to more accumulated fatigue damage in the transverse matrix cracks. With more distributed intra-laminar cracking, tractions in the interface decrease and therefore less inter-laminar fatigue damage accumulates \cite{hofmanNumerical2024}. Since the largest stiffness drops are associated with interface delamination (see \Cref{fig:open-hole_damage-evol}), the fatigue life of the laminate is longer. 
 The interface shear strength shows the most influence on the \emph{global} S-N curve, where an increase in shear strength results in a longer fatigue life. The slope of the underlying \emph{local} S-N curve remains the same with increasing strength, while the curve is shifted upwards, thus resulting in less fatigue damage at the same stress level compared to a lower interfacial strength. Finally, decreasing the interfacial mode-II fracture energy results in a shorter fatigue life, although the effect is minimal in the range of typical values for the mode-II fracture energy (0.75 - \SI{1.0}{\newton\per\milli\meter}).

Previously, the static version of the open-hole simulation \cite{vandermeerComputationalModelingComplex2012} indicated that fracture energy is a more important parameter than strength, since laminate failure is governed by delamination propagation. In elementary static crack propagation tests, Turon's static mixed-mode cohesive zone model, which is the basis of the fatigue formulation by \Davila~\cite{Davila2020}, ensures correct energy dissipation, independent of strength \cite{Turon2010, Turon2018b}. However, with the fatigue damage extension, it has been shown that crack propagation rates do in fact depend on the static strength values \cite{lecinana2022global,lecinanaRobust2023}. The results of the present study confirm these previous findings.

\subsubsection{Sub-laminate scaled specimen}

The \emph{sub-laminate scaled specimen} is simulated with five maximum \emph{global} stress levels $\sigma^{\mathrm{max}}(\mathrm{MPa})=\{523.3, 494.2, 465.1, 407.0, 377.9\}$, corresponding to $90, 85, 80, 70$ and $65\%$ severity levels, respectively. The \emph{global} load ratio $R_\mathrm{glob}$ is $0.1$.

The S-N curve is shown in \Cref{fig:open-hole-sublam_S-N}, where fatigue life is again defined as the number of cycles to reach a $15\%$ reduction of the initial effective stiffness. 
It can be observed that a good match is obtained in terms of fatigue life, except for the experimental specimens with the highest peak load.
It is reported in Ref.~\cite{nixon-pearsonInvestigationDamageDevelopment2015} that the highest load level ($90\%$ severity) resulted in a pull-out failure mode (which cannot be captured with the present numerical framework where fiber failure is not considered), while the second highest load level ($85\% $ severity) showed both delamination and pull-out failures. 
However, the three lowest load levels resulted in all cases in a delamination-type failure mode in the experiments.
For the specimens with a delamination-type failure mode, the simulation results are in excellent accordance with the experiments.

\begin{figure}
  \centering 
  \vspace{-3.0cm}
  \begin{tikzpicture}
    \newlength{\myfigurewidth}
    \setlength{\myfigurewidth}{0.4\columnwidth} 
    
    \node at (3.4,11.0) {\includegraphics[clip, width=0.72\myfigurewidth, trim = 6.5cm 0.1cm 6.5cm 0.2cm]{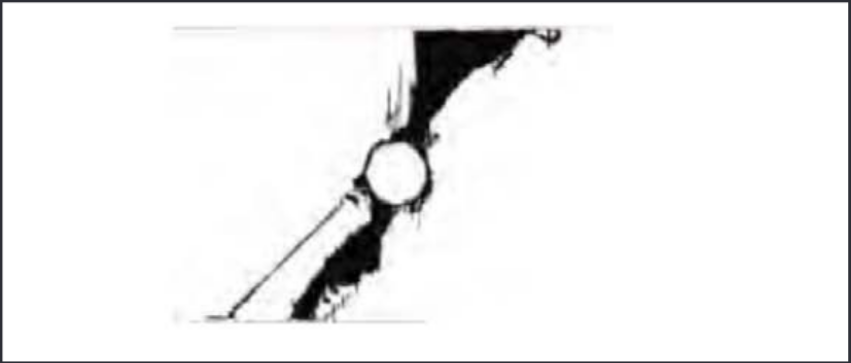}};

    \node at (3,8.1) {\includegraphics[clip, width=0.72\myfigurewidth, trim = 6.5cm 0.1cm 6.5cm 0.2cm]{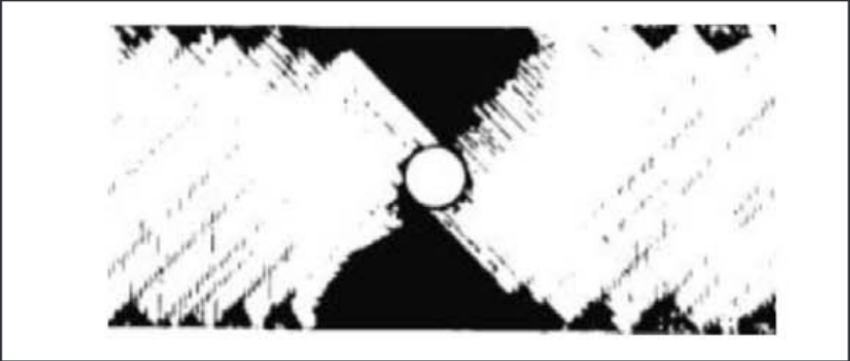}};

    \node at (3,5.1) {\includegraphics[clip, width=0.72\myfigurewidth, trim = 6.5cm 0.1cm 6.5cm 0.2cm]{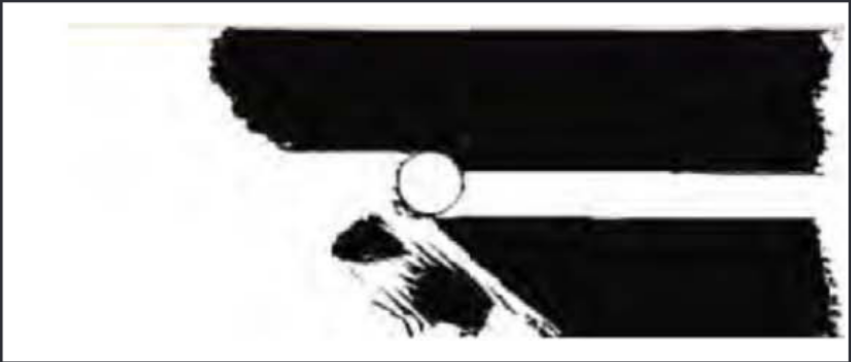}};

    \node at (3.2,2.0) {\includegraphics[clip, width=0.72\myfigurewidth, trim = 6.5cm 0.1cm 6.5cm 1.5cm]{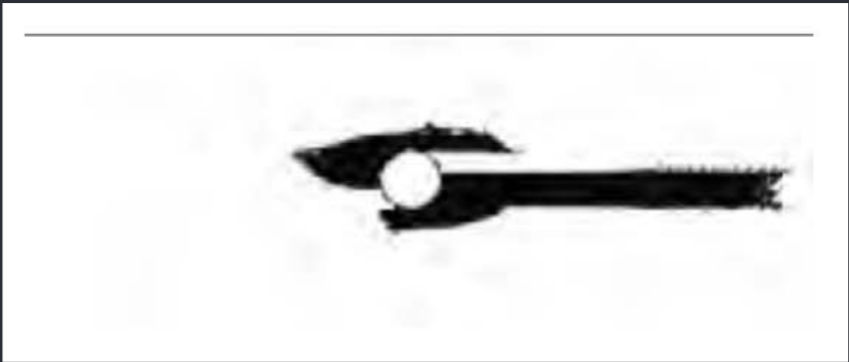}};

    \node at (3.2,-1.0) {\includegraphics[clip, width=0.72\myfigurewidth, trim = 6.5cm 0.1cm 6.5cm 0.2cm]{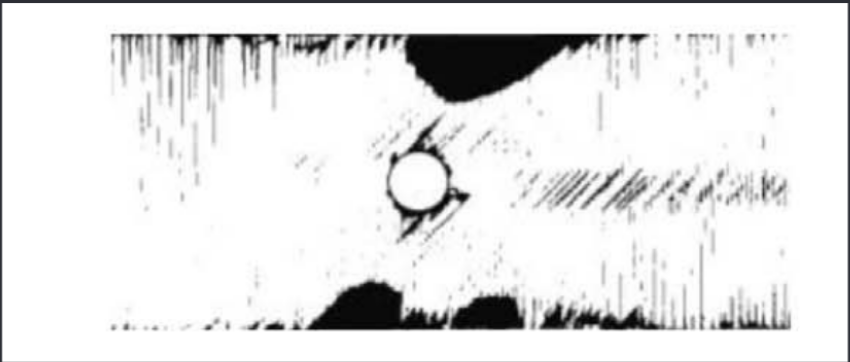}};

    \node at (3.0,-7.0) {\includegraphics[clip, width=0.72\myfigurewidth, trim = 6.5cm 0.1cm 6.5cm 0.2cm]{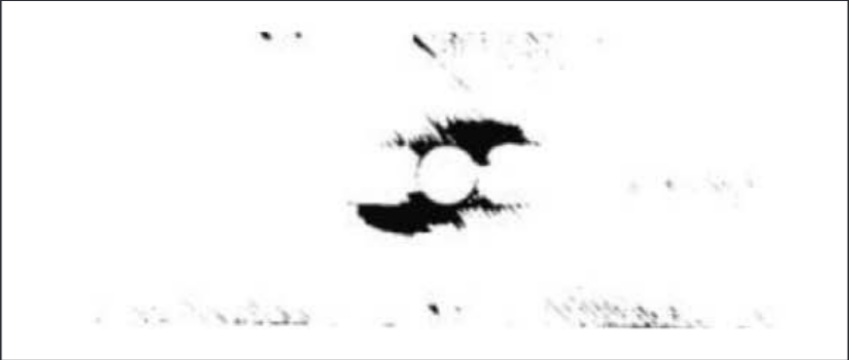}};

    \node at (-3,11.0) {\includegraphics[clip, width=\myfigurewidth, trim = 0 0.0cm 0 0cm]{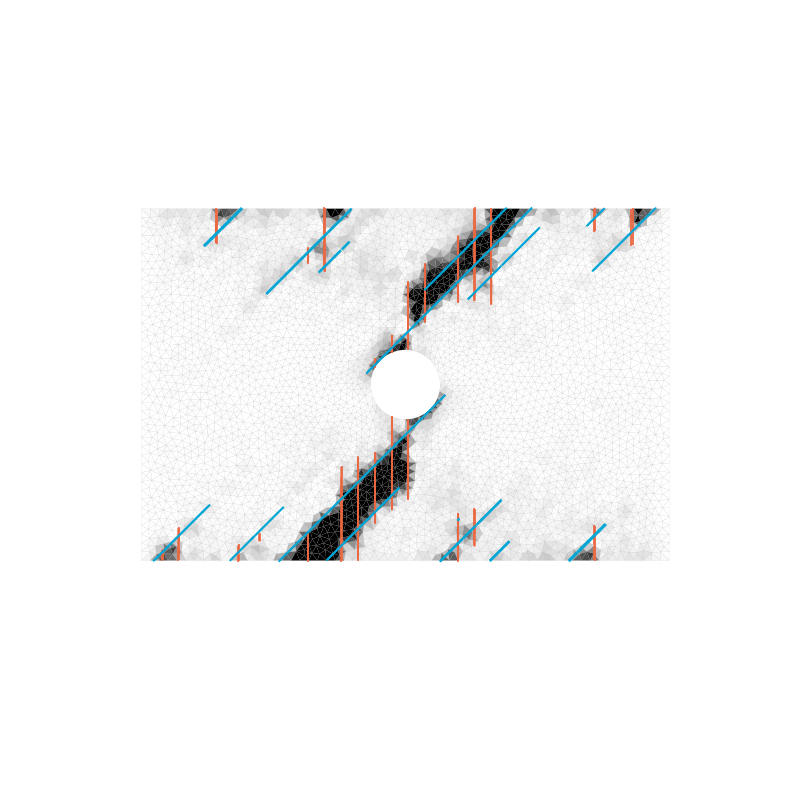}};

    \node at (-3,8.0) {\includegraphics[clip, width=\myfigurewidth, trim = 0 0.0cm 0 0cm]{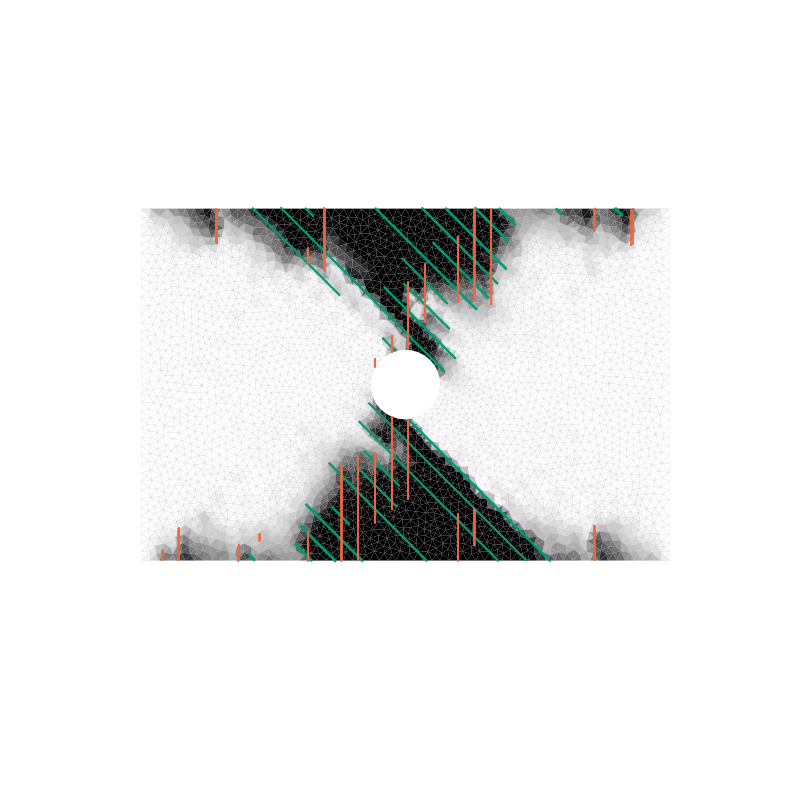}};

    \node at (-3,5.0) {\includegraphics[clip, width=\myfigurewidth, trim = 0 0.0cm 0 0cm]{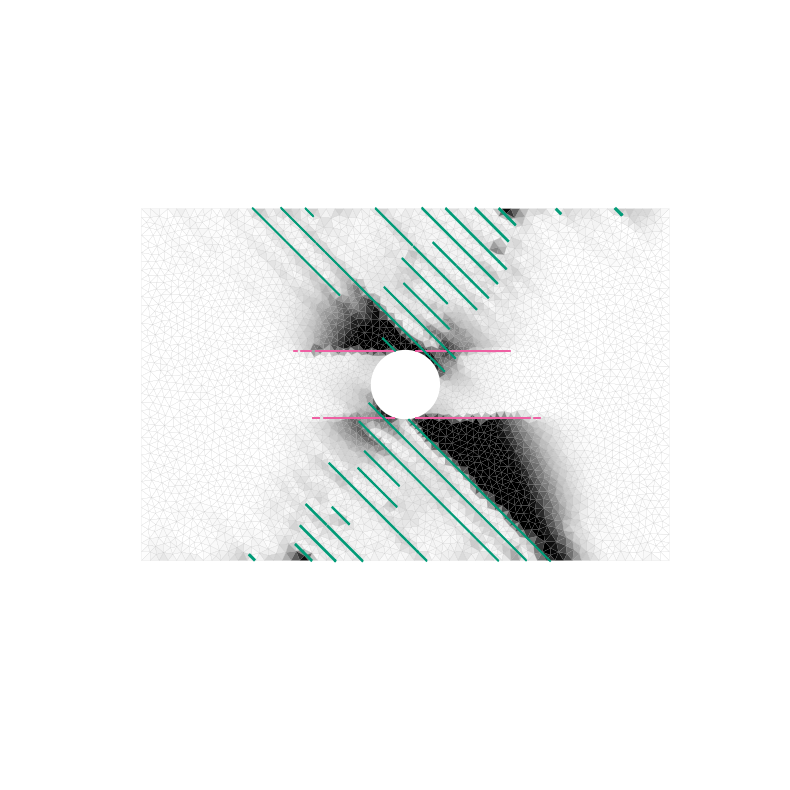}};

    \node at (-3,2.0) {\includegraphics[clip, width=\myfigurewidth, trim = 0 0.0cm 0 0cm]{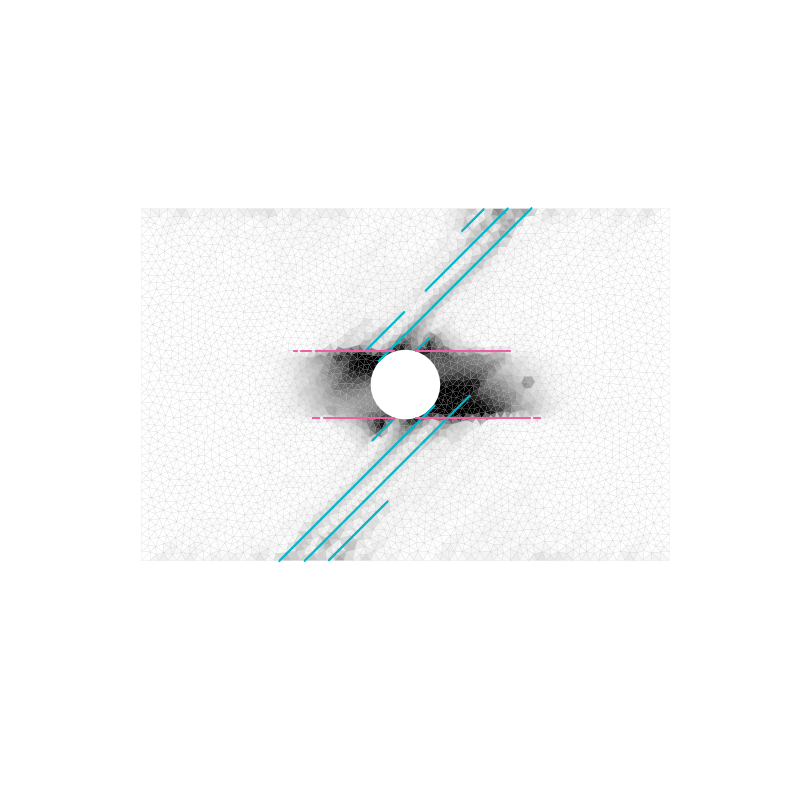}};

    \node at (-3,-1.0) {\includegraphics[clip, width=\myfigurewidth, trim = 0 0.0cm 0 0cm]{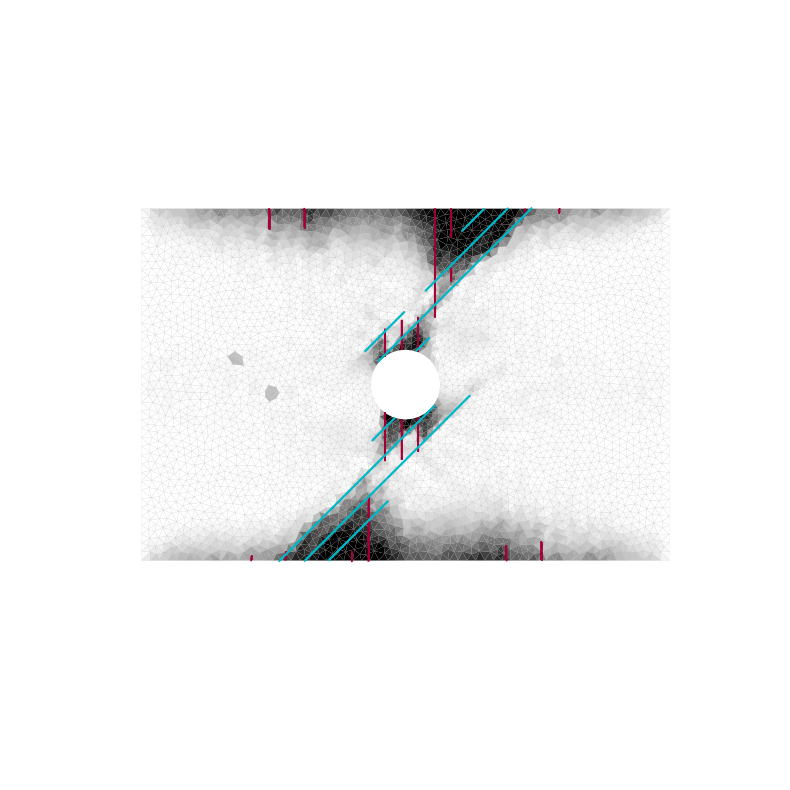}};

    \node at (-3,-4.0) {\includegraphics[clip, width=\myfigurewidth, trim = 0 0.0cm 0 0cm]{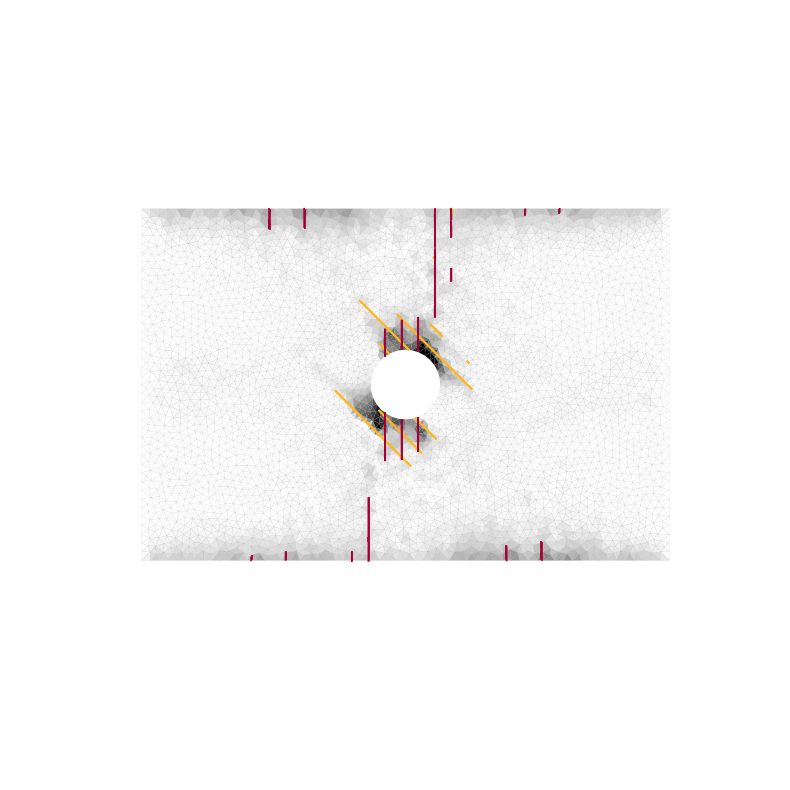}};

    \node at (-3,-7.0) {\includegraphics[clip, width=\myfigurewidth, trim = 0 0.0cm 0 0cm]{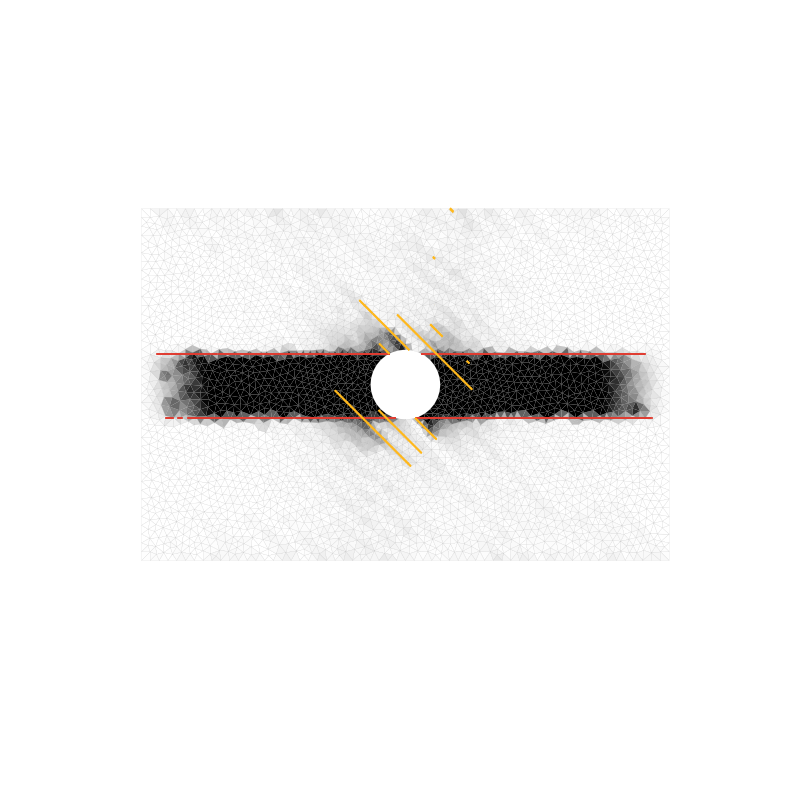}};

    \node[] at ( -7,11){outer 45/90};
    \node[] at ( -7,8){outer 90/-45};
    \node[] at ( -7,5){outer -45/0};
    \node[] at ( -7,2){outer 0/45};
    \node[] at ( -7,-1){inner 45/90};
    \node[] at ( -7,-4){inner 90/-45};
    \node[] at ( 3,-4){missing};
    \node[] at ( -7,-7){inner -45/0};
  \end{tikzpicture}

  \caption{Sub-laminate scaled specimen: damage in XFEM matrix cracks and interface delamination vs experimental CT scans (taken from \cite{Tao2018}) at $9\%$ stiffness loss. Only fully damaged matrix cracks ($\Dam=1$) are depicted}
  \label{fig:open-hole-sublam_damage}
\end{figure}
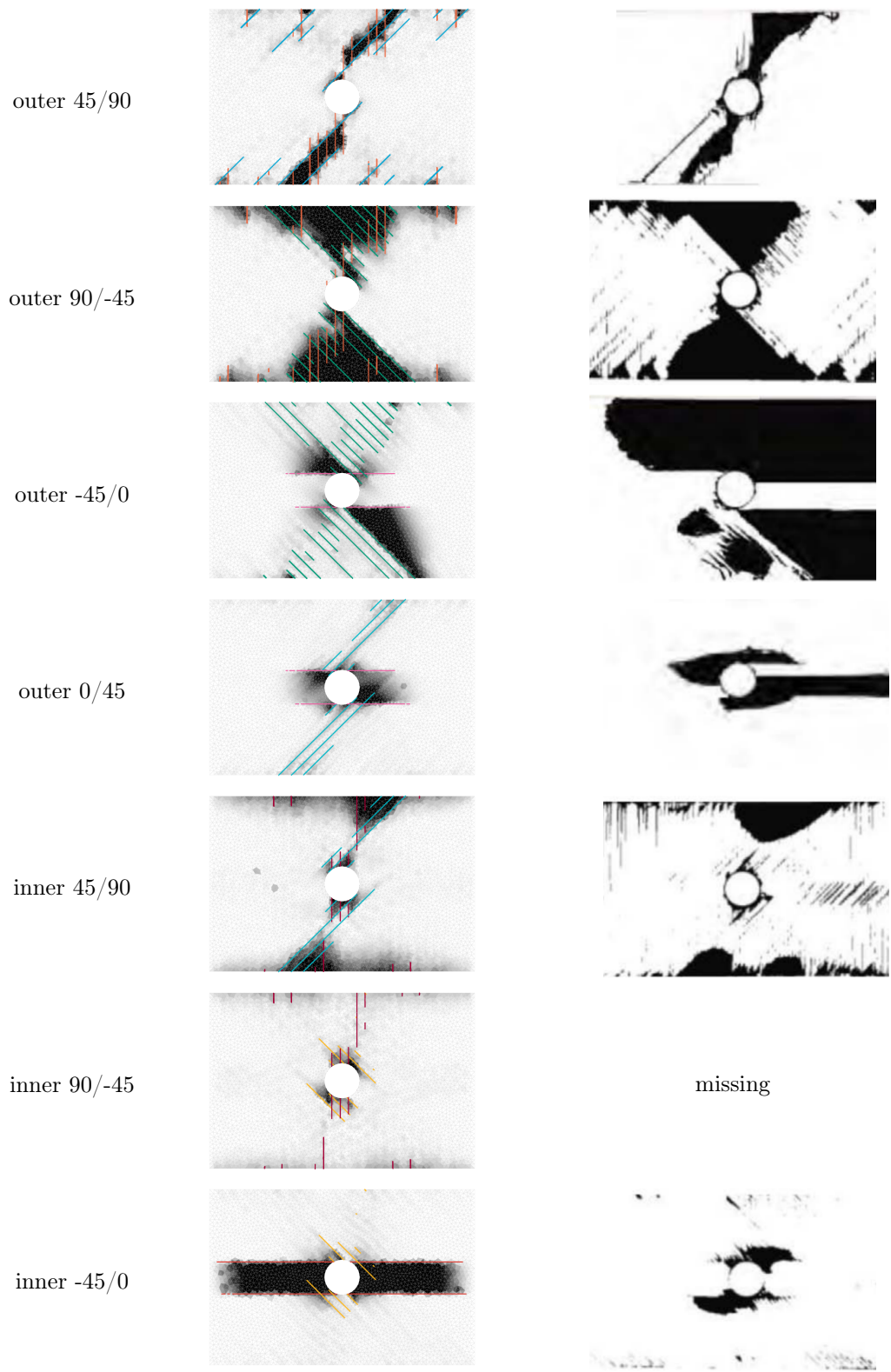

The simulation damage patterns are compared to the experimental patterns under severity $80\%$, at
approximately $9\%$ stiffness loss (see \Cref{fig:open-hole-sublam_damage}). The patterns are in good agreement for the outer $45/90$, $90/\mathrm{-}45$ and inner $45/90$ interfaces. 
Similar to the experimental observations \cite{nixon-pearsonInvestigationDamageDevelopment2015, Tao2018}, damage grows from the free edge towards the hole in the inner $45/90$ interface.
Also a more dispersed damage pattern compared to the \emph{ply-level scaled specimen}, with increased free-edge delamination, can be observed.
For the outer $\mathrm{-}45/0$ and $0/45$ interfaces, slightly underdeveloped delamination is predicted, while the inner $\mathrm{-45/0}$ interface shows overdeveloped delamination. Given the scatter in experimental damage patterns \cite{nixon-pearsonInvestigationDamageDevelopment2015, Tao2018}, the simulated patterns are overall in good correspondence with the experimental ones. 

\begin{figure}
   \centering
  \begin{tikzpicture}

  \node[] at (0,0)
    {
      \input{figures/sub-laminate/S-N.tex}
    };
  
  \end{tikzpicture}
  \caption{S-N curve of the sub-laminate scaled specimen. Experimental values are extracted from \cite{nixon-pearsonInvestigationDamageDevelopment2015}. The symbols ${\tiny\textcolor{red}{\square}}$ and $\textcolor{red}{\bullet}$ indicate experimental specimens with pull-out and delamination-type failure modes, respectively}
  \label{fig:open-hole-sublam_S-N}
\end{figure}

%% file: inkscape/open-hole-dimensions.pdf_tex
\begingroup%
  \makeatletter%
  \providecommand\color[2][]{%
    \errmessage{(Inkscape) Color is used for the text in Inkscape, but the package 'color.sty' is not loaded}%
    \renewcommand\color[2][]{}%
  }%
  \providecommand\transparent[1]{%
    \errmessage{(Inkscape) Transparency is used (non-zero) for the text in Inkscape, but the package 'transparent.sty' is not loaded}%
    \renewcommand\transparent[1]{}%
  }%
  \providecommand\rotatebox[2]{#2}%
  \newcommand*\fsize{\dimexpr\f@size pt\relax}%
  \newcommand*\lineheight[1]{\fontsize{\fsize}{#1\fsize}\selectfont}%
  \ifx\svgwidth\undefined%
    \setlength{\unitlength}{340.23271936bp}%
    \ifx\svgscale\undefined%
      \relax%
    \else%
      \setlength{\unitlength}{\unitlength * \real{\svgscale}}%
    \fi%
  \else%
    \setlength{\unitlength}{\svgwidth}%
  \fi%
  \global\let\svgwidth\undefined%
  \global\let\svgscale\undefined%
  \makeatother%
  \begin{picture}(1,0.2265676)%
    \lineheight{1}%
    \setlength\tabcolsep{0pt}%
    \put(0,0){\includegraphics[width=\unitlength,page=1]{open-hole-dimensions.pdf}}%
    \put(0.32339382,0.01152182){\color[rgb]{0,0,0}\makebox(0,0)[lt]{\lineheight{1.25}\smash{\begin{tabular}[t]{l}$64$\end{tabular}}}}%
    \put(-0.0014465,0.11389214){\color[rgb]{0,0,0}\makebox(0,0)[lt]{\lineheight{1.25}\smash{\begin{tabular}[t]{l}$16$\end{tabular}}}}%
    \put(0,0){\includegraphics[width=\unitlength,page=2]{open-hole-dimensions.pdf}}%
    \put(0.62452859,0.20039972){\color[rgb]{0.04705882,0.1372549,0.25098039}\makebox(0,0)[lt]{\lineheight{1.25}\smash{\begin{tabular}[t]{l}$\sigma^{\mathrm{max}}$\end{tabular}}}}%
    \put(0,0){\includegraphics[width=\unitlength,page=3]{open-hole-dimensions.pdf}}%
    \put(0.33824929,0.21332792){\color[rgb]{0,0,0}\makebox(0,0)[lt]{\lineheight{1.25}\smash{\begin{tabular}[t]{l}$24$\end{tabular}}}}%
    \put(0,0){\includegraphics[width=\unitlength,page=4]{open-hole-dimensions.pdf}}%
    \put(0.37894406,0.14730194){\color[rgb]{0,0,0}\makebox(0,0)[lt]{\lineheight{1.25}\smash{\begin{tabular}[t]{l}$3.175$\end{tabular}}}}%
  \end{picture}%
\endgroup%

%% file: figures/ply-level/S-N-locR.tex
\begin{tikzpicture}

\definecolor{crimson2246049}{RGB}{224,60,49}
\definecolor{darkcyan0118194}{RGB}{0,118,194}
\definecolor{darkcyan0155119}{RGB}{0,155,119}
\definecolor{darkgray176}{RGB}{176,176,176}
\definecolor{gray}{RGB}{128,128,128}

\begin{axis}[
height=6cm,
legend cell align={left},
legend style={fill opacity=0.8, draw opacity=1, text opacity=1, at={(1,1)}, draw=none},
log basis x={10},
minor xtick={},
minor ytick={},
tick align=outside,
tick pos=left,
width=8cm,
x grid style={darkgray176},
xlabel={\(\displaystyle N\) [-]},
xmin=1, xmax=100000000,
xmode=log,
xtick style={color=black},
xtick={0.01,1,100,10000,1000000,100000000,10000000000},
xticklabels={
  \(\displaystyle {10^{-2}}\),
  \(\displaystyle {10^{0}}\),
  \(\displaystyle {10^{2}}\),
  \(\displaystyle {10^{4}}\),
  \(\displaystyle {10^{6}}\),
  \(\displaystyle {10^{8}}\),
  \(\displaystyle {10^{10}}\)
},
y grid style={darkgray176},
ylabel={stress [MPa]},
ymin=150, ymax=470,
ytick style={color=black},
ytick={150,200,250,300,350,400,450,500},
yticklabels={
  \(\displaystyle {150}\),
  \(\displaystyle {200}\),
  \(\displaystyle {250}\),
  \(\displaystyle {300}\),
  \(\displaystyle {350}\),
  \(\displaystyle {400}\),
  \(\displaystyle {450}\),
  \(\displaystyle {500}\)
}
]
\addplot [semithick, crimson2246049, mark=*, mark size=1.25, mark options={solid}, only marks]
table {%
79722.6492655018 209
200578.652799832 209
277256.827517136 209
347776.73606127 209
383247.904659031 209
8679.37747021351 251
11804.7349131027 251
13008.748002353 251
15048.8874339328 251
16583.7848782625 251
17981.7949256128 251
20467.842020279 251
27838.1082193806 251
36656.2156999394 251
1166.22898574719 293
1224.25944214336 293
1665.10308244606 293
2455.60309582591 293
2750.21920888487 293
3130.44679474198 293
3449.7338387812 293
5516.3526382235 293
7625.17052832109 293
220.143822243671 334
234.869168649772 334
246.556037366555 334
741.220250458929 334
765.608999639563 334
991.939218778981 334
};
\addlegendentry{experiments}
\addplot [semithick, darkcyan0155119, mark=asterisk, mark size=4, mark options={solid,fill opacity=0}, only marks]
table {%
397.350307752287 334.4
2848.0358684358 292.6
23545.1155672878 251
249283.891030978 209
};
\addlegendentry{exp. mean}
\addplot [line width=1pt, darkcyan0118194, mark=square, mark size=1.25, mark options={solid,fill opacity=0}]
table {%
125570.948470609 209
14854.0619833096 250.8
2949.15547050009 292.6
703.10552880584 334.4
};
\addlegendentry{simulation}
\draw (axis cs:8,330) node[
  scale=0.9,
  anchor=base west,
  text=gray,
  rotate=0.0
]{$80\%$};
\draw (axis cs:40,280) node[
  scale=0.9,
  anchor=base west,
  text=gray,
  rotate=0.0
]{$70\%$};
\draw (axis cs:300,240) node[
  scale=0.9,
  anchor=base west,
  text=gray,
  rotate=0.0
]{$60\%$};
\draw (axis cs:2000,200) node[
  scale=0.9,
  anchor=base west,
  text=gray,
  rotate=0.0
]{$50\%$};
\end{axis}

\end{tikzpicture}

%% file: figures/ply-level/k-N.tex
\begin{tikzpicture}

\definecolor{darkcyan0118194}{RGB}{0,118,194}
\definecolor{darkgray176}{RGB}{176,176,176}

\begin{axis}[
height=6cm,
log basis x={10},
minor xtick={0.02,0.03,0.04,0.05,0.06,0.07,0.08,0.09,0.2,0.3,0.4,0.5,0.6,0.7,0.8,0.9,2,3,4,5,6,7,8,9,20,30,40,50,60,70,80,90,200,300,400,500,600,700,800,900,2000,3000,4000,5000,6000,7000,8000,9000,20000,30000,40000,50000,60000,70000,80000,90000,200000,300000,400000,500000,600000,700000,800000,900000,2000000,3000000,4000000,5000000,6000000,7000000,8000000,9000000,20000000,30000000,40000000,50000000,60000000,70000000,80000000,90000000},
minor ytick={},
tick align=outside,
tick pos=left,
width=8cm,
x grid style={darkgray176},
xlabel={\(\displaystyle N\) [-]},
xmin=0.298410879685173, xmax=1000000,
xmode=log,
xtick style={color=black},
xtick={0.01,0.1,1,10,100,1000,10000,100000,1000000,10000000},
xticklabels={
  \(\displaystyle {10^{-2}}\),
  \(\displaystyle {10^{-1}}\),
  \(\displaystyle {10^{0}}\),
  \(\displaystyle {10^{1}}\),
  \(\displaystyle {10^{2}}\),
  \(\displaystyle {10^{3}}\),
  \(\displaystyle {10^{4}}\),
  \(\displaystyle {10^{5}}\),
  \(\displaystyle {10^{6}}\),
  \(\displaystyle {10^{7}}\)
},
y grid style={darkgray176},
ylabel={normalized \(\displaystyle E_{\mathrm{eff}}\) [N/mm]},
ymin=0.84, ymax=1.03,
ytick style={color=black},
ytick={0.85,0.9,0.95,1},
yticklabels={
  \(\displaystyle {0.85}\),
  \(\displaystyle {0.90}\),
  \(\displaystyle {0.95}\),
  \(\displaystyle {1.00}\)
}
]
\addplot [semithick, darkcyan0118194]
table {%
0.5 1
0.75 0.999826459570623
1.75 0.999417854682469
2.75 0.99907719887181
4.75 0.998656621431473
8.75 0.998112040754346
12.75 0.997682385152894
20.75 0.997092223380809
36.75 0.996286606586104
52.75 0.995661645656542
84.75 0.994685265154489
148.75 0.993130523497
212.75 0.992106553566109
340.75 0.990724227493693
596.75 0.988975871261636
852.75 0.987636242332563
1108.75 0.986568806661596
1620.75 0.98484270885099
2132.75 0.983468220757128
3156.75 0.981477352990806
4385.55 0.979598284901387
5122.83 0.978370581733754
5860.11 0.977089372090361
6597.39 0.974441673135028
7039.758 0.973227088312684
7358.26296 0.972380068521937
7549.365936 0.971814631627501
7740.468912 0.970669126789472
7836.0204 0.970317639376889
8218.226352 0.969051067916397
8600.432304 0.967704614899902
8791.53528 0.967063013986203
9555.947184 0.964697139934973
10320.359088 0.962228500463287
11084.770992 0.959163214933251
11543.4181344 0.957050994171667
11818.60641984 0.955452040849686
12093.79470528 0.953308448817824
12102.3943392 0.953247058220336
12110.99397312 0.953185029260427
12145.3925088 0.952942447572227
12227.948994432 0.952346312551823
12310.505480064 0.951444748912017
12310.827966336 0.951442270533528
12313.407856512 0.951423077790591
12315.987746688 0.951403151425122
12336.626868096 0.951260123267668
12419.183353728 0.950695388340166
12501.73983936 0.950139045940798
12831.96578189 0.947836121922889
13162.19172442 0.94528706379022
13492.41766695 0.942125311229256
13822.64360948 0.93710570097237
13987.75658074 0.933822391171726
14152.869552 0.926925101779062
14235.426037632 0.92383401495444
14400.539008892 0.917512416381579
14565.651980152 0.908805195950521
14648.208465784 0.904087041405255
14730.764951416 0.897246107867883
14813.321437048 0.887350703630387
14833.960558456 0.885204014500208
14875.238801272 0.881169886618297
14916.517044088 0.876639537334701
14941.283989778 0.873833384194295
14956.144157192 0.872124451724833
14971.004324606 0.870429662297734
15000.724659434 0.866612014750235
15015.584826848 0.864756559613093
15030.444994262 0.862917730649247
15045.305161676 0.861290731247743
15075.025496504 0.858362966840398
15104.745831332 0.855848427501901
15122.578032229 0.854530319597397
15143.976673305 0.853171158469834
15165.375314381 0.852026679346191
15208.172596533 0.850295328876459
};
\addplot [semithick, red, mark=*, mark size=1, mark options={solid}]
table {%
0.5 1
};
\addplot [semithick, red, mark=*, mark size=1, mark options={solid}]
table {%
4385.55 0.979598284901387
};
\addplot [semithick, red, mark=*, mark size=1, mark options={solid}]
table {%
12419.183353728 0.950695388340166
};
\addplot [semithick, red, mark=*, mark size=1, mark options={solid}]
table {%
15208.172596533 0.850295328876459
};
\draw (axis cs:0.7,1.004) node[
  scale=0.9,
  anchor=base west,
  text=black,
  rotate=0.0
]{$N_1$};
\draw (axis cs:4485.55,0.983598284901387) node[
  scale=0.9,
  anchor=base west,
  text=black,
  rotate=0.0
]{$N_2$};
\draw (axis cs:13419.183353728,0.954695388340166) node[
  scale=0.9,
  anchor=base west,
  text=black,
  rotate=0.0
]{$N_3$};
\draw (axis cs:16208.172596533,0.854295328876459) node[
  scale=0.9,
  anchor=base west,
  text=black,
  rotate=0.0
]{$N_4$};
\end{axis}

\end{tikzpicture}

%% file: figures/ply-level/all-k-N.tex
\begin{tikzpicture}

\definecolor{darkcyan0155119}{RGB}{0,155,119}
\definecolor{darkgray176}{RGB}{176,176,176}
\definecolor{darkturquoise0166214}{RGB}{0,166,214}
\definecolor{firebrick165052}{RGB}{165,0,52}
\definecolor{tomato23610466}{RGB}{236,104,66}

\begin{axis}[
height=6cm,
legend cell align={left},
legend style={
  fill opacity=0.8,
  draw opacity=1,
  text opacity=1,
  at={(0,0)},
  anchor=south west,
  draw=none
},
log basis x={10},
minor xtick={0.02,0.03,0.04,0.05,0.06,0.07,0.08,0.09,0.2,0.3,0.4,0.5,0.6,0.7,0.8,0.9,2,3,4,5,6,7,8,9,20,30,40,50,60,70,80,90,200,300,400,500,600,700,800,900,2000,3000,4000,5000,6000,7000,8000,9000,20000,30000,40000,50000,60000,70000,80000,90000,200000,300000,400000,500000,600000,700000,800000,900000,2000000,3000000,4000000,5000000,6000000,7000000,8000000,9000000,20000000,30000000,40000000,50000000,60000000,70000000,80000000,90000000},
minor ytick={},
tick align=outside,
tick pos=left,
width=8cm,
x grid style={darkgray176},
xlabel={\(\displaystyle N\) [-]},
xmin=0.268375328579348, xmax=236453.486991419,
xmode=log,
xtick style={color=black},
xtick={0.01,0.1,1,10,100,1000,10000,100000,1000000,10000000},
xticklabels={
  \(\displaystyle {10^{-2}}\),
  \(\displaystyle {10^{-1}}\),
  \(\displaystyle {10^{0}}\),
  \(\displaystyle {10^{1}}\),
  \(\displaystyle {10^{2}}\),
  \(\displaystyle {10^{3}}\),
  \(\displaystyle {10^{4}}\),
  \(\displaystyle {10^{5}}\),
  \(\displaystyle {10^{6}}\),
  \(\displaystyle {10^{7}}\)
},
y grid style={darkgray176},
ylabel={normalized \(\displaystyle E_{\mathrm{eff}}\) [N/mm]},
ymin=0.85, ymax=1.03,
ytick style={color=black},
ytick={0.85,0.9,0.95,1},
yticklabels={
  \(\displaystyle {0.85}\),
  \(\displaystyle {0.90}\),
  \(\displaystyle {0.95}\),
  \(\displaystyle {1.00}\)
}
]
\addplot [semithick, darkturquoise0166214]
table {%
0.5 1
2.5 0.99947183916863
13.813708499 0.998546464463668
25.127416998 0.998125661408551
89.127416998 0.996927844090322
451.166088968 0.994504017574868
813.204760938 0.993329473047451
2261.359448838 0.990527995327887
8053.978200338 0.986123012024075
13846.596951838 0.983971219420529
37017.071957838 0.979206777197605
55017.071957838 0.976833110426051
61497.071957838 0.975505373763548
64856.303957838 0.974718475317718
83858.989771838 0.96730635582936
95260.601260838 0.960423646282498
102226.304736038 0.956081332767939
108244.672538638 0.951606545188851
110411.284947538 0.948453587536233
115611.154728938 0.933907428316219
116261.138451618 0.931780417502293
116911.122174298 0.928914109872196
118749.553766198 0.91904331736096
122426.416949898 0.895795681345807
122597.964678598 0.892220757087413
122889.090828028 0.885749909210479
122995.807239098 0.883276138538071
123018.857983889 0.882479158665081
123030.807489989 0.882124265752139
123051.456236529 0.881552671430919
123072.104983069 0.880990829140845
123188.911932679 0.877887269945891
123469.248611749 0.871378769883355
123749.585290819 0.864365972452979
124225.332411109 0.855746334242315
125570.948470609 0.846637975939998
126916.564530109 0.842954131595424
};
\addlegendentry{$50\%$}
\addplot [semithick, darkcyan0155119]
table {%
0.5 1
0.75 0.999826459570623
1.75 0.999417854682469
2.75 0.99907719887181
4.75 0.998656621431473
8.75 0.998112040754346
12.75 0.997682385152894
20.75 0.997092223380809
36.75 0.996286606586104
52.75 0.995661645656542
84.75 0.994685265154489
148.75 0.993130523497
212.75 0.992106553566109
340.75 0.990724227493693
596.75 0.988975871261636
852.75 0.987636242332563
1108.75 0.986568806661596
1620.75 0.98484270885099
2132.75 0.983468220757128
3156.75 0.981477352990806
4385.55 0.979598284901387
5122.83 0.978370581733754
5860.11 0.977089372090361
6597.39 0.974441673135028
7039.758 0.973227088312684
7358.26296 0.972380068521937
7549.365936 0.971814631627501
7740.468912 0.970669126789472
7836.0204 0.970317639376889
8218.226352 0.969051067916397
8600.432304 0.967704614899902
8791.53528 0.967063013986203
9555.947184 0.964697139934973
10320.359088 0.962228500463287
11084.770992 0.959163214933251
11543.4181344 0.957050994171667
11818.60641984 0.955452040849686
12093.79470528 0.953308448817824
12102.3943392 0.953247058220336
12110.99397312 0.953185029260427
12145.3925088 0.952942447572227
12227.948994432 0.952346312551823
12310.505480064 0.951444748912017
12310.827966336 0.951442270533528
12313.407856512 0.951423077790591
12315.987746688 0.951403151425122
12336.626868096 0.951260123267668
12419.183353728 0.950695388340166
12501.73983936 0.950139045940798
12831.96578189 0.947836121922889
13162.19172442 0.94528706379022
13492.41766695 0.942125311229256
13822.64360948 0.93710570097237
13987.75658074 0.933822391171726
14152.869552 0.926925101779062
14235.426037632 0.92383401495444
14400.539008892 0.917512416381579
14565.651980152 0.908805195950521
14648.208465784 0.904087041405255
14730.764951416 0.897246107867883
14813.321437048 0.887350703630387
14833.960558456 0.885204014500208
14875.238801272 0.881169886618297
14916.517044088 0.876639537334701
14941.283989778 0.873833384194295
14956.144157192 0.872124451724833
14971.004324606 0.870429662297734
15000.724659434 0.866612014750235
15015.584826848 0.864756559613093
15030.444994262 0.862917730649247
15045.305161676 0.861290731247743
15075.025496504 0.858362966840398
15104.745831332 0.855848427501901
15122.578032229 0.854530319597397
15143.976673305 0.853171158469834
15165.375314381 0.852026679346191
15208.172596533 0.850295328876459
};
\addlegendentry{$60\%$}
\addplot [semithick, tomato23610466]
table {%
0.5 1
1.9142135624 0.99892783946726
7.5710678119 0.996952569522957
13.2279220614 0.99570023530603
13.93502884259 0.995575032152343
19.59188309209 0.994731533424299
25.24873734159 0.994066651359483
57.24873734159 0.991540075048108
185.24873734159 0.986915422435892
313.24873734159 0.984524459309004
825.24873734159 0.978498492721653
1439.64873734159 0.968649482875417
1808.28873734159 0.959551992875856
2073.70953734159 0.952711185458535
2235.86578962159 0.948357647364237
2398.02204190159 0.941792665038749
2535.61618462159 0.935352075745296
2565.87843304759 0.933693879631183
2596.14068147359 0.9307214792634
2656.66517832559 0.925527585392238
2759.37865548559 0.910653378302727
2796.35550726459 0.902496281257504
2822.97884054559 0.894297434536988
2826.49213802939 0.89311208580651
2830.00543551319 0.891791268733489
2838.43734947429 0.88866343309716
2845.72252313669 0.885979686478718
2853.00769679909 0.883293291885378
2882.14839144809 0.871019936783676
2887.48935850859 0.868569226513552
2892.83032556909 0.865951389446389
2905.64864651409 0.860144009377614
2927.40205850709 0.852837378303049
2949.15547050009 0.848380442475402
3036.16911847209 0.842498618133181
};
\addlegendentry{$70\%$}
\addplot [semithick, firebrick165052]
table {%
0.5 1
1.5 0.998691545327241
5.5 0.996232170776556
9.5 0.994829456051562
25.5 0.991504788210394
89.5 0.985608244701361
153.5 0.982061780545541
186.6776 0.98042996971859
215.3430464 0.978904733761956
244.0084928 0.976041225358949
272.6739392 0.973496292159325
287.0066624 0.972395425371086
301.3393856 0.971318026229222
382.417511741 0.964743319428162
431.95140312 0.960489078107706
461.671737948 0.957103723106146
487.350107239 0.954233815011579
524.326959018 0.948770081757448
561.303810797 0.941884798927073
598.954889798 0.934328391008333
637.292489722 0.921700694134146
651.094025695 0.915180405591193
659.5259396561 0.909988736132092
664.6773353549 0.9060478057358
669.8287310537 0.897550925961328
670.9414325246 0.895721931476559
672.0744227326 0.893885737631
673.2074129406 0.891999340233874
677.7393737725 0.885403349264657
679.6971808519 0.882567714863184
680.8718650995 0.88071729007314
681.23723888788 0.879989107495204
681.285454744167 0.879863317313595
681.295869369125 0.879782408868287
681.302232082794 0.879773065084687
681.30424722694 0.879769328912931
681.85969389137 0.863819540723416
681.99855555748 0.863350816835876
682.55400222191 0.862085473875716
683.10944888634 0.861009869201414
685.33123554404 0.857866667734256
694.21838217494 0.851115883003871
703.10552880584 0.848424146478206
753.37882199084 0.843268168712847
};
\addlegendentry{$80\%$}
\addplot [semithick, darkturquoise0166214, dash pattern=on 1.5pt off 2.475pt, forget plot]
table {%
0.5 1
2.5 0.999262413756691
13.813708499 0.997694637700995
25.127416998 0.996933843511194
89.127416998 0.994327312661507
345.127416998 0.989519899501343
601.127416998 0.986808749256189
1625.127416998 0.980598000969487
5721.127416998 0.969713918984333
7195.687416998 0.96739544717385
8997.423553498 0.963680841921228
11545.463233498 0.958552060989469
14093.502913498 0.954165092452831
21300.447459298 0.943076328780254
26489.447532298 0.934360048938748
29602.847576098 0.929334550314213
32292.825213898 0.923328207474782
32411.706553208 0.923080055098807
32530.587892518 0.922832255620689
33481.638606998 0.92086116719509
36709.611772398 0.913421940891581
38646.395671598 0.907556886735787
39650.424444998 0.903378230016429
40152.438831688 0.900607025632987
40654.453218378 0.897875223755893
42358.346548378 0.883737657617219
44403.018544378 0.853730912516186
};
\addplot [semithick, darkcyan0155119, dash pattern=on 1.5pt off 2.475pt, forget plot]
table {%
0.5 1
1.9142135624 0.999090368428007
9.9142135624 0.996830386983353
17.9142135624 0.995714135626603
63.1690475584 0.991761642627755
191.1690475584 0.986551297026394
319.1690475584 0.983682964618697
1043.2463914984 0.974668159651984
1780.5263914984 0.96803356928536
2517.8063914984 0.961512047838354
3143.4092166584 0.956813392819796
4644.8559970584 0.943718117176869
6146.3027774584 0.930167143005983
7227.3444593584 0.917958354167252
7507.5504632984 0.91379855688287
7675.6740656684 0.910679068862149
8348.1684751284 0.845697886312526
8369.1839254244 0.844499183534876
8390.1993757204 0.843470826743773
};
\addplot [semithick, tomato23610466, dash pattern=on 1.5pt off 2.475pt, forget plot]
table {%
0.5 1
1.9142135624 0.998564513747121
7.5710678119 0.99566526993196
13.2279220614 0.993962637288375
35.8553390594 0.98997838839181
126.3650070514 0.982866592841155
216.8746750434 0.979240214123018
578.9133470134 0.96595125878347
610.9133470134 0.964824848351976
642.9133470134 0.963750481340044
898.9133470134 0.953986621268263
1079.9326829934 0.947279777493144
1260.9520189734 0.941059003838092
1521.6198627934 0.931995790414672
1617.1713507934 0.928285833199692
1674.5022435934 0.925789415502138
1757.0587292254 0.920003783482982
1787.3209776514 0.917800685629859
1805.4783267064 0.916353545393763
1842.4551784854 0.912475713444761
1905.2069768214 0.903752370168592
1942.8580558224 0.89611290109407
1969.9668327034 0.88746769640997
2046.6420325524 0.856819121725203
2123.3172324014 0.84213351833255
};
\addplot [semithick, firebrick165052, dash pattern=on 1.5pt off 2.475pt, forget plot]
table {%
0.5 1
1.5 0.998349106102975
5.5 0.995027891465847
9.5 0.993201147590516
32.127416998 0.987207514240203
86.433217793 0.980323568985632
119.01669827 0.977435918823029
197.217051415 0.9688017356447
216.767139701 0.966658341434081
236.317227987 0.964673839960157
302.672427987 0.957737653344464
396.512851761 0.945757520203513
430.29540432 0.941215282509468
442.678877165 0.939128391190882
443.22615451645 0.939028185646627
443.7734318679 0.938926925217164
449.9651682903 0.937744781910623
484.9909187833 0.93012530359514
520.0166692763 0.921751107737347
541.4153103523 0.915321712910964
563.2041292193 0.906168458072204
576.2774205393 0.897802090815158
589.3507118593 0.886008350088739
600.4437673933 0.8551295022601
611.5368229273 0.845338775950925
};
\end{axis}

\end{tikzpicture}

%% file: figures/ply-level/S-N-locR-globR.tex
\begin{tikzpicture}

\definecolor{crimson2246049}{RGB}{224,60,49}
\definecolor{darkcyan0118194}{RGB}{0,118,194}
\definecolor{darkcyan0155119}{RGB}{0,155,119}
\definecolor{darkgray176}{RGB}{176,176,176}
\definecolor{gray}{RGB}{128,128,128}

\begin{axis}[
height=6cm,
legend cell align={left},
legend style={fill opacity=0.8, draw opacity=1, text opacity=1, at={(1,1)}, draw=none},
log basis x={10},
minor xtick={},
minor ytick={},
tick align=outside,
tick pos=left,
width=8cm,
x grid style={darkgray176},
xlabel={\(\displaystyle N\) [-]},
xmin=1, xmax=100000000,
xmode=log,
xtick style={color=black},
xtick={0.01,1,100,10000,1000000,100000000,10000000000},
xticklabels={
  \(\displaystyle {10^{-2}}\),
  \(\displaystyle {10^{0}}\),
  \(\displaystyle {10^{2}}\),
  \(\displaystyle {10^{4}}\),
  \(\displaystyle {10^{6}}\),
  \(\displaystyle {10^{8}}\),
  \(\displaystyle {10^{10}}\)
},
y grid style={darkgray176},
ylabel={stress [MPa]},
ymin=150, ymax=470,
ytick style={color=black},
ytick={150,200,250,300,350,400,450,500},
yticklabels={
  \(\displaystyle {150}\),
  \(\displaystyle {200}\),
  \(\displaystyle {250}\),
  \(\displaystyle {300}\),
  \(\displaystyle {350}\),
  \(\displaystyle {400}\),
  \(\displaystyle {450}\),
  \(\displaystyle {500}\)
}
]
\addplot [semithick, crimson2246049, mark=*, mark size=1.25, mark options={solid}, only marks]
table {%
79722.6492655018 209
200578.652799832 209
277256.827517136 209
347776.73606127 209
383247.904659031 209
8679.37747021351 251
11804.7349131027 251
13008.748002353 251
15048.8874339328 251
16583.7848782625 251
17981.7949256128 251
20467.842020279 251
27838.1082193806 251
36656.2156999394 251
1166.22898574719 293
1224.25944214336 293
1665.10308244606 293
2455.60309582591 293
2750.21920888487 293
3130.44679474198 293
3449.7338387812 293
5516.3526382235 293
7625.17052832109 293
220.143822243671 334
234.869168649772 334
246.556037366555 334
741.220250458929 334
765.608999639563 334
991.939218778981 334
};
\addlegendentry{experiments}
\addplot [line width=1pt, darkcyan0118194, mark=square, mark size=1.25, mark options={solid,fill opacity=0}]
table {%
125570.948470609 209
14854.0619833096 250.8
2949.15547050009 292.6
703.10552880584 334.4
};
\addlegendentry{\emph{local} $R$}
\addplot [line width=1pt, darkcyan0155119, mark=square, mark size=1.25, mark options={solid,fill opacity=0}]
table {%
46447.690540378 209
8348.1684751284 250.8
2123.3172324014 292.6
611.5368229273 334.4
};
\addlegendentry{\emph{global} $R$}
\draw (axis cs:8,330) node[
  scale=0.9,
  anchor=base west,
  text=gray,
  rotate=0.0
]{$80\%$};
\draw (axis cs:40,280) node[
  scale=0.9,
  anchor=base west,
  text=gray,
  rotate=0.0
]{$70\%$};
\draw (axis cs:300,240) node[
  scale=0.9,
  anchor=base west,
  text=gray,
  rotate=0.0
]{$60\%$};
\draw (axis cs:2000,200) node[
  scale=0.9,
  anchor=base west,
  text=gray,
  rotate=0.0
]{$50\%$};
\end{axis}

\end{tikzpicture}

%% file: figures/ply-level/sens-matrix_g1-S-N.tex
\begin{tikzpicture}

\definecolor{crimson2246049}{RGB}{224,60,49}
\definecolor{darkcyan0118194}{RGB}{0,118,194}
\definecolor{darkgray176}{RGB}{176,176,176}
\definecolor{orange25518428}{RGB}{255,184,28}

\begin{axis}[
height=6cm,
legend cell align={left},
legend style={fill opacity=0.8, draw opacity=1, text opacity=1, at={(1,1)}, draw=none},
log basis x={10},
minor xtick={},
minor ytick={},
tick align=outside,
tick pos=left,
width=8cm,
x grid style={darkgray176},
xmin=1, xmax=100000000,
xmode=log,
xtick style={color=black},
xtick={0.01,1,100,10000,1000000,100000000,10000000000},
xticklabels={
  \(\displaystyle {10^{-2}}\),
  \(\displaystyle {10^{0}}\),
  \(\displaystyle {10^{2}}\),
  \(\displaystyle {10^{4}}\),
  \(\displaystyle {10^{6}}\),
  \(\displaystyle {10^{8}}\),
  \(\displaystyle {10^{10}}\)
},
y grid style={darkgray176},
ylabel={stress [MPa]},
ymin=150, ymax=470,
ytick style={color=black},
ytick={150,200,250,300,350,400,450,500},
yticklabels={
  \(\displaystyle {150}\),
  \(\displaystyle {200}\),
  \(\displaystyle {250}\),
  \(\displaystyle {300}\),
  \(\displaystyle {350}\),
  \(\displaystyle {400}\),
  \(\displaystyle {450}\),
  \(\displaystyle {500}\)
}
]
\addplot [semithick, crimson2246049, mark=*, mark size=1.25, mark options={solid}, only marks, forget plot]
table {%
79722.6492655018 209
200578.652799832 209
277256.827517136 209
347776.73606127 209
383247.904659031 209
8679.37747021351 251
11804.7349131027 251
13008.748002353 251
15048.8874339328 251
16583.7848782625 251
17981.7949256128 251
20467.842020279 251
27838.1082193806 251
36656.2156999394 251
1166.22898574719 293
1224.25944214336 293
1665.10308244606 293
2455.60309582591 293
2750.21920888487 293
3130.44679474198 293
3449.7338387812 293
5516.3526382235 293
7625.17052832109 293
220.143822243671 334
234.869168649772 334
246.556037366555 334
741.220250458929 334
765.608999639563 334
991.939218778981 334
};
\addplot [line width=1pt, black, mark=square, mark size=1.25, mark options={solid,fill opacity=0}]
table {%
125570.948470609 209
14854.0619833096 250.8
2949.15547050009 292.6
703.10552880584 334.4
};
\addlegendentry{intra $G_{Ic}$=1.0 Nmm$^{-1}$}
\addplot [line width=1pt, darkcyan0118194, mark=diamond, mark size=1.25, mark options={solid,fill opacity=0}]
table {%
89017.46108553 209
12401.88280624 250.8
2463.85297273833 292.6
};
\addlegendentry{intra $G_{Ic}$=0.75 Nmm$^{-1}$}
\addplot [line width=1pt, orange25518428, mark=o, mark size=1.25, mark options={solid,fill opacity=0}]
table {%
65020.718640314 209
8691.83351267332 250.8
1795.397854351 292.6
515.212915897115 334.4
};
\addlegendentry{intra $G_{Ic}$=0.5 Nmm$^{-1}$}
\end{axis}

\end{tikzpicture}

%% file: figures/ply-level/sens-matrix_f2t-S-N.tex
\begin{tikzpicture}

\definecolor{crimson2246049}{RGB}{224,60,49}
\definecolor{darkcyan0118194}{RGB}{0,118,194}
\definecolor{darkgray176}{RGB}{176,176,176}
\definecolor{orange25518428}{RGB}{255,184,28}

\begin{axis}[
height=6cm,
legend cell align={left},
legend style={fill opacity=0.8, draw opacity=1, text opacity=1, at={(1,1)}, draw=none},
log basis x={10},
minor xtick={},
minor ytick={},
tick align=outside,
tick pos=left,
width=8cm,
x grid style={darkgray176},
xmin=1, xmax=100000000,
xmode=log,
xtick style={color=black},
xtick={0.01,1,100,10000,1000000,100000000,10000000000},
xticklabels={
  \(\displaystyle {10^{-2}}\),
  \(\displaystyle {10^{0}}\),
  \(\displaystyle {10^{2}}\),
  \(\displaystyle {10^{4}}\),
  \(\displaystyle {10^{6}}\),
  \(\displaystyle {10^{8}}\),
  \(\displaystyle {10^{10}}\)
},
ymin=150, ymax=470,
ytick=\empty
]
\addplot [semithick, crimson2246049, mark=*, mark size=1.25, mark options={solid}, only marks, forget plot]
table {%
79722.6492655018 209
200578.652799832 209
277256.827517136 209
347776.73606127 209
383247.904659031 209
8679.37747021351 251
11804.7349131027 251
13008.748002353 251
15048.8874339328 251
16583.7848782625 251
17981.7949256128 251
20467.842020279 251
27838.1082193806 251
36656.2156999394 251
1166.22898574719 293
1224.25944214336 293
1665.10308244606 293
2455.60309582591 293
2750.21920888487 293
3130.44679474198 293
3449.7338387812 293
5516.3526382235 293
7625.17052832109 293
220.143822243671 334
234.869168649772 334
246.556037366555 334
741.220250458929 334
765.608999639563 334
991.939218778981 334
};
\addplot [line width=1pt, black, mark=square, mark size=1.25, mark options={solid,fill opacity=0}]
table {%
125570.948470609 209
14854.0619833096 250.8
2949.15547050009 292.6
703.10552880584 334.4
};
\addlegendentry{intra $f_{n}$=95 MPa}
\addplot [line width=1pt, darkcyan0118194, mark=diamond, mark size=1.25, mark options={solid,fill opacity=0}]
table {%
201605.755086166 209
26241.46812405 250.8
4875.9687432019 292.6
1372.85742585873 334.4
};
\addlegendentry{intra $f_{n}$=75 MPa}
\addplot [line width=1pt, orange25518428, mark=o, mark size=1.25, mark options={solid,fill opacity=0}]
table {%
316317.923426824 209
37611.237065991 250.8
7373.0923857178 292.6
2021.810499847 334.4
};
\addlegendentry{intra $f_{n}$=60 MPa}
\end{axis}

\end{tikzpicture}

%% file: figures/ply-level/sens-matrix_g2-S-N.tex
\begin{tikzpicture}

\definecolor{crimson2246049}{RGB}{224,60,49}
\definecolor{darkcyan0118194}{RGB}{0,118,194}
\definecolor{darkgray176}{RGB}{176,176,176}
\definecolor{orange25518428}{RGB}{255,184,28}

\begin{axis}[
height=6cm,
legend cell align={left},
legend style={fill opacity=0.8, draw opacity=1, text opacity=1, at={(1,1)}, draw=none},
log basis x={10},
minor xtick={},
minor ytick={},
tick align=outside,
tick pos=left,
width=8cm,
x grid style={darkgray176},
xmin=1, xmax=100000000,
xmode=log,
xtick style={color=black},
xtick={0.01,1,100,10000,1000000,100000000,10000000000},
xticklabels={
  \(\displaystyle {10^{-2}}\),
  \(\displaystyle {10^{0}}\),
  \(\displaystyle {10^{2}}\),
  \(\displaystyle {10^{4}}\),
  \(\displaystyle {10^{6}}\),
  \(\displaystyle {10^{8}}\),
  \(\displaystyle {10^{10}}\)
},
y grid style={darkgray176},
ylabel={stress [MPa]},
ymin=150, ymax=470,
ytick style={color=black},
ytick={150,200,250,300,350,400,450,500},
yticklabels={
  \(\displaystyle {150}\),
  \(\displaystyle {200}\),
  \(\displaystyle {250}\),
  \(\displaystyle {300}\),
  \(\displaystyle {350}\),
  \(\displaystyle {400}\),
  \(\displaystyle {450}\),
  \(\displaystyle {500}\)
}
]
\addplot [semithick, crimson2246049, mark=*, mark size=1.25, mark options={solid}, only marks, forget plot]
table {%
79722.6492655018 209
200578.652799832 209
277256.827517136 209
347776.73606127 209
383247.904659031 209
8679.37747021351 251
11804.7349131027 251
13008.748002353 251
15048.8874339328 251
16583.7848782625 251
17981.7949256128 251
20467.842020279 251
27838.1082193806 251
36656.2156999394 251
1166.22898574719 293
1224.25944214336 293
1665.10308244606 293
2455.60309582591 293
2750.21920888487 293
3130.44679474198 293
3449.7338387812 293
5516.3526382235 293
7625.17052832109 293
220.143822243671 334
234.869168649772 334
246.556037366555 334
741.220250458929 334
765.608999639563 334
991.939218778981 334
};
\addplot [line width=1pt, black, mark=square, mark size=1.25, mark options={solid,fill opacity=0}]
table {%
125570.948470609 209
14854.0619833096 250.8
2949.15547050009 292.6
703.10552880584 334.4
};
\addlegendentry{intra $G_{IIc}$=1.0 Nmm$^{-1}$}
\addplot [line width=1pt, darkcyan0118194, mark=diamond, mark size=1.25, mark options={solid,fill opacity=0}]
table {%
161654.446222803 209
21824.5094171744 250.8
3883.3443272855 292.6
900.99120709938 334.4
};
\addlegendentry{intra $G_{IIc}$=0.75 Nmm$^{-1}$}
\addplot [line width=1pt, orange25518428, mark=o, mark size=1.25, mark options={solid,fill opacity=0}]
table {%
163555.500163666 209
19312.5545375554 250.8
3467.42551561321 292.6
897.398292279732 334.4
};
\addlegendentry{intra $G_{IIc}$=0.5 Nmm$^{-1}$}
\end{axis}

\end{tikzpicture}

%% file: figures/ply-level/sens-matrix_f6-S-N.tex
\begin{tikzpicture}

\definecolor{crimson2246049}{RGB}{224,60,49}
\definecolor{darkcyan0118194}{RGB}{0,118,194}
\definecolor{darkgray176}{RGB}{176,176,176}
\definecolor{orange25518428}{RGB}{255,184,28}

\begin{axis}[
height=6cm,
legend cell align={left},
legend style={fill opacity=0.8, draw opacity=1, text opacity=1, at={(1,1)}, draw=none},
log basis x={10},
minor xtick={},
minor ytick={},
tick align=outside,
tick pos=left,
width=8cm,
x grid style={darkgray176},
xmin=1, xmax=100000000,
xmode=log,
xtick style={color=black},
xtick={0.01,1,100,10000,1000000,100000000,10000000000},
xticklabels={
  \(\displaystyle {10^{-2}}\),
  \(\displaystyle {10^{0}}\),
  \(\displaystyle {10^{2}}\),
  \(\displaystyle {10^{4}}\),
  \(\displaystyle {10^{6}}\),
  \(\displaystyle {10^{8}}\),
  \(\displaystyle {10^{10}}\)
},
ymin=150, ymax=470,
ytick=\empty
]
\addplot [semithick, crimson2246049, mark=*, mark size=1.25, mark options={solid}, only marks, forget plot]
table {%
79722.6492655018 209
200578.652799832 209
277256.827517136 209
347776.73606127 209
383247.904659031 209
8679.37747021351 251
11804.7349131027 251
13008.748002353 251
15048.8874339328 251
16583.7848782625 251
17981.7949256128 251
20467.842020279 251
27838.1082193806 251
36656.2156999394 251
1166.22898574719 293
1224.25944214336 293
1665.10308244606 293
2455.60309582591 293
2750.21920888487 293
3130.44679474198 293
3449.7338387812 293
5516.3526382235 293
7625.17052832109 293
220.143822243671 334
234.869168649772 334
246.556037366555 334
741.220250458929 334
765.608999639563 334
991.939218778981 334
};
\addplot [line width=1pt, black, mark=square, mark size=1.25, mark options={solid,fill opacity=0}]
table {%
125570.948470609 209
14854.0619833096 250.8
2949.15547050009 292.6
703.10552880584 334.4
};
\addlegendentry{intra $f_{s}$=107 MPa}
\addplot [line width=1pt, darkcyan0118194, mark=diamond, mark size=1.25, mark options={solid,fill opacity=0}]
table {%
149308.264044376 209
20260.482493462 250.8
3992.3828995465 292.6
};
\addlegendentry{intra $f_{s}$=90 MPa}
\addplot [line width=1pt, orange25518428, mark=o, mark size=1.25, mark options={solid,fill opacity=0}]
table {%
117109.970550928 209
15743.591120298 250.8
2977.411173849 292.6
754.18756818692 334.4
};
\addlegendentry{intra $f_{s}$=75 MPa}
\end{axis}

\end{tikzpicture}

%% file: figures/ply-level/sens-interface_g2-S-N.tex
\begin{tikzpicture}

\definecolor{crimson2246049}{RGB}{224,60,49}
\definecolor{darkcyan0118194}{RGB}{0,118,194}
\definecolor{darkgray176}{RGB}{176,176,176}
\definecolor{orange25518428}{RGB}{255,184,28}

\begin{axis}[
height=6cm,
legend cell align={left},
legend style={fill opacity=0.8, draw opacity=1, text opacity=1, at={(1,1)}, draw=none},
log basis x={10},
minor xtick={},
minor ytick={},
tick align=outside,
tick pos=left,
width=8cm,
x grid style={darkgray176},
xlabel={\(\displaystyle N\) [-]},
xmin=1, xmax=100000000,
xmode=log,
xtick style={color=black},
xtick={0.01,1,100,10000,1000000,100000000,10000000000},
xticklabels={
  \(\displaystyle {10^{-2}}\),
  \(\displaystyle {10^{0}}\),
  \(\displaystyle {10^{2}}\),
  \(\displaystyle {10^{4}}\),
  \(\displaystyle {10^{6}}\),
  \(\displaystyle {10^{8}}\),
  \(\displaystyle {10^{10}}\)
},
y grid style={darkgray176},
ylabel={stress [MPa]},
ymin=150, ymax=470,
ytick style={color=black},
ytick={150,200,250,300,350,400,450,500},
yticklabels={
  \(\displaystyle {150}\),
  \(\displaystyle {200}\),
  \(\displaystyle {250}\),
  \(\displaystyle {300}\),
  \(\displaystyle {350}\),
  \(\displaystyle {400}\),
  \(\displaystyle {450}\),
  \(\displaystyle {500}\)
}
]
\addplot [semithick, crimson2246049, mark=*, mark size=1.25, mark options={solid}, only marks, forget plot]
table {%
79722.6492655018 209
200578.652799832 209
277256.827517136 209
347776.73606127 209
383247.904659031 209
8679.37747021351 251
11804.7349131027 251
13008.748002353 251
15048.8874339328 251
16583.7848782625 251
17981.7949256128 251
20467.842020279 251
27838.1082193806 251
36656.2156999394 251
1166.22898574719 293
1224.25944214336 293
1665.10308244606 293
2455.60309582591 293
2750.21920888487 293
3130.44679474198 293
3449.7338387812 293
5516.3526382235 293
7625.17052832109 293
220.143822243671 334
234.869168649772 334
246.556037366555 334
741.220250458929 334
765.608999639563 334
991.939218778981 334
};
\addplot [line width=1pt, black, mark=square, mark size=1.25, mark options={solid,fill opacity=0}]
table {%
125570.948470609 209
14854.0619833096 250.8
2949.15547050009 292.6
703.10552880584 334.4
};
\addlegendentry{inter $G_{IIc}$=1.0 Nmm$^{-1}$}
\addplot [line width=1pt, darkcyan0118194, mark=diamond, mark size=1.25, mark options={solid,fill opacity=0}]
table {%
73948.850384633 209
8799.684070351 250.8
};
\addlegendentry{inter $G_{IIc}$=0.75 Nmm$^{-1}$}
\addplot [line width=1pt, orange25518428, mark=o, mark size=1.25, mark options={solid,fill opacity=0}]
table {%
40463.3528925048 209
};
\addlegendentry{inter $G_{IIc}$=0.5 Nmm$^{-1}$}
\end{axis}

\end{tikzpicture}

%% file: figures/ply-level/sens-interface_f6-S-N.tex
\begin{tikzpicture}

\definecolor{crimson2246049}{RGB}{224,60,49}
\definecolor{darkcyan0118194}{RGB}{0,118,194}
\definecolor{darkgray176}{RGB}{176,176,176}
\definecolor{orange25518428}{RGB}{255,184,28}

\begin{axis}[
height=6cm,
legend cell align={left},
legend style={fill opacity=0.8, draw opacity=1, text opacity=1, at={(1,1)}, draw=none},
log basis x={10},
minor xtick={},
minor ytick={},
tick align=outside,
tick pos=left,
width=8cm,
x grid style={darkgray176},
xlabel={\(\displaystyle N\) [-]},
xmin=1, xmax=100000000,
xmode=log,
xtick style={color=black},
xtick={0.01,1,100,10000,1000000,100000000,10000000000},
xticklabels={
  \(\displaystyle {10^{-2}}\),
  \(\displaystyle {10^{0}}\),
  \(\displaystyle {10^{2}}\),
  \(\displaystyle {10^{4}}\),
  \(\displaystyle {10^{6}}\),
  \(\displaystyle {10^{8}}\),
  \(\displaystyle {10^{10}}\)
},
ymin=150, ymax=470,
ytick=\empty
]
\addplot [semithick, crimson2246049, mark=*, mark size=1.25, mark options={solid}, only marks, forget plot]
table {%
79722.6492655018 209
200578.652799832 209
277256.827517136 209
347776.73606127 209
383247.904659031 209
8679.37747021351 251
11804.7349131027 251
13008.748002353 251
15048.8874339328 251
16583.7848782625 251
17981.7949256128 251
20467.842020279 251
27838.1082193806 251
36656.2156999394 251
1166.22898574719 293
1224.25944214336 293
1665.10308244606 293
2455.60309582591 293
2750.21920888487 293
3130.44679474198 293
3449.7338387812 293
5516.3526382235 293
7625.17052832109 293
220.143822243671 334
234.869168649772 334
246.556037366555 334
741.220250458929 334
765.608999639563 334
991.939218778981 334
};
\addplot [line width=1pt, black, mark=square, mark size=1.25, mark options={solid,fill opacity=0}]
table {%
125570.948470609 209
14854.0619833096 250.8
2949.15547050009 292.6
703.10552880584 334.4
};
\addlegendentry{inter $f_{s}$=45 MPa}
\addplot [line width=1pt, orange25518428, mark=o, mark size=1.25, mark options={solid,fill opacity=0}]
table {%
1352893.06505748 209
127735.485737436 250.8
21421.4754828262 292.6
};
\addlegendentry{inter $f_{s}$=75 MPa}
\addplot [line width=1pt, darkcyan0118194, mark=diamond, mark size=1.25, mark options={solid,fill opacity=0}]
table {%
2876511.27697296 209
260039.325667157 250.8
43815.173316927 292.6
};
\addlegendentry{inter $f_{s}$=90 MPa}
\end{axis}

\end{tikzpicture}

%% file: figures/sub-laminate/S-N.tex
\begin{tikzpicture}

\definecolor{crimson2246049}{RGB}{224,60,49}
\definecolor{darkcyan0118194}{RGB}{0,118,194}
\definecolor{darkcyan0155119}{RGB}{0,155,119}
\definecolor{darkgray176}{RGB}{176,176,176}

\begin{axis}[
height=6cm,
legend cell align={left},
legend style={fill opacity=0.8, draw opacity=1, text opacity=1, at={(1,1)}, draw=none},
log basis x={10},
minor xtick={},
minor ytick={},
tick align=outside,
tick pos=left,
width=8cm,
x grid style={darkgray176},
xlabel={\(\displaystyle N\) [-]},
xmin=1, xmax=100000000,
xmode=log,
xtick style={color=black},
xtick={0.01,1,100,10000,1000000,100000000,10000000000},
xticklabels={
  \(\displaystyle {10^{-2}}\),
  \(\displaystyle {10^{0}}\),
  \(\displaystyle {10^{2}}\),
  \(\displaystyle {10^{4}}\),
  \(\displaystyle {10^{6}}\),
  \(\displaystyle {10^{8}}\),
  \(\displaystyle {10^{10}}\)
},
y grid style={darkgray176},
ylabel={stress [MPa]},
ymin=350, ymax=650,
ytick style={color=black},
ytick={350,400,450,500,550,600,650},
yticklabels={
  \(\displaystyle {350}\),
  \(\displaystyle {400}\),
  \(\displaystyle {450}\),
  \(\displaystyle {500}\),
  \(\displaystyle {550}\),
  \(\displaystyle {600}\),
  \(\displaystyle {650}\)
}
]
\addplot [forget plot, semithick, crimson2246049, mark=square, mark size=1.25, mark options={solid,fill opacity=0}, only marks]
table {%
297.635144163131 523.3
688.74217883767 523.3
1182.69971981976 523.3
1451.92211356594 494.2
6333.14502484398 494.2
};
\addplot [forget plot,semithick, crimson2246049, mark=*, mark size=1.25, mark options={solid}, only marks]
table {%
14384.4988828766 494.2
32671.5727019967 494.2
8534.42151923193 465.1
13104.1515441274 465.1
16087.0989377905 465.1
17991.2247441816 465.1
21678.6577313655 465.1
25639.3424249546 465.1
28144.4465311249 465.1
81457.5181939366 407
92813.539706044 407
101881.930602041 407
120495.730777731 407
178242.861525581 407
185014.80305063 377.9
227130.419822196 377.9
323684.167795652 377.9
822204.325259919 377.9
587801.607227491 377.9
};
\addplot [semithick, darkcyan0155119, mark=asterisk, mark size=4, mark options={solid,fill opacity=0}, only marks]
table {%
826.913947983772 523.3
8329.80664765827 494.2
21861.6299963404 465.1
119176.458663437 407
233482.302915653 377.9
};
\addlegendentry{exp. mean}
\addplot [line width=1pt, darkcyan0118194, mark=square, mark size=1.25, mark options={solid,fill opacity=0}]
table {%
195040.091571689 377.9
82575.651318218 407
22701.2077393304 465.09375
11633.4908583201 494.175
6743.4579339426 523.2975
};
\addlegendentry{simulation}
\end{axis}

\end{tikzpicture}

%% file: sections/conclusion.tex
\section{Conclusion}

A previously developed progressive fatigue failure framework has been extended with an adaptive cycle jump approach to capture non-uniform local stress ratios in multidirectional laminates with thermal residual stresses. 
The local stress ratio is regularly computed by explicitly applying a load cycle to assess the minimum and maximum stress in every integration point, before cycle jumps take place.

The progressive fatigue failure framework is applied to the simulation of two open-hole quasi-isotropic laminates. The model is capable of predicting the fatigue life as a consequence of interacting intra- and inter-laminar damage with excellent agreement. Furthermore, the experimentally observed damage evolution and final failure modes are accurately captured. Moreover, the implicit fatigue damage update, together with the adaptive stepping scheme, allows for simulating a large amount of fatigue cycles while accounting for local stress ratios in an efficient manner.
In addition, it is demonstrated that computing the local stress ratios, instead of just using the global load ratio, is relevant for this type of problems.

The numerical model requires only static material properties and a few fatigue-related parameters, calibrated on elementary fracture tests. 
The effect of local stress ratio and mode-mixity is internally accounted for in the constitutive relations of the fatigue cohesive zone model, which poses a significant advantage over Paris-type models that require Paris data for different mode-mixities and stress ratios and separate S-N curves for crack initiation. 
However, the progressive fatigue failure model shows sensitivity to the static material properties, in particular to the inter-laminar strength. Previously it has been shown, with the static version of the open-hole tests, that fracture energy is a more important parameter than strength. Conversely, the present investigation indicates that static strength is an important parameter in the fatigue simulations with the embedded fatigue cohesive zone model. 

The progressive fatigue failure framework has been validated for the case of multidirectional laminates with complex failure processes and thermal residual stresses, demonstrating an important step towards efficient virtual testing of composite structural elements under high-cycle fatigue loading.